\documentclass[twocolumn,pre,tightenlines,superscriptaddress,longbibliography]{revtex4-1}
\usepackage{hyperref}
\usepackage[T1]{fontenc}
\usepackage[utf8]{inputenc}
\usepackage{graphicx}
\usepackage{float}
\usepackage{units}
\usepackage{amsmath}
\usepackage{esint}
\usepackage{epsfig}

\usepackage[caption=false]{subfig}
\makeatother

\begin{document}
\title{Instantaneous frequencies in the Kuramoto model}
\author{Julio D. da Fonseca}
\affiliation{Departamento de F\'isica, Universidade Estadual Paulista, Bela Vista,
13506-900 Rio Claro, SP, Brazil}
\author{Edson D. Leonel}
\affiliation{Departamento de F\'isica, Universidade Estadual Paulista, Bela Vista,
13506-900 Rio Claro, SP, Brazil}
\author{Hugues Chaté}
\affiliation{Service de Physique de l'Etat Condens\'e, CEA, CNRS Universit\'e Paris-Saclay, CEA-Saclay, 91191 Gif-sur-Yvette, France}
\affiliation{Computational Science Research Center, Beijing 100094, China}
\affiliation{Sorbonne Universit\'e, CNRS, Laboratoire de Physique Th\'eorique de la Mati\`ere Condens\'ee, 75005 Paris, France}

\date{\today}
\begin{abstract}
Using the main results of the Kuramoto theory of globally coupled phase oscillators combined
with methods from probability and generalized function theory in a geometric analysis, 
we extend Kuramoto's results and obtain a mathematical description of the
instantaneous frequency (phase-velocity) distribution.
Our result is validated against numerical simulations, and we illustrate it in cases where the natural
frequencies have normal and Beta distributions. In both cases, we
vary the coupling strength and compare systematically the distribution
of time-averaged frequencies (a known result of Kuramoto theory) to that of instantaneous
frequencies, focussing on their qualitative differences near the synchronized frequency and in their tails.  
For a class of natural frequency distributions with power-law tails, which
includes the Cauchy-Lorentz distribution, we analyze rare events by means of
an asymptotic formula obtained from a power series expansion of the instantaneous frequency distribution. 
\end{abstract}
\maketitle

\section{Introduction}

In large ensembles of interacting oscillatory units, order emerges
when a group starts showing the same frequency. This
frequency adjustment, called synchronization \citep{Pikovsky}, is
a crucial and ubiquitous phenomenon in many areas of science, such as
neurosciences \citep{Breakspear10},
semiconductor laser arrays \citep{Kozyreff}, cardiac pacemaker cells \citep{Winfree80},
power grids \citep{Motter13}, cell metabolism \citep{Bier},
Josephson junction arrays \citep{Wiesenfeld}, and chemical oscillators \citep{Kiss02}.

Arthur Winfree was the first to present a theoretical description
of synchronization in a model of biological oscillators \citep{Winfree67}.
Influenced by Winfree's work, Yoshiki Kuramoto introduced his famous minimal model of coupled
oscillators \citep{KuramotoA} and its mathematical analysis \citep{KuramotoB,KuramotoC,KuramotoD,KuramotoE}.
Together these seminal works occupy a proeminent place in synchronization research.
Since Kuramoto's early works, a large number of Kuramoto-like models
appeared in later studies discussing effects such as those resulting
from phase shifts in the coupling function \citep{Sakaguchi}, noise \citep{Sakaguchi88},
periodic external fields \citep{Aguiar19}, higher modes \citep{Pikovsky14},
finite size \citep{Chate15}, and complex coupling networks \citep{Rodrigues}.

The Kuramoto model consists in an infinitely large ensemble of oscillators coupled globally. 
The oscillators are reduced to their phase, they are characterized by their individual natural frequency,
and their dynamics is first-order in time. The remarkable discovery of Kuramoto is that when coupled 
strongly-enough, the oscillators can overcome their nominal frequency quenched disorder and synchronize.

Kuramoto showed that his model exhibits a transition
between an incoherent state, where instantaneous frequencies are completely
desynchronized, and a partially synchronized state, in which some
oscillators share the same instantaneous frequency, and are thus phase locked. 
The theoretical framework developed by Kuramoto to analyze his model,
hereafter "Kuramoto theory", comprises a set of assumptions and analytical
results describing stationary collective states \citep{daFonseca18}.
Kuramoto assumed that these states would be characterized by phase
distributions 
\footnote{For the sake of simplicity, here we always use the terms ``distribution''
or ``PDF'' when referring to a probability density function.} with stationary profiles. 
These profiles might be uniform (incoherent state)
or be a steadily rotating traveling wave profile (synchronized state).

The most common characterization of synchronization in the Kuramoto model is, however,
not in terms of phase distributions but simply in terms of a scalar order parameter 
(which is zero in
the incoherent state and takes finite values when oscillators synchronize). 
Here we pursue yet another, finer, description in terms of 
the {\it distribution of instantaneous frequencies}. 
Although synchronization is a direct manifestation of
the way instantaneous frequencies are distributed, this problem
was not, to our knowledge, addressed so far. Not even by Kuramoto,
who instead solved the problem of the distribution of ``coupling-modified
frequencies'' \citep{KuramotoB,KuramotoC}, i.e. instantaneous frequencies
averaged over an infinitely long time.

The main goal of this paper is to extend Kuramoto theory by presenting
a detailed derivation of the instantaneous frequency distribution
without any time-averaging procedure. Our work is essentially based
on Kuramoto results and provides a mathematical description of the
instantaneous frequency distribution in stationary states, whether
incoherent or with synchronized oscillators.

This paper is structured as follows. 
In Section~\ref{sec:Some-results-from-Kuramoto} we briefly discuss important aspects of Kuramoto
theory, including the Kuramoto model, the order parameter, phase distributions
of synchronized and desynchronized oscillators, and the distribution
of time-averaged frequencies. 
In Section~\ref{sec:G}, results from Kuramoto
theory, together with methods from probability and generalized function
theory, are used to develop a geometric analysis that solves the problem
of the instantaneous frequency distribution. 
In Section~\ref{sec:Applications},
we illustrate the properties of this distribution in two cases: in the
first one, we consider the classic case of an unbounded normal distribution of natural frequencies;
in the second, we adopt a symmetric Beta distribution of natural
frequencies, defined on a bounded interval. 
In Section~\ref{sec:Rare-events}, we obtain a power series
expansion of the instantaneous frequency distribution and an asymptotic
formula, which allow us to study rare events (large instantaneous
frequency occurrences) in cases where the natural
frequency distribution has power-law tails, e.g., Cauchy-Lorentz
distributions. Our conclusions and open problems left for further investigation
are presented in Sec.~\ref{sec:Conclusion}. In the Appendix, the
formula of the instantaneous frequency distribution is compared to
instantaneous frequency histograms obtained from numerical simulations
of the Kuramoto model.

\section{Kuramoto theory \label{sec:Some-results-from-Kuramoto}}

In this section we present some results from the theoretical framework
developed by Kuramoto to analyze his coupled oscillator model. For
details about how these results can be obtained we refer the reader
to Refs. \citep{KuramotoB,daFonseca18}.

The Kuramoto model consists of an ensemble of $N$ all-to-all coupled oscillators
with randomly distributed natural frequencies $\omega_i$ ($i=1,2,\dots,N$) whose phases $\theta_i$ evolve according to: 
\begin{equation}
\dot{\theta_{i}}=\omega_{i}+\frac{K}{N}\sum_{j=1}^{N}\sin(\theta_{j}-\theta_{i}),\label{eq:Kuramoto-model}
\end{equation}
where $K$ is the coupling strength. 
In Kuramoto theory, $N$ is assumed to be
a infinitely large number and the natural frequencies $\omega_i$ are randomly distributed
according to a given probability density function $g(\omega)$.

Collective states in the Kuramoto model are usually analyzed by using
measures which quantify the level of synchronization. One such quantity, proposed by Kuramoto, is 
\begin{equation}
R=\lim_{N\rightarrow+\infty}\left|\frac{1}{N}\sum_{j=1}^{N}\exp\left(i\theta_{j}\right)\right|,\label{eq:sigma}
\end{equation}
which we call here order parameter. If the oscillator state
is represented by $\exp\left(i\theta\right)$, $R$ is the magnitude
of a complex number representing the mean state of oscillators.
A fundamental surmise in Kuramoto's theory is that stationary collective states, 
when they exist, are characterized by time-independent values
of the order parameter reached after sufficiently long time. This
is equivalent to the assumption that the distribution of phases has a well-defined, steady profile.
In an ordered state, the synchronized oscillators adopt a common frequency denoted $\Omega$ hereafter.

In order to simplify notations, from here on we use the definitions
\begin{equation}
a=KR\label{eq:extended-order-parameter} \;\;\;\;{\rm and}\;\;\;\;
\tilde{\chi}=\frac{\chi-\Omega}{a},
\end{equation}
where $\chi$ is a generic quantity.

Kuramoto' analysis considers the use of a frame rotating with angular
velocity equal to $\Omega$ . The dynamics of an oscillator of natural frequency $\omega$ 
can then be written
\begin{equation}
\dot{\psi}=\omega-\Omega-a\sin\psi,\label{eq:inst-freq-rotating}
\end{equation}
where $\dot{\psi}=\dot{\theta}-\Omega$ and $\psi=\theta-\Omega t$
are descriptions in the rotating
frame of the oscillator's instantaneous frequency $\dot{\theta}$
and phase $\theta$.

\begin{figure*}
\begin{minipage}[t]{0.33333\linewidth}%
\subfloat[$K=0.8$]{\includegraphics[width=1\linewidth]{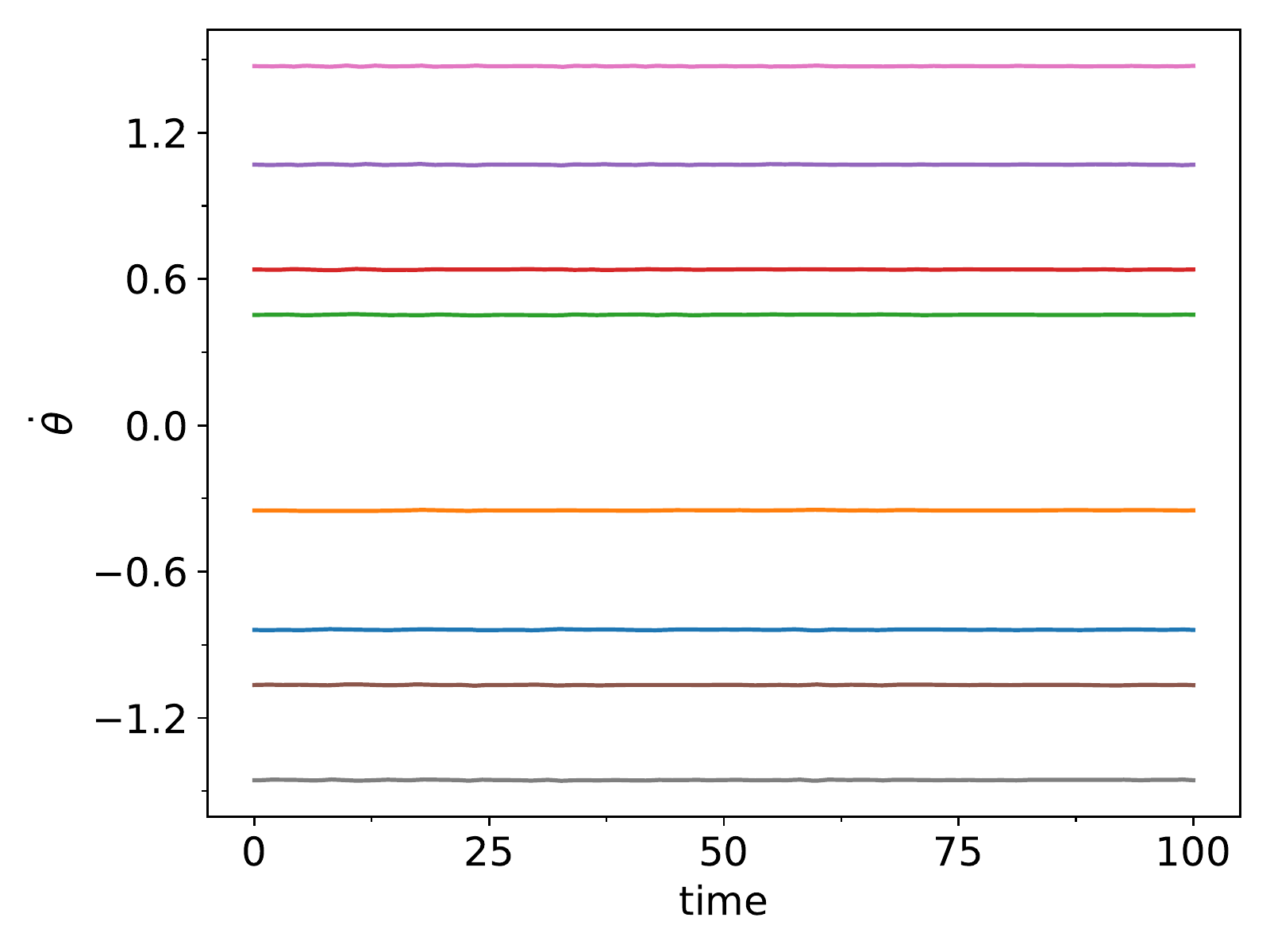}

}

\subfloat[$K=1.8$]{\includegraphics[width=1\linewidth]{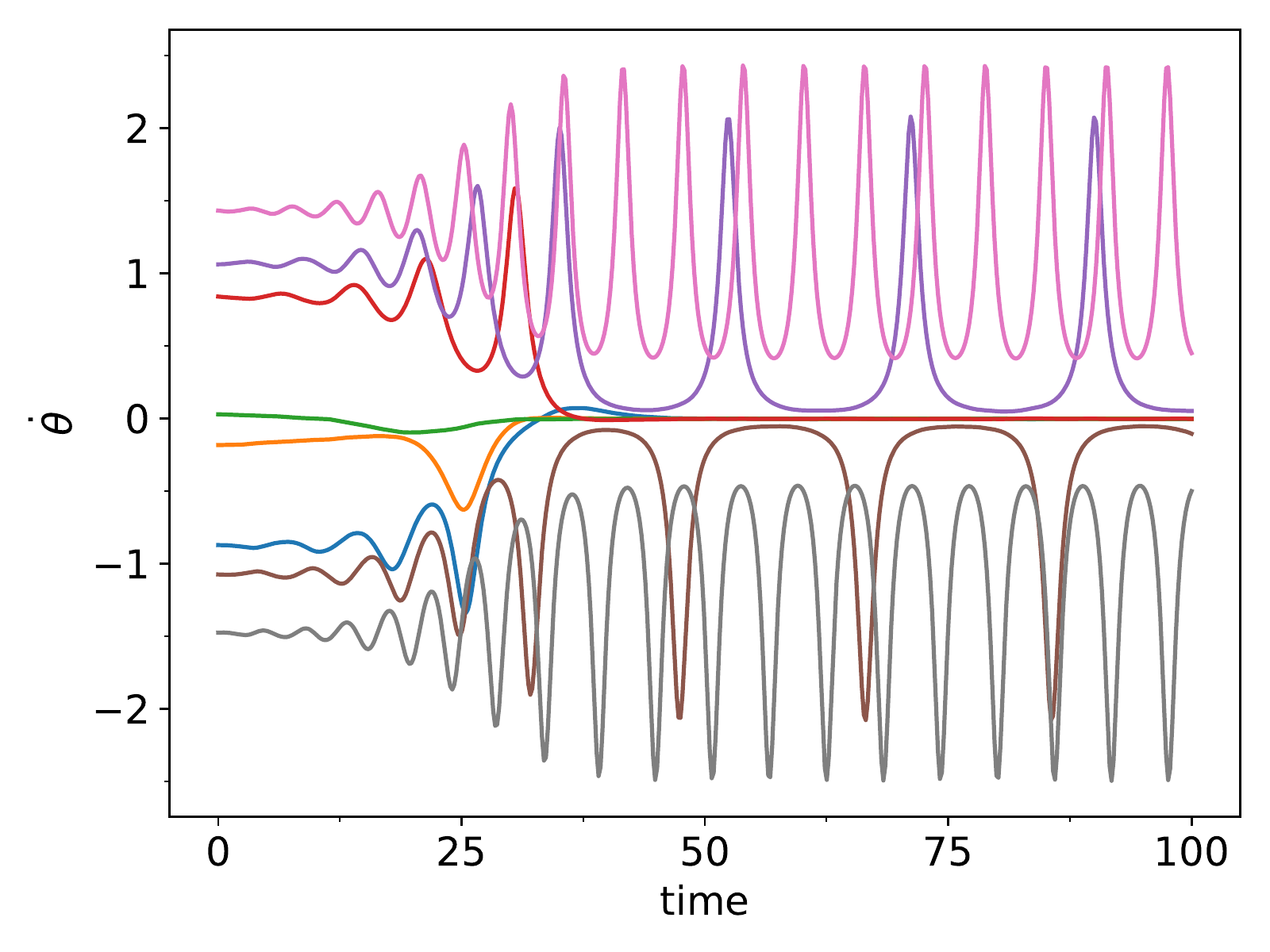}}%
\end{minipage}%
\begin{minipage}[t]{0.33333\linewidth}%
\subfloat[$K=0.8$]{\includegraphics[width=1\linewidth]{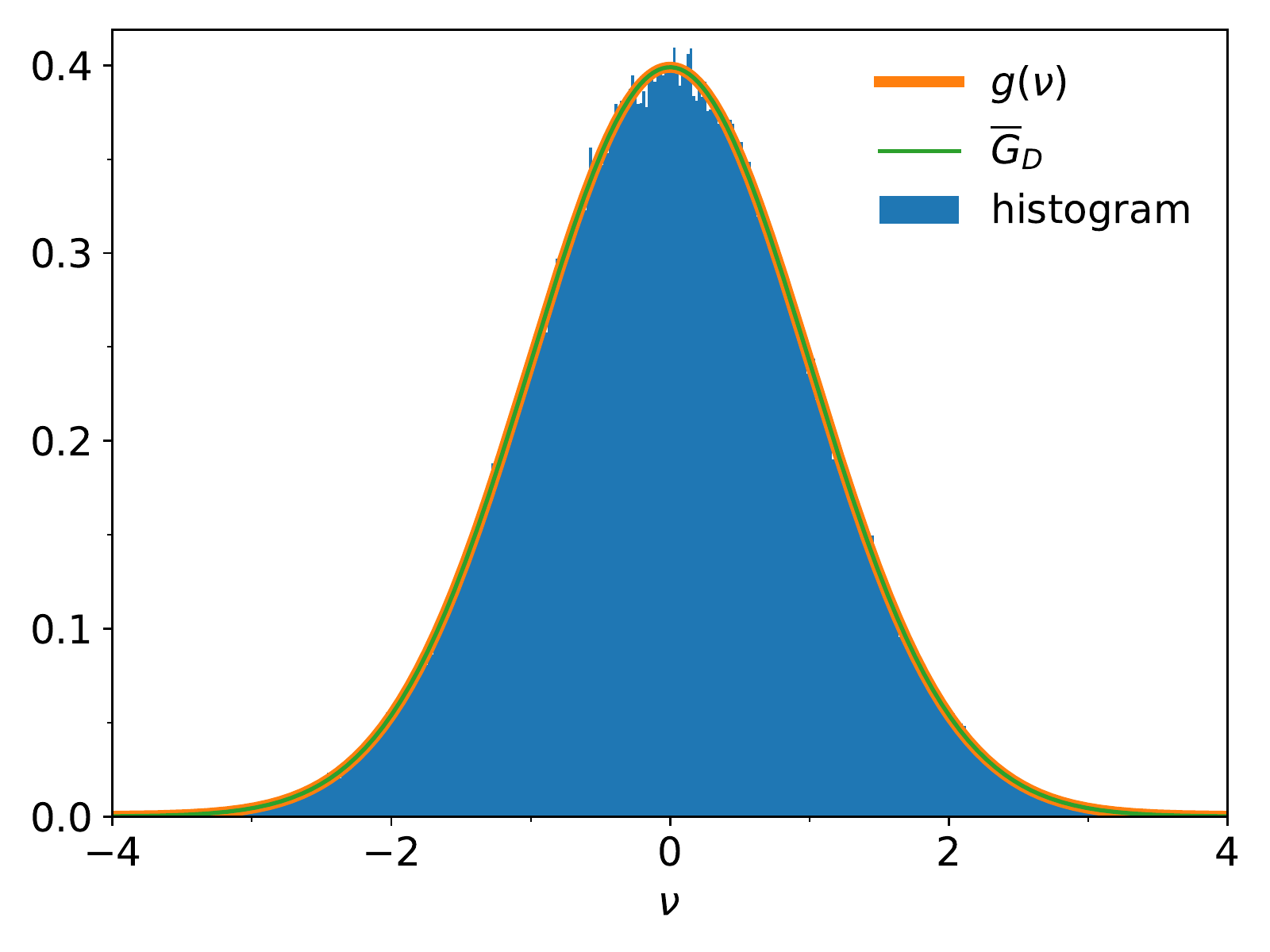}}

\subfloat[$K=1.8$]{\includegraphics[width=1\linewidth]{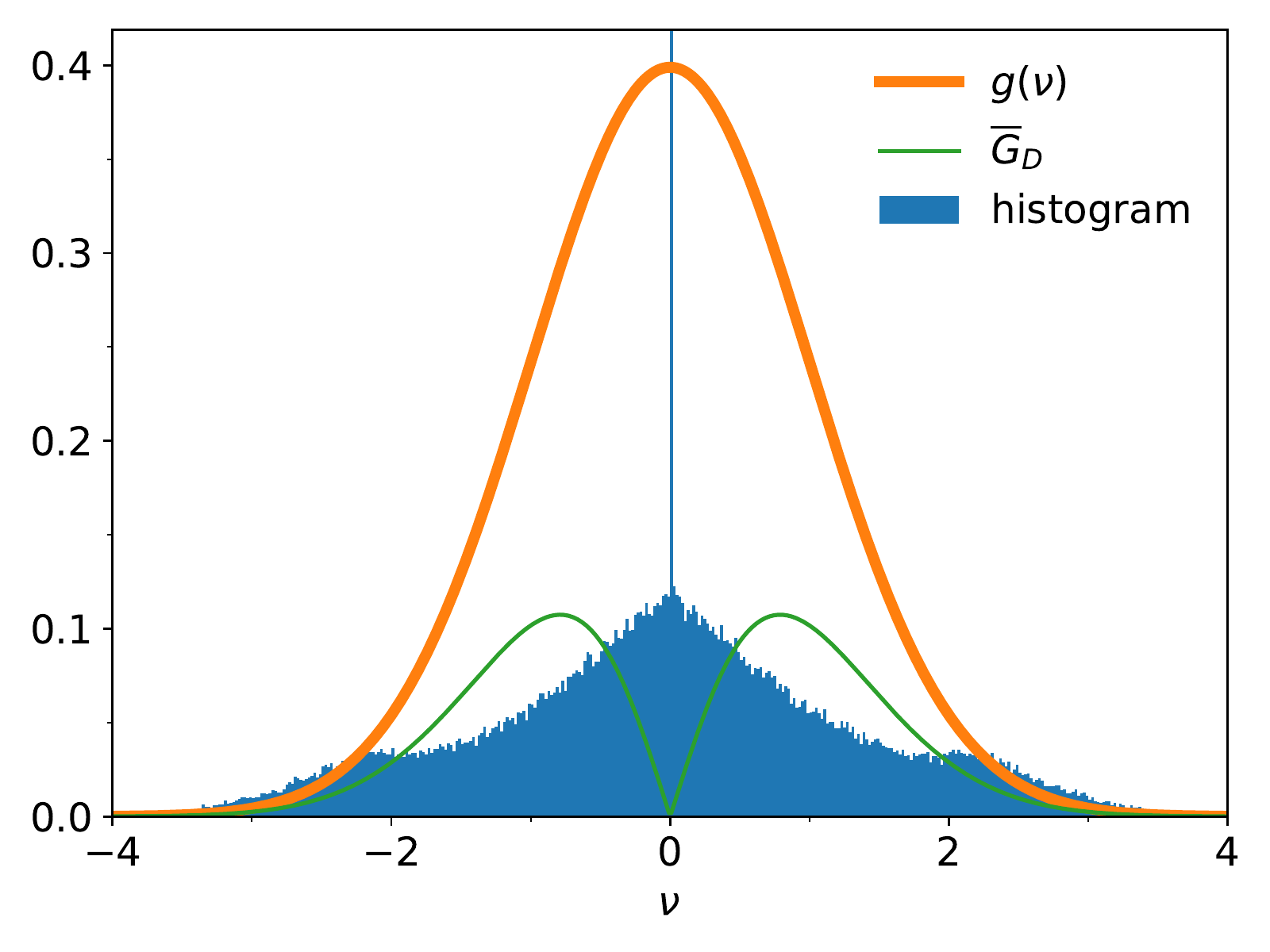}}%
\end{minipage}%
\begin{minipage}[t]{0.33333\linewidth}%
\subfloat[]{\includegraphics[width=1\linewidth]{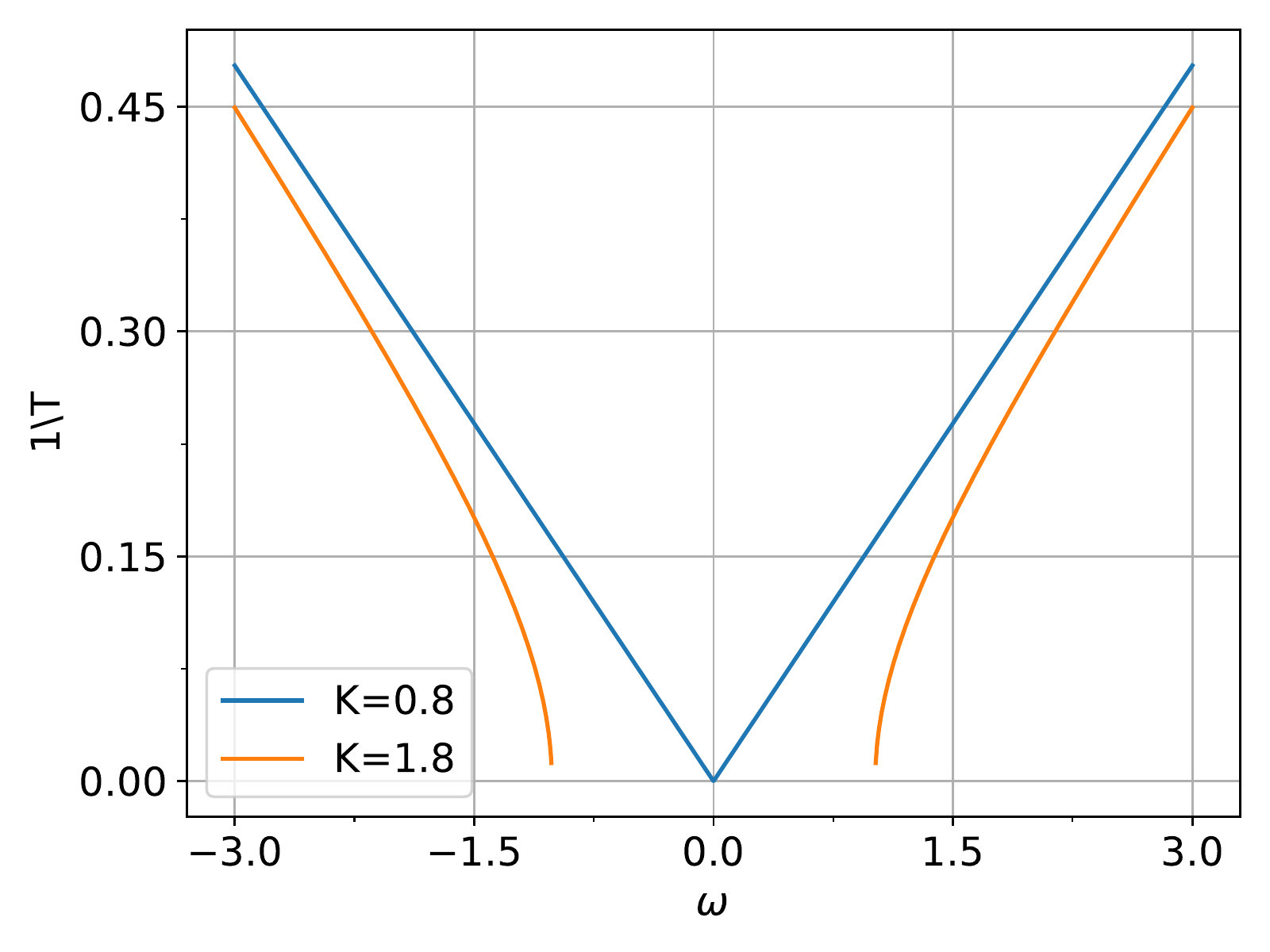}

}

\subfloat[]{\includegraphics[width=1\linewidth]{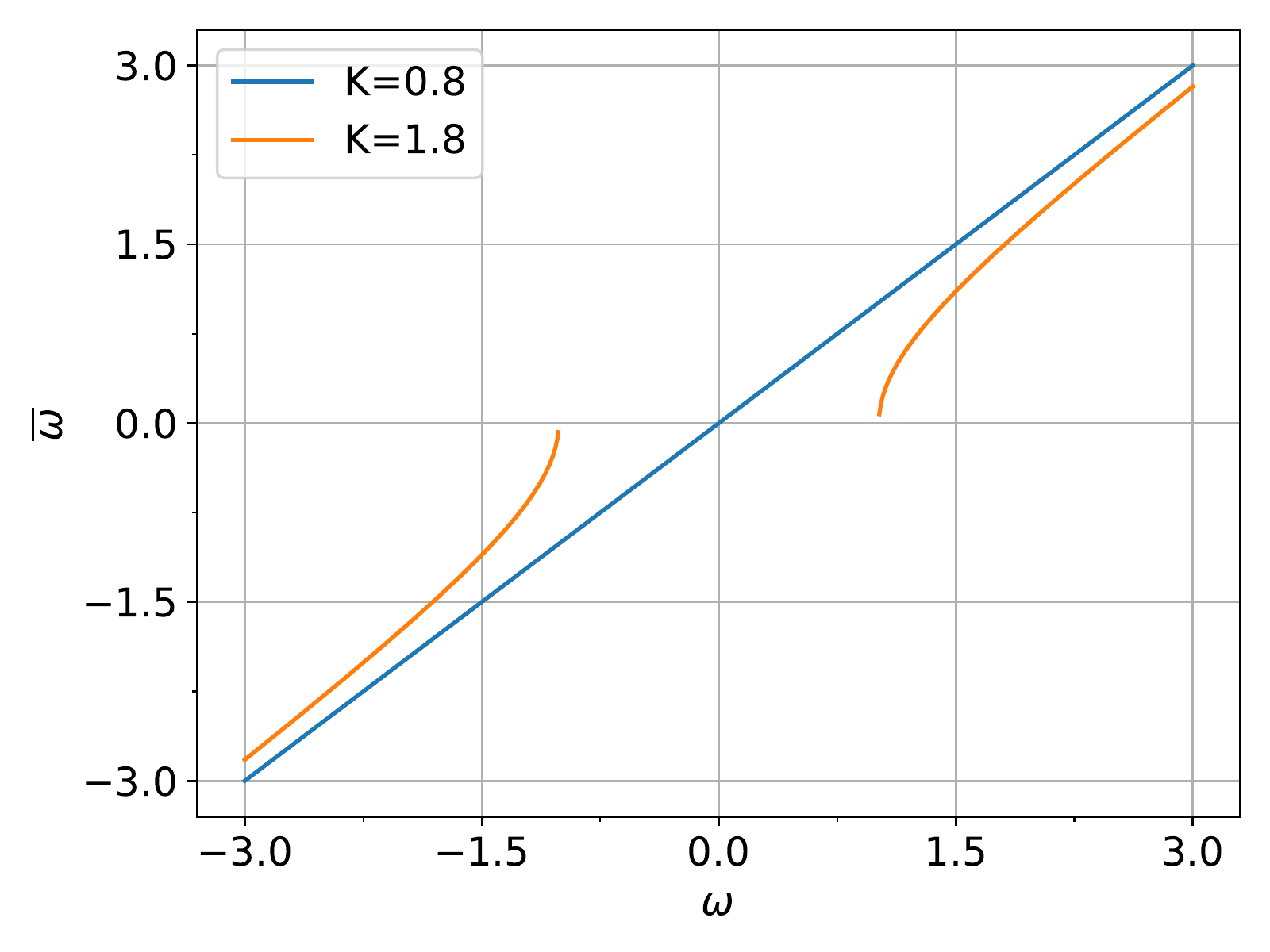}

}%
\end{minipage}
\caption{
(a) Time-series of instantaneous frequencies for $K<K_{c}$. 
(b) Same as (a), but for $K>K_{c}$. 
(c) $g$ and $\overline{G}_{D}$ compared to normalized histograms of instantaneous frequencies. 
(d) Same as (c), but for $K>K_{c}$. 
(e) Oscillation frequency, i.e. number of phase-cycles per time unit ($T^{-1}$), defined by Eq.(\ref{eq:period}).
(f) Time-averaged instantaneous frequency ($\overline{\omega}$), defined
by Eq.~(\ref{eq:bar-omega}). Numerical results were obtained using a normal
distribution of natural frequencies and a number of oscillators $N=5\times10^{5}$. See Appendix for numerical details.}
\label{fig:introduction}
\end{figure*}

As Kuramoto did in his early works, here we distinguish two groups
of oscillators: synchronized and desynchronized oscillators. Synchronized
oscillators are those whose natural frequencies satisfy $\left|\tilde{\omega}\right|\leq1$,
which means that (\ref{eq:inst-freq-rotating}) has a stable fixed point given by
\begin{equation}
\psi^{*}\left(\omega\right)=\arcsin\left(\tilde{\omega}\right).\label{eq:attractor}
\end{equation}
Desynchronized oscillators have natural frequencies such that $\left|\tilde{\omega}\right|>1$.
In this case, Eq.~(\ref{eq:inst-freq-rotating}) has no fixed point:
phases show oscillatory dynamics, which look like relaxation oscillations: they change rapidly at some stage of their period, then evolve much more slowly for the rest of the cycle (see Fig.~\ref{fig:introduction}(b) for some examples).

According to Eq.~(\ref{eq:inst-freq-rotating}), if $R=0$, instantaneous
and natural frequencies are the same. For heterogeneous natural frequencies,
$R=0$ characterizes the incoherent state, where oscillators are out
of synchrony. Assuming that the natural frequency distribution, $g$,
is symmetric and unimodal, as Kuramoto does in his early works, then
$\Omega$ coincides with the center of symmetry of $g$. When oscillators
synchronize, the order parameter has a positive value $R=\frac{a}{K}$,
obtained by finding the value of $a$ which solves 
\begin{equation}
K\intop_{-\frac{\pi}{2}}^{+\frac{\pi}{2}}d\psi g(\Omega+a\sin\psi)\cos^{2}\psi=1.\label{eq:sigmaVsK}
\end{equation}
Eq.~(\ref{eq:sigmaVsK}) can be used if the assumption of a symmetric
and unimodal $g$ holds. For more general profiles, the order parameter
has to be computed using a more general (and more difficult to solve)
equation whose solution is defined in terms of both $a$ and $\Omega$
(see Ref. \citep{Sakaguchi}). In this paper, our analytical results
do not depend on how $a$ is computed or on any specific assumption
about $g$. However, we will here consider Eq.(\ref{eq:sigmaVsK})
in the numerical examples, since it is simpler and more widely known
\footnote{Our opinion is that there are few works based on Kuramoto theory discussing
asymmetry effects of the natural frequency distribution, such as, for example, Refs. \citep{Basnarkov08,Terada17}. Eq.~(\ref{eq:sigmaVsK})
or alternative equivalent forms are more commonly used.}.

If $g$ is symmetric and unimodal, the critical coupling strength, i.e. the value of $K$ marking
the transition between the desynchronized and the partially synchronized states, is given by 
\begin{equation}
K_{c}=\frac{2}{\pi g(\Omega)},\label{eq:Kc}
\end{equation}
Again, a more general expression has to be used for more general forms
of $g$ \citep{Sakaguchi}.

Let $p(\psi,\omega)$ denote the joint probability density involving
the oscillator's phase in the rotating frame and the oscillator's
natural frequency. Applying Bayes' rule, $p(\psi,\omega)$ can be expressed
\begin{equation}
p(\psi,\omega)=p(\psi|\omega)g(\omega),\label{eq:joint_prob-1}
\end{equation}
where $p(\psi|\omega)$ is the conditional phase density for a given
natural frequency $\omega$.

A detailed discussion about how to obtain the conditional density
$p(\psi|\omega)$ can be found in Ref. \citep{daFonseca18}. Here
we only show $p(\psi|\omega)$ in its final possible forms. For $\left|\tilde{\omega}\right|\leq1$,
\begin{equation}
p(\psi|\omega)=\delta[\psi-\psi^{*}\left(\omega\right)],\label{eq:cond_prob_sync}
\end{equation}
where $\psi^{*}\left(\omega\right)$ is the stable fixed point of
Eq.~(\ref{eq:inst-freq-rotating}), given by Eq.~(\ref{eq:attractor}).
For $\left|\tilde{\omega}\right|>1$, Eq.~(\ref{eq:inst-freq-rotating})
has no fixed point, and the density $p(\psi|\omega)$ can be be written
as

\begin{equation}
p(\psi|\omega)=\frac{\omega-\Omega}{2\pi\dot{\psi}}\sqrt{1-\frac{1}{\tilde{\omega}^{2}}},\label{eq:cond_prob_desync}
\end{equation}
where $\dot{\psi}$ is given by Eq.~(\ref{eq:inst-freq-rotating}).

From formulas equivalent to Eqs. (\ref{eq:cond_prob_sync}) and (\ref{eq:cond_prob_desync}),
Kuramoto obtained the phase distribution, given by 
\begin{equation}
n(\psi)=n_{S}(\psi)+n_{D}(\psi),\label{eq:n-1}
\end{equation}
where $n_{S}(\psi)$, the phase distribution of synchronized oscillators,
is 
\begin{align}
n_{S}(\psi)= & \begin{cases}
g(\Omega+a\sin\psi)a\cos\psi,\quad & \left|\psi\right|\leq\frac{\pi}{2}\\
0, & \left|\psi\right|>\frac{\pi}{2}
\end{cases}\label{eq:ns-3}
\end{align}
and $n_{D}(\psi)$ denotes the phase distribution of desynchronized
oscillators, given by 
\begin{equation}
n_{D}(\psi)=\frac{1}{2\pi}\intop_{\left|x\right|>a}\frac{xg(\Omega+x)}{x-a\sin\psi}\sqrt{1-\left(\frac{a}{x}\right)^{2}}\,dx.\label{eq:nd-2}
\end{equation}
For $R>0$, the phase distribution profile shape is fixed 
and travels a distance of $\Omega t$
during a time interval $t$ in the non-rotating frame (where phases
are described by $\theta)$.
This is the scenario with synchronized oscillators: a phase distribution as a steadily traveling
wave. In the incoherent state, $R=0$, whence, according to Eqs. (\ref{eq:ns-3})
and (\ref{eq:nd-2}), phases are uniformly distributed, viz., $n(\psi)=\frac{1}{2\pi}$.

Kuramoto also obtained the distribution of ``coupling-modified
frequencies'' \citep{KuramotoB,KuramotoC}, which are the instantaneous
frequencies averaged over an infinitely long time, as thoroughly discussed
in Ref.\citep{daFonseca18}. The infinite-time average $\overline{\omega}$ 
of an oscillator's instantaneous frequency $\dot{\theta}$, can be defined as 
\begin{equation}
\overline{\omega}=\lim_{T\rightarrow+\infty}\,\frac{1}{T}\intop_{0}^{T}\dot{\theta}(t)\,dt.\label{eq:avg-freq}
\end{equation}
Kuramoto showed that the coupling-modified frequencies are distributed
accordingly to 
\begin{equation}
\overline{G}(\nu)=\delta\left(\nu-\Omega\right)S\left(K\right)+\overline{G}_{D}(\nu),\label{eq:Gbar}
\end{equation}
where 
\begin{equation}
S\left(K\right)=\intop_{\Omega-a}^{\Omega+a}g(\omega)d\omega,\label{eq:r_bar_S}
\end{equation}
\begin{equation}
\overline{G}_{D}(\nu)=\frac{\left|\tilde{\nu}\right|}{\sqrt{1+\tilde{\nu}^{2}}}g\left(\Omega+\frac{\nu-\Omega}{\left|\tilde{\nu}\right|}\sqrt{1+\tilde{\nu}^{2}}\right)\label{eq:GbarD}
\end{equation}
for $\nu\neq\Omega$, and $\overline{G}_{D}(\Omega)=0$.

Equation (\ref{eq:Gbar}) states that: (i) $\overline{G}(\nu)$ exhibits
a singularity at the synchronization frequency $\Omega$; (ii) $\overline{G}(\nu)$
goes linearly to zero for $\nu$ near $\Omega$; and (iii) $\lim_{\epsilon\rightarrow0^{+}}\intop_{\Omega-\epsilon}^{\Omega+\epsilon}\overline{G}(\nu)d\nu=S\left(K\right)$,
i.e. the probability of an oscillator having a time-averaged frequency
arbitrarily near $\Omega$ is given by $S\left(K\right)$, which represents
the fraction of synchronized oscillators.

Figure~\ref{fig:introduction} summarizes these findings, together with some numerical
illustration of the object of central interest here, the distribution of instantaneous frequencies.
In Figs. \ref{fig:introduction}(a) and (b) we show time series
of the instantaneous frequencies of eight oscillators selected from a total ensemble of $N=5\times 10^5$ oscillators
with their natural frequencies distributed according to a normal (Gaussian) distribution centered at $\Omega=0$.
For this case $K_{c}\simeq1.5957$. In Fig.
\ref{fig:introduction}(a), we set $K=0.8<K_{c}$ in the desynchronized regime ($a=0$), and all oscillators quickly
keep their natural frequency.
In Fig.~\ref{fig:introduction}(b), $K=1.8>K_{c}$, in the synchronized regime ($a>0$): 
the 4 oscillators with their natural frequency
$|\tilde{\omega}|<1$ synchronize to $\Omega=0$, while the others stay desynchronized 
and their instantaneous frequencies exhibit relaxation-oscillation-like dynamics.

In Figs.~\ref{fig:introduction}(c) and (d) we use the same $K$ values as in (a) and (b) 
and show: i) $g$ the Gaussian distribution of natural frequencies; 
ii) $\overline{G}_{D}$ the continuous part of the distribution of time-averaged
instantaneous frequencies given by Eq.~\eqref{eq:GbarD}; and iii)
numerically-determined normalized histograms of instantaneous frequencies (see Appendix for details). 
As expected, all these distributions coincide in the subcritical case shown in Fig.~\ref{fig:introduction}(c).
For $K=1.8>K_c$ (Fig.~\ref{fig:introduction}(d)),
the histogram of instantaneous frequencies shows
a peak located at $\Omega=0$ which correspond to the synchronized oscillators. 
Note that the instantaneous frequencies (shown in the histogram) are distributed in a
qualitatively different way from their time-averaged counterparts. Remarkable differences
are the accumulation of desynchronized oscillators' instantaneous frequencies near zero
and the fact that the tails of the instantaneous frequency distribution are ``fatter" 
than those of the time-averaged and nominal frequencies.

Coming back to the periodic dynamics of individual desynchronized oscillators in the synchronized regime 
(Fig.~\ref{fig:introduction}(b)),
we see from Eq.~(\ref{eq:inst-freq-rotating}) that the amplitude of oscillations of $\dot{\theta}$ is $a$, and the 
minimum and maximum values reached are 
 $\dot{\theta}_{min}=\omega-a$ and $\dot{\theta}_{max}=\omega+a$, so that the middle value is the natural frequency of the oscillator.
As discussed in Ref. \citep{daFonseca18}, a desynchronized oscillator's phase
makes a complete $2\pi$ turn during a time-interval $T(\omega)$, which is given by
\begin{equation}
T(\omega)=\frac{2\pi}{\sqrt{\left(\omega-\Omega\right)^{2}-a^{2}}}.\label{eq:period}
\end{equation}
The time-averaged instantaneous frequency, defined by (\ref{eq:avg-freq}), can be given
in terms of $T$ as 
\begin{align}
\overline{\omega} & =\begin{cases}
\Omega+\frac{2\pi}{T}, & \omega>\Omega+a\\
\Omega-\frac{2\pi}{T}, & \omega<\Omega-a.
\end{cases}\label{eq:bar-omega}
\end{align}
If $a=0$, we can write (\ref{eq:period}) and (\ref{eq:bar-omega})
as $\frac{1}{T}=\frac{\left|\omega-\Omega\right|}{2\pi}$, and $\overline{\omega}=\omega$.
If $a>0$, (\ref{eq:period}) is the same as $\frac{1}{T}=\frac{a}{2\pi}\sqrt{\widetilde{\omega}^{2}-1}$,
and (\ref{eq:bar-omega}) can be written as $\overline{\omega}=\Omega\pm a\sqrt{\widetilde{\omega}^{2}-1}$
for $\pm\widetilde{\omega}>1$.

These formulas allow us to infer the following properties of the instantaneous frequency of 
non-synchronized oscillators: 
i) as $\widetilde{\omega}\rightarrow\pm1^{\pm}$, we have $\frac{1}{T}\rightarrow0$,
which means slow oscillations, or oscillations with large-time periods;
ii) if $\widetilde{\omega}\rightarrow1^{+}$, then $\overline{\omega}\rightarrow\Omega^{+}$
and $\dot{\theta}_{min}=\omega-a\rightarrow\Omega^{+}$, i.e. both
the time-averaged and minimum instantaneous frequencies have values close to and greater
than $\Omega$, the frequency of synchronized oscillators (Similar properties are of course valid 
if $\widetilde{\omega}\rightarrow1^{-}$);
iii) as $\left|\widetilde{\omega}\right|\rightarrow\infty$, then
we have $\frac{1}{T}\rightarrow\infty$ and $\overline{\omega}\rightarrow\omega$,
viz. fast oscillations with time-averages becoming close to their
oscillation centers. 

Fig.~\ref{fig:introduction}(b) illustrates all these properties. 
The long periods of time spent near $\Omega$ by the instantaneous frequency of oscillators with 
$|\tilde{\omega}|\gtrsim 1$ explains why the
histogram in Fig.~\ref{fig:introduction}(d) exhibits frequent
occurrences near zero. 

Figs. \ref{fig:introduction}(e) and (f) show plots of $\frac{1}{T}$
and $\overline{\omega}$ as functions of $\omega$ created using Eqs.
(\ref{eq:period}) and (\ref{eq:bar-omega}). The synchronized frequency $\Omega$ has been set to zero
and we adopted the same values of $K$ as used in Figs.\ref{fig:introduction}(a)-(d).
For $K=0.8$ ($a=0$), the graphs are shown in blue. The curves in
orange correspond to the case where $K=1.8$ ($a\simeq1$).

In summary, many of the properties shown in Fig.~\ref{fig:introduction} are
straight consequences of Kuramoto theory. Those regarding instantaneous frequencies
can be qualitatively explained by it. This is the case of the accumulation of instantaneous 
frequencies near the synchronization frequency (Fig.~\ref{fig:introduction}(d)). 
We now proceed to the core of this work, which is the calculation of the full analytical expression 
of the distribution of instantaneous frequencies.

\section{Distribution of instantaneous frequencies \label{sec:G}}

Our goal in this section is to obtain, based on the results discussed
in Section \ref{sec:Some-results-from-Kuramoto}, the distribution
of instantaneous frequencies in the Kuramoto model. This distribution
is a probability density function $G(\nu$), which means that $G(\nu)\,d\nu$
is the probability of an oscillator showing its fixed frame instantaneous
frequency , $\dot{\theta},$ in the interval $\left[\nu,\,\nu+d\nu\right)$.

By using the random variable transformation theorem \citep{Gillepsie83},
we have 
\begin{equation}
G(\nu)=\int_{-\infty}^{+\infty}\int_{-\pi}^{+\pi}\delta[\nu-\dot{\theta}(\omega,\psi)]p(\psi,\omega)\,d\psi d\omega,\label{eq:G-1}
\end{equation}
where $\delta$ denotes the delta function, $\dot{\theta}(\omega,\psi)$
is a random variable transformation, given by 
\begin{equation}
\dot{\theta}(\omega,\psi)=\omega-a\sin\psi,\label{eq:inst-freq-fixed}
\end{equation}
and $p(\psi,\omega)$ is the joint probability density involving the
phase in the rotating frame and the natural frequency. Equation (\ref{eq:inst-freq-fixed})
comes from Eq.(\ref{eq:inst-freq-rotating}), which decribes instantaneous
frequencies in the rotating frame. 

The delta function in Eq.~(\ref{eq:G-1}), $\delta[\nu-\dot{\theta}(\omega,\psi)]$,
is concentrated in a curve embedded in the two-dimensional space defined
by the integration variables $\psi$ and $\omega$. In order to calculate
the double integral in Eq.~(\ref{eq:G-1}), we use a method proposed
by Seeley \citep{Seeley62,Jager69}, which generalizes the usual concept
of one-dimensional delta functions to delta functions concentrated
in manifolds with an arbitrary number of dimensions.

Let $\delta(P)$ denote a delta function concentrated in a $n-1$-dimensional
manifold $M$ embedded in a $n$-dimensional space $V$. The manifold
$M$ is defined by $P(\boldsymbol{x})=0$, where $P(\boldsymbol{x})$
is a function at $\boldsymbol{x}=(x_{1},\ldots,x_{n})$, i.e. a point
in $V$ with coordinates $x_{1},\ldots,x_{n}$. The delta function
$\delta(P)$ can be defined by 
\begin{equation}
\delta(P)=\lim_{c\rightarrow0^{+}}\frac{\Theta(P+c)-\Theta(P)}{c},\label{eq:deltaP-def}
\end{equation}
where $\Theta(\,.\,)$ is a Heaviside step function such that $\text{\ensuremath{\Theta(P)=1}}$
for $P\geq0$, and $\text{\ensuremath{\Theta(P)=0}}$ for $P<0$.

Consider a function $\varphi(\boldsymbol{x})$ defined in $V$. From
Eq.~(\ref{eq:deltaP-def}), we have 
\begin{equation}
\intop_{V}\delta(P)\varphi(\boldsymbol{x})d^{n}\boldsymbol{x}=\lim_{c\rightarrow0^{+}}\intop_{-c\leq P<0}\frac{\varphi(\boldsymbol{x})}{c}d^{n}\boldsymbol{x},\label{eq:delta-vol-integral}
\end{equation}
where $d^{n}\boldsymbol{x}=dx_{1}.\ldots.dx_{n}.$

Let $\gamma$ be the distance between the manifolds $M_{P=0}$ and
$M_{P=-c}$, defined by the equations $P=0$ and $P=-c$, respectively.
Then, for $P(\boldsymbol{x})=0$, we have $P\left[\boldsymbol{x}+\gamma\hat{n}(\boldsymbol{x})\right]=-c$,
i.e., if $\boldsymbol{x}\in M_{P=0}$, then $\boldsymbol{x}+\gamma\hat{n}(\boldsymbol{x})\in M_{P=-c}$,
where $\hat{n}(\boldsymbol{x})$ is the unit vector normal to $M_{P=0}$
at $\boldsymbol{x}$. As $c\rightarrow0^{+}$, a first-order expansion
of $P\left[\boldsymbol{x}+\gamma\hat{n}(\boldsymbol{x})\right]$ results
in $P(\boldsymbol{x})-\gamma\hat{n}\nabla P=-c$, from which we obtain
\begin{equation}
\gamma=\frac{c}{\left|\nabla P\right|},\label{eq:gamma}
\end{equation}
since $P(\boldsymbol{x})=0$ and $\hat{n}(\boldsymbol{x})=\frac{\nabla P}{\left|\nabla P\right|}$.

Changing the infinitesimal volume element $d^{n}\boldsymbol{x}$ by
$\gamma dS$, where $\gamma$ is given by Eq.~(\ref{eq:gamma}) and
$dS$ is an infinitesimal surface element of $M_{P=0}$, Eq.~(\ref{eq:delta-vol-integral})
can be rewritten as 
\begin{equation}
\intop_{V}\delta(P)\varphi(\boldsymbol{x})d^{n}\boldsymbol{x}=\intop_{M_{P=0}}\frac{\varphi(\boldsymbol{x})}{\left|\nabla P\right|}dS,\label{eq:surface-integral}
\end{equation}
which means that the volume integral in the right-side of Eq.~(\ref{eq:delta-vol-integral})
can be changed by a surface integral on $M_{P=0}$.

Suppose that $M_{P=0}$ is a curve, and $V,$ a two-dimensional space.
In this particular case, Eq.~(\ref{eq:surface-integral}) can be written
in the form 
\begin{equation}
\intop_{V}\delta\left[P\left(\boldsymbol{x}\right)\right]\varphi\left(\boldsymbol{x}\right)d^{2}\boldsymbol{x}=\intop_{M_{P=0}}\frac{\varphi\left(\boldsymbol{x}\right)}{\left|\nabla P\right|}dl,\label{eq:line-integral}
\end{equation}
where $\boldsymbol{x}=(x_{1},x_{2})$, $d^{2}\boldsymbol{x}=dx_{1}dx_{2}$,
and $dl$ is an infinitesimal line element of $M_{P=0}$. Let $P(x_{1},x_{2})$
be defined by $P(x_{1},x_{2})=f(x_{1})-x_{2}$, where $f(x_{1})$
is a continuous function, and the range of $x_{1}$ is the the interval
$[a,b)$. Then, the curve $M_{P(x_{1},x_{2})=0}$ is the graph of
$x_{2}=f(x_{1})$ with $a\leq x_{1}<b$. Suppose that $C_{M}$ is
a curve corresponding to a part of $M_{P(x_{1},x_{2})=0}$. This curve
can be defined as a subset of $\text{\ensuremath{M_{P(x_{1},x_{2})=0}}}$
by $C_{M}=\{(x_{1},x_{2})\in V\mid x_{2}=f(x_{1})\,\text{and}\,\psi_{a}\leq x_{1}\leq\psi_{b}<b\}$.
An integral along $C_{M}$, analogous to the right-hand side of Eq.
(\ref{eq:line-integral}), can be written as 
\begin{equation}
\intop_{_{C}}\frac{\varphi(x_{1},x_{2})}{\left|\nabla P(x_{1},x_{2})\right|}dl=\intop_{\psi_{a}}^{\psi_{b}}\varphi\left[x_{1},f(x_{1})\right]dx_{1}.\label{eq:definite-integral}
\end{equation}

We are now able to compute the right-hand side of Eq.~(\ref{eq:G-1}).
Let us first put Eq.~(\ref{eq:G-1}) in the form 
\begin{equation}
G(\nu)=\intop_{V}\delta\left[P_{\nu}(\psi,\omega)\right]p(\psi,\omega)\,d\psi d\omega,\label{eq:G-2}
\end{equation}
where the integration manifold, $V$, is an infinite-length cylinder
$V=[-\frac{\pi}{2},+\frac{3\pi}{2})\times(-\infty,+\infty)$, $P_{\nu}(\psi,\omega)$
is given by 
\begin{equation}
P_{\nu}(\psi,\omega)=F_{\nu}(\psi)-\omega,\label{eq:P-nu}
\end{equation}
with 
\begin{equation}
F_{\nu}(\psi)=a\sin\psi+\nu,\label{eq:f}
\end{equation}
and $-\frac{\pi}{2}\leq\psi<+\frac{3\pi}{2}$. 

\begin{figure}
\begin{centering}
\includegraphics[width=0.90\linewidth]{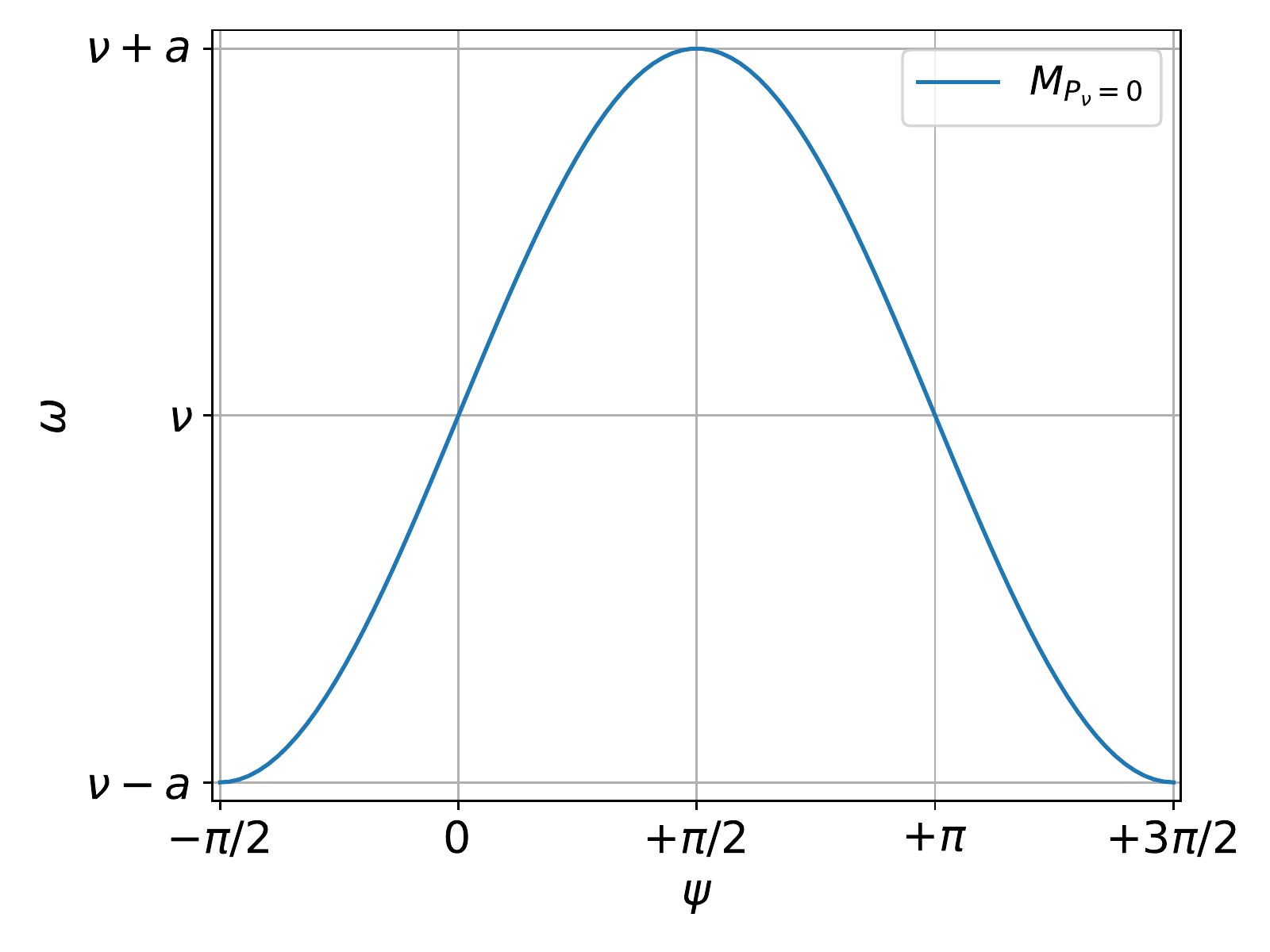} 
\par\end{centering}
\caption{Graph of $F_{\nu}$, Eq.~(\ref{eq:f}), representing the curve $M_{P_{\nu}=0}$,
Eq.~(\ref{eq:MPnu}). The minimum is $\left(-\frac{\pi}{2},\nu-a\right)$,
and the point of maximum is $\left(+\frac{\pi}{2},\nu+a\right).$ }
\label{Fig:MPnu} 
\end{figure}

Then, $M_{P_{\nu}=0}$ is a closed curve in $V$ defined by 
\begin{equation}
M_{P_{\nu}=0}=\left\{ (\psi,\omega)\in V\mid\omega=F_{\nu}(\psi)\right\} .\label{eq:MPnu}
\end{equation}
This curve is represented by the graph of $F_{\nu}$, shown in Fig.~\ref{Fig:MPnu}. 
The position of the curve $M_{P_{\nu}=0}$ in the integration manifold
$V$ is determined by the value of $\nu$, which is the argument of
$G$. And the height of $M_{P_{\nu}=0}$, as shown in Fig.~\ref{Fig:MPnu},
is $2a$. The curve shifts by varying $\nu$ and stretches as the
product $a$ increases.

\begin{figure}
\includegraphics[width=0.7\linewidth]{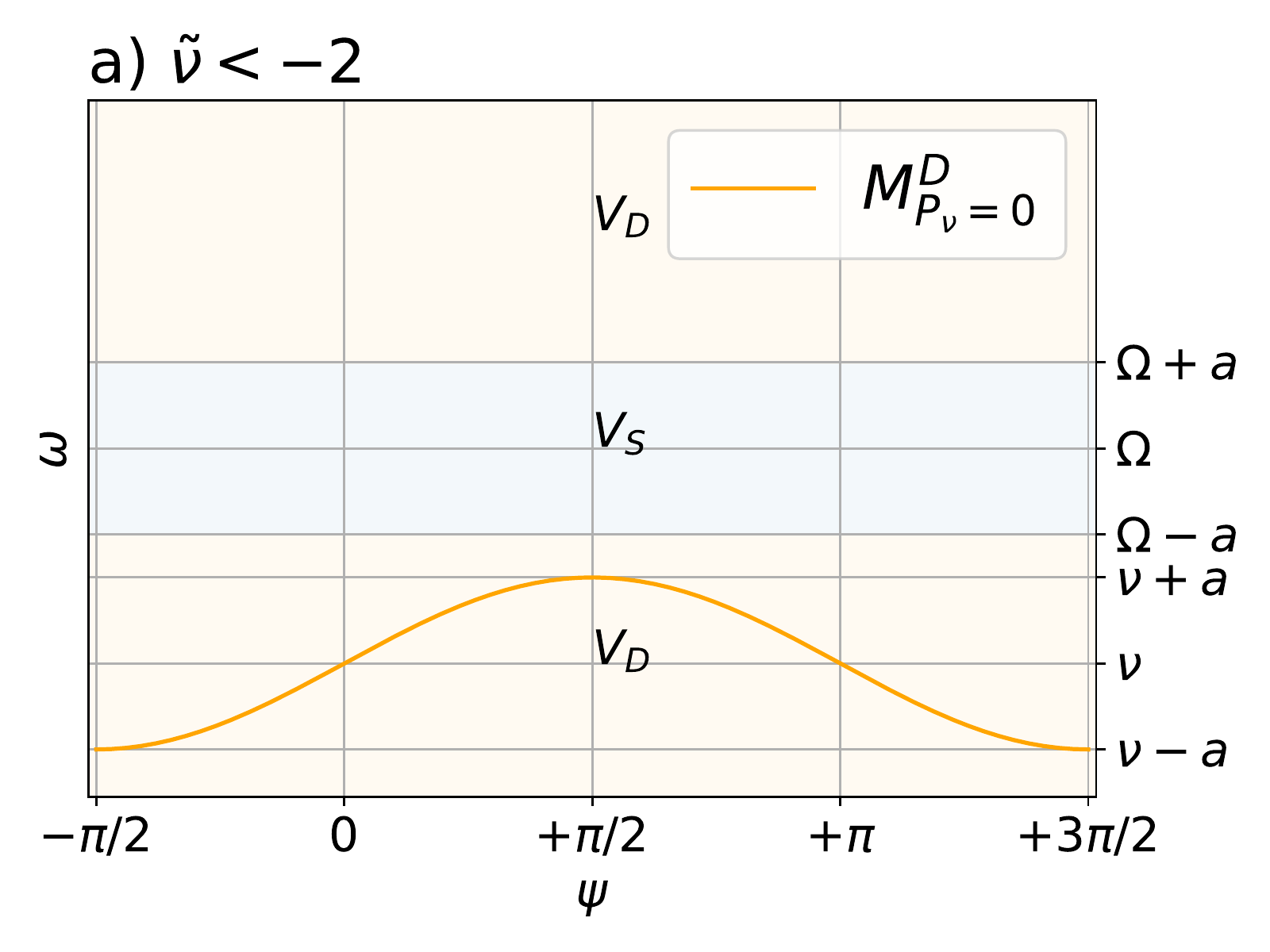}
\includegraphics[width=0.7\linewidth]{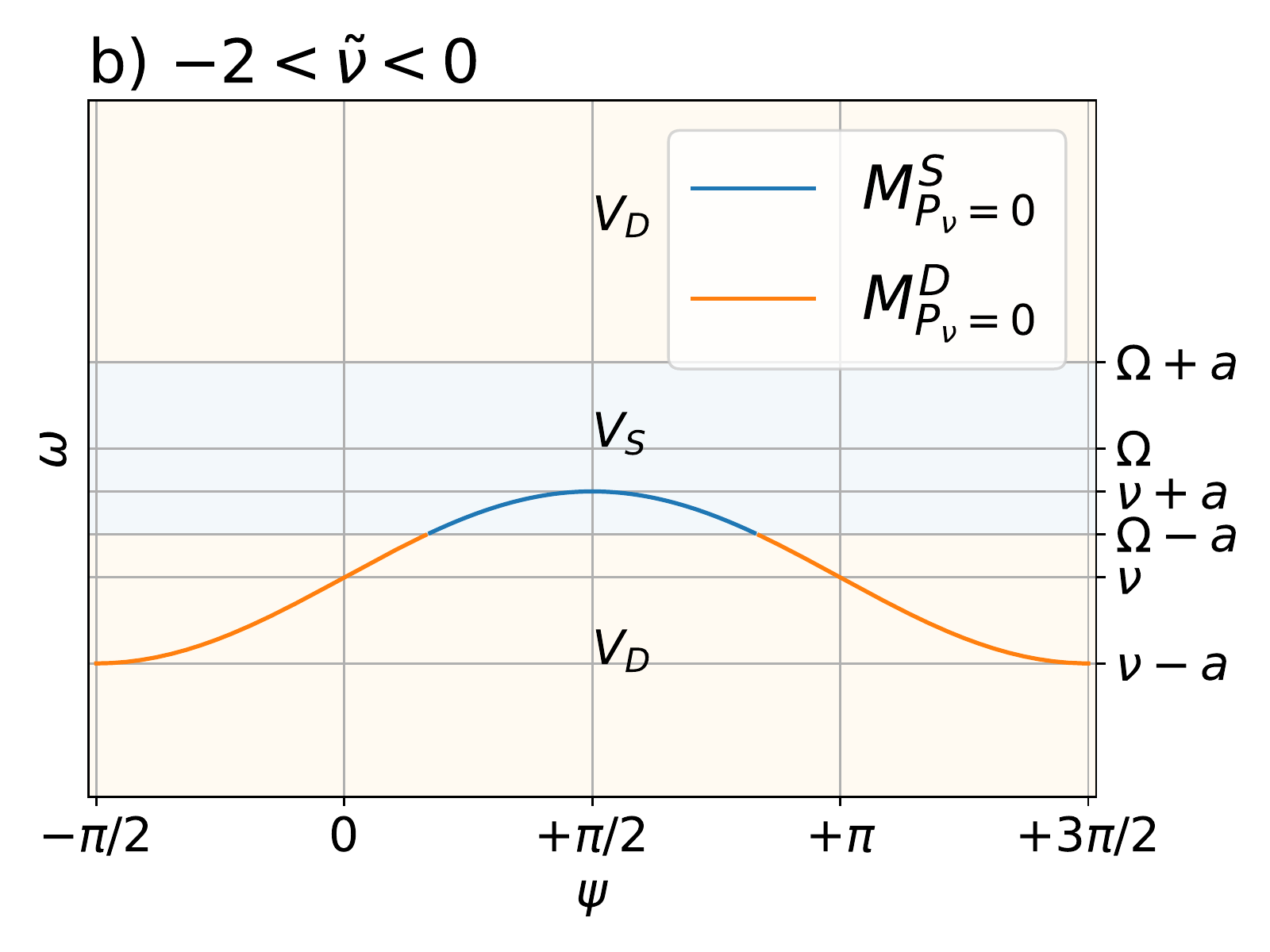}
\includegraphics[width=0.7\linewidth]{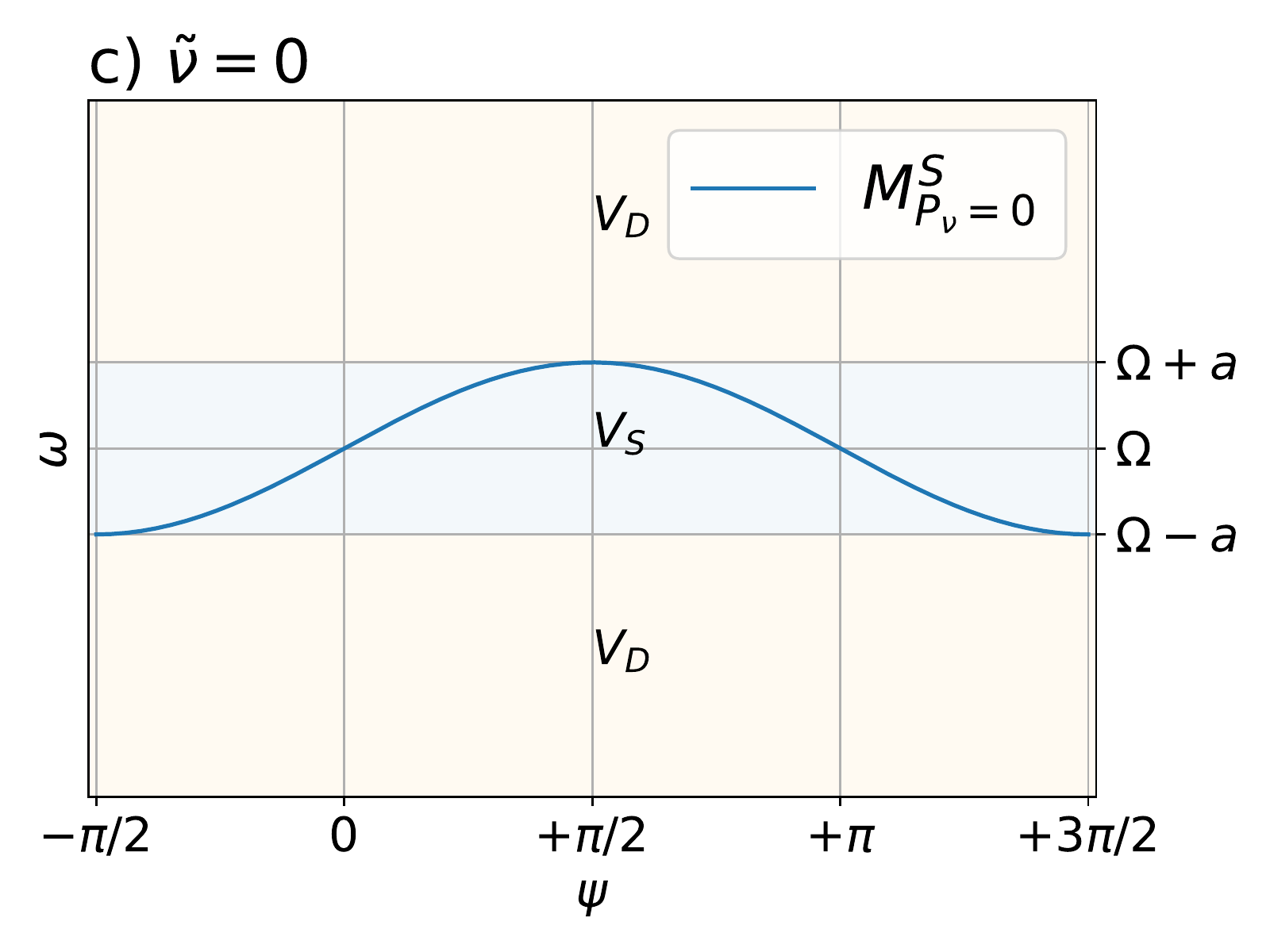}
\includegraphics[width=0.7\linewidth]{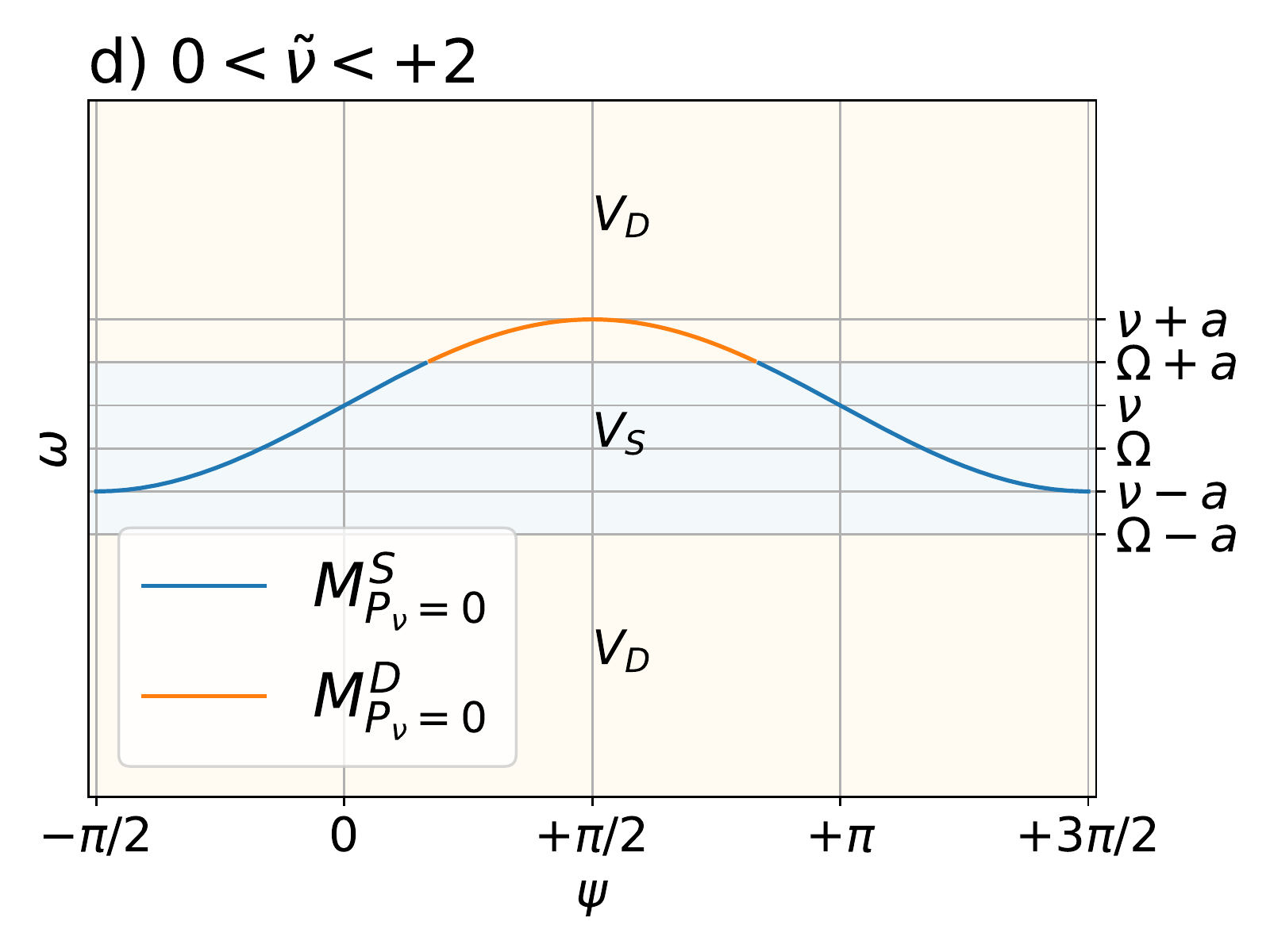}
\includegraphics[width=0.7\linewidth]{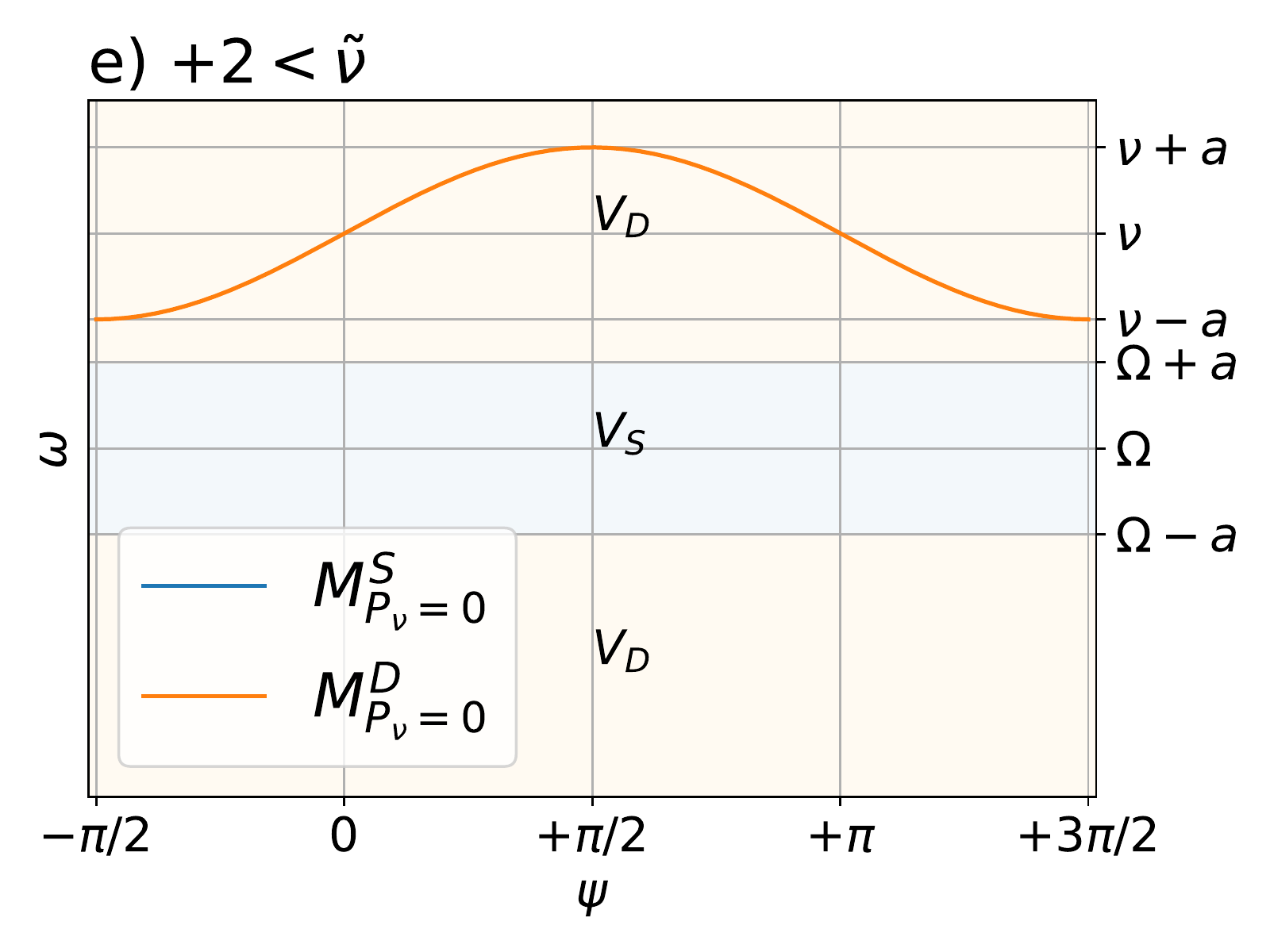}
\caption{The curve $M_{P_{\nu}=0}$ in different locations depending on the
value of $\nu$. The part of $M_{P_{\nu}=0}$ immersed in $V_{D}$
is $M_{P_{\nu}=0}^{D}$, and the intersection between $M_{P_{\nu}=0}$
and $V_{S}$ is $M_{P_{\nu}=0}^{S}$. }
\label{MP_SD} 
\end{figure}

Using the relation (\ref{eq:line-integral}), we obtain from Eq.~(\ref{eq:G-2})
the formula 
\begin{equation}
G(\nu)=\intop_{M_{P_{\nu}=0}}\frac{p(\psi,\omega)}{\left|\nabla P_{\nu}(\psi,\omega)\right|}dl.\label{eq:G-3}
\end{equation}
We can calculate the line integral in Eq.~(\ref{eq:G-3}) by use of
a geometric analysis based on dividing $V$ in two disjoints regions,
$V_{S}=\left[-\frac{\pi}{2},+\frac{3\pi}{2}\right)\times\left[-a+\Omega,\Omega+a\right]$
and $V_{D}=V-V_{D}$. A sketch of both regions and $M_{P_{\nu}=0}$
in different locations is shown in Figs.\ref{MP_SD}(a)-(e). 
Depending on the location and height of $M_{P_{\nu}=0}$, this curve
is completely inside $V_{D}$ (Figs.~\ref{MP_SD} (a) and (e)) and can also be
partly or entirely in $V_{S}$ (Figs.~\ref{MP_SD}(b),(c) and (d)).
We denote the parts in $V_{D}$ by $M_{P_{\nu}=0}^{D}$, which are
the blue curves, and those in $V_{S}$ by $M_{P_{\nu}=0}^{S}$, represented
by the yellow curves.
 
Using Eq.~(\ref{eq:joint_prob-1}), we have that Eq.~(\ref{eq:G-3})
is the same as 
\begin{equation}
G(\nu)=G_{S}(\nu)+G_{D}(\nu),\label{eq:G-4}
\end{equation}
where 
\begin{equation}
G_{S}(\nu)=\intop_{M_{P_{\nu}=0}^{S}}\frac{p(\psi|\omega)g(\omega)}{\left|\nabla P_{\nu}(\psi,\omega)\right|}dl,\label{eq:G_S}
\end{equation}
and 
\begin{equation}
G_{D}(\nu)=\intop_{M_{P_{\nu}=0}^{D}}\frac{p(\psi|\omega)g(\omega)}{\left|\nabla P_{\nu}(\psi,\omega)\right|}dl.\label{eq:G_D}
\end{equation}
If $M_{P_{\nu}=0}$ has no part inside $V_{s}$, then the curve $M_{P_{\nu}=0}^{S}$
does not exist and $G_{S}(\nu)=0$. Similarly, if $M_{P_{\nu}=0}^{D}$
is an empty set, then we also have $G_{D}(\nu)=0$.

We consider first the case in which $M_{P_{\nu}=0}^{S}$ exists. A
point $(\psi,\omega)$ in $M_{P_{\nu}=0}^{S}$ satifies the conditions:
i) $\omega=F_{\nu}(\psi)$; ii) $-a+\Omega\leq\omega\leq\Omega+a$,
as can also be seen in Figs. \ref{MP_SD}(b),(c) and (d). Condition
ii) means that the density $p(\psi|\omega)$ in Eq.~(\ref{eq:G_S})
is defined by Eq.~(\ref{eq:cond_prob_sync}). Then 
\begin{equation}
G_{S}(\nu)=\intop_{M_{P_{\nu}=0}^{S}}\delta\left[\psi-\psi^{*}\left(\omega\right)\right]\frac{g(\omega)}{\left|\nabla P_{\nu}(\psi,\omega)\right|}dl,\label{eq:G_S-1}
\end{equation}
where $\psi^{*}\left(\omega\right)$ is given by Eq.~(\ref{eq:attractor}).
Integration along $M_{P_{\nu}=0}^{S}$ can be done in three steps:
the first one is integration along $M_{P_{\nu}=0}^{S_{L}}$, i.e.
the subset of $M_{P_{\nu}=0}^{S}$ whose projection in the $\psi$-axis
is contained in the closed interval $\left[-\frac{\pi}{2},+\frac{\pi}{2}\right]$;
the second step is integration along $M_{P_{\nu}=0}^{S_{R}}$, which
is the subset of $M_{P_{\nu}=0}^{S}$ whose projection in the $\psi$-axis
is contained in the open interval $\left(-\frac{\pi}{2},+\frac{3\pi}{2}\right)$;
the last step consists in summing the results of both integrations.
For a point $\left(\psi,\omega\right)$ in $M_{P_{\nu}=0}^{S_{R}}$,
we have $+\frac{\pi}{2}<\psi<+\frac{3\pi}{2}$ and $-\frac{\pi}{2}\leq\psi^{*}\left(\omega\right)\leq+\frac{\pi}{2}$
. Then, $\psi-\psi^{*}\left(\omega\right)>0$ and $\delta\left[\psi-\psi^{*}\left(\omega\right)\right]=0$,
which means that the integral along $M_{P_{\nu}=0}^{S_{R}}$ is zero,
and the integral along $M_{P_{\nu}=0}^{S}$ in Eq.~(\ref{eq:G_S-1})
reduces to the integral along $M_{P_{\nu}=0}^{S_{L}}$.

If $M_{P_{\nu}=0}^{S_{L}}$ is a non-empty set, the projection of
$M_{P_{\nu}=0}^{S_{L}}$ in the $\psi$-axis can be represented by
the interval $\left[\psi_{1}^{S}(\nu),\,\psi_{2}^{S}(\nu)\right]$,
and we can use (\ref{eq:definite-integral}) and (\ref{eq:G_S-1})
to obtain 
\begin{equation}
G_{S}(\nu)=\int_{\psi_{1}^{S}(\nu)}^{\psi_{2}^{S}(\nu)}\delta\left\{ \psi-\psi^{*}\left[F_{\nu}(\psi)\right]\right\} g\left[F_{\nu}(\psi)\right]\,d\psi,\label{eq:G_S_2}
\end{equation}
where $F_{\nu}$ is defined by Eq.~(\ref{eq:f}), and the integration
limits, $\psi_{1}^{S}(\nu)$ and $\psi_{2}^{S}(\nu)$, are given in
Table \ref{tab:T1}. 

\begin{table}
\begin{centering}
\begin{ruledtabular}
\begin{tabular}{ccc}
 & $-2<\tilde{\nu}\leq0$ & $0<\tilde{\nu}<+2$\tabularnewline
$\psi_{1}^{S}(\nu)$ & {\small{}{}-$\arcsin\left(1+\tilde{\nu}\right)$} & {\small{}{}$-\frac{\pi}{2}$}\tabularnewline
$\psi_{2}^{S}(\nu)$ & {\small{}{}$+\frac{\pi}{2}$} & {\small{}{}$\arcsin\left(1-\tilde{\nu}\right)$}\tabularnewline
\end{tabular}
\end{ruledtabular}
\par\end{centering}
\caption{Integration limits of Eq.~(\ref{eq:G_S_2}), given by $\psi_{1}^{S}(\nu)$
and $\psi_{2}^{S}(\nu)$. For $\left|\tilde{\nu}\right|\protect\geq2$,
$\psi_{1}^{S}(\nu)$ and $\psi_{2}^{S}(\nu)$ are not defined, and
$G_{S}(\nu)=0$.}
\label{tab:T1}
\end{table}

For $\left|\tilde{\nu}\right|\geq2$, $\psi_{1}^{S}(\nu)$ and $\psi_{2}^{S}(\nu)$
are not defined, and $G_{S}(\nu)=0$. This case is illustrated by
Figs. \ref{MP_SD}(a) and (e), which show $M_{P_{\nu}=0}$ entirely
outside $V_{S}$. The case $-2<\tilde{\nu}\leq0$ is related to configurations
with $M_{P_{\nu}=0}$ partly or completely inside $V_{S}$, such as
those depicted in Figs. \ref{MP_SD}(b)-(c). When $0<\tilde{\nu}<+2$,
there is still partial embedding of $M_{P_{\nu}=0}$ in $V_{s}$(See
Fig.~\ref{MP_SD}(d)).

From the definitions of $\psi^{*}$ and $F_{\nu}$ (see Eqs. (\ref{eq:attractor})
and (\ref{eq:f})), the delta function in the integrand of Eq.~(\ref{eq:G_S_2})
is given by 
\begin{equation}
\delta\left\{ \psi-\psi^{*}\left[F_{\nu}(\psi)\right]\right\} =\delta\left[\psi-\arcsin\left(\sin\psi+\tilde{\nu}\right)\right],\label{eq:delta-Gs}
\end{equation}
which is singular when 
\begin{equation}
\psi=\arcsin\left(\sin\psi+\tilde{\nu}\right).\label{eq:singularity}
\end{equation}
By applying the sine function on both sides of Eq.~(\ref{eq:singularity}),
we see that the singularity occurs if and only if $\nu=\Omega$ independently
of the value taken by $\psi$. So Eq.~(\ref{eq:delta-Gs}) only makes
sense in an integral where $\nu$ is the integration variable and will
remain as a delta function despite of integration in Eq.~(\ref{eq:G_S_2}).

From Eq.~(\ref{eq:delta-Gs}), $\delta\left\{ \psi-\psi^{*}\left[F_{\nu}(\psi)\right]\right\} $
in its Gaussian form reads 
\begin{equation}
\delta\left\{ \psi-\psi^{*}\left[F_{\nu}(\psi)\right]\right\} =L\left[\psi-\arcsin\left(\sin\psi+\tilde{\nu}\right)\right],\label{eq:delta-gaussian}
\end{equation}
where
\begin{equation}
L(x)=\lim_{\epsilon\rightarrow0^{+}}\frac{1}{\epsilon\sqrt{\pi}}\exp\left[-\left(\frac{x}{\epsilon}\right)^{2}\right].
\end{equation}
This limit can be analyzed in an arbitrarily small open neighborhood
$\mathcal{{N}}$ of radius $\epsilon$ centered in the singularity
point, defined by $\nu=\Omega$. As expected for delta functions in
non-singular points, if $v$ is not in $\mathcal{{N}}$, i.e. $\left|\nu-\Omega\right|\geq\epsilon$,
the limit in (\ref{eq:delta-gaussian}) is zero, since the Gaussian
is $O(\epsilon^{2m-1})$ as $\epsilon\rightarrow0^{+}$ for any positive
integer $m$. If $v$ is inside $\mathcal{{N}},$i.e. $\left|\nu-\Omega\right|<\epsilon$,
then 
\begin{equation}
\arcsin\left(\sin\psi+\tilde{\nu}\right)=\psi+\frac{\tilde{\nu}}{\cos\psi}+O\left(\tilde{\nu}^{2}\right),\label{eq:arcsin-asym}
\end{equation}
for $\epsilon\longrightarrow0^{+}$. By redefining $\epsilon$ as
$\epsilon\cos\psi$ and substituting (\ref{eq:arcsin-asym}) in (\ref{eq:delta-gaussian}),
we obtain 
\begin{equation}
\delta\left\{ \psi-\psi^{*}\left[F_{\nu}(\psi)\right]\right\} =\cos\psi L(\tilde{\nu}),\label{eq:limit-delta}
\end{equation}
where $L(\tilde{\nu})$ is the Gaussian representation of $\delta\left(\tilde{\nu}\right)$.
Substituting (\ref{eq:limit-delta}) in (\ref{eq:G_S_2}), we have
\begin{equation}
G_{S}(\nu)=\delta\left(\tilde{\nu}\right)\int_{\psi_{a}^{S}(\nu)}^{\psi_{b}^{S}(\nu)}g\left(a\sin\psi+\nu\right)\cos\psi\,d\psi,\label{eq:G_S_2-1}
\end{equation}
which is the same as
\begin{equation}
G_{S}(\nu)=a\delta\left(\nu-\Omega\right)\int_{-\frac{\pi}{2}}^{+\frac{\pi}{2}}g\left(a\sin\psi+\Omega\right)\cos\psi\,d\psi.\label{eq:GS-final_form}
\end{equation}
Changing the integration variable $\psi$ to $\omega=a\sin\psi+\Omega$,
Eq.~(\ref{eq:GS-final_form}) results in
\begin{equation}
G_{S}(\nu)=\delta\left(\nu-\Omega\right)S(K),\label{eq:GS-final_form-1}
\end{equation}
where $S(K)$, as mentioned in Sec.\ref{sec:Some-results-from-Kuramoto},
is the fraction of synchronized oscillators, defined by Eq.~(\ref{eq:r_bar_S}).
It is worth mentioning that $G_{S}(\nu)$ is identical to the singular
term in the distribution of time-averaged frequencies, as shown by
Eq.~(\ref{eq:Gbar}).

A similar geometric analysis can be used to calculate $G_{D}(\nu)$
from Eq.~(\ref{eq:G_D}), where integration is now performed along
the curve $M_{P_{\nu}=0}^{D}$. As mentioned before, this curve corresponds
to the part of $M_{P_{\nu}=0}$ in $V_{D}.$ A point $(\psi,\omega)$
in $M_{P_{\nu}=0}^{D}$ satifies the conditions: i) $\omega=F_{\nu}(\psi)$;
ii) $\omega<\Omega-a$ or $\omega>\Omega+a$ (see orange curves in
Figs. \ref{MP_SD}(a),(b),(d), and (e)). From condition i), (\ref{eq:f}),
and (\ref{eq:inst-freq-rotating}) , we have that $\dot{\psi}=\nu-\Omega$.
From condition ii), $p(\psi|\omega)$ is defined by (\ref{eq:cond_prob_desync}),
and Eq.~(\ref{eq:G_D}) can then be rewritten as 
\begin{equation}
G_{D}(\nu)=\frac{1}{2\pi\tilde{\nu}}\intop_{M_{P_{\nu}=0}^{D}}\frac{\tilde{\omega}g(\omega)}{\left|\nabla P_{\nu}(\psi,\omega)\right|}\sqrt{1-\frac{1}{\tilde{\omega}^{2}}}dl.\label{eq:G_D-1}
\end{equation}
For $\nu=\Omega$, $M_{P_{\nu}=0}^{D}$ is an empty set, as shown
in Fig.~\ref{MP_SD}(c), and $G_{D}(\nu)=0$.

Let $U(x)$ be defined as
\begin{equation}
U(x)=\sqrt{x^{2}-1}.
\end{equation}
From Eq.(\ref{eq:definite-integral}) and condition i), (\ref{eq:G_D-1})
reads 
\begin{equation}
G_{D}(\nu)=\frac{1}{\pi\left|\tilde{\nu}\right|}\int_{\psi_{1}^{D}(\nu)}^{\psi_{2}^{D}(\nu)}g\left[F_{\nu}(\psi)\right]U\left[\tilde{F}_{\nu}(\psi)\right]d\psi,\label{eq:G_D-3}
\end{equation}
where $\psi_{1}^{D}(\nu)$ and $\psi_{2}^{D}(\nu)$ are given in Table~\ref{tab:T2}.
\begin{table}
\begin{ruledtabular}
\begin{tabular}{ccccc}
 & $\tilde{\nu}\leq-2$  & $-2<\tilde{\nu}<0$  & $0<\tilde{\nu}<+2$  & $+2\leq\tilde{\nu}$ \tabularnewline
$\psi_{1}^{D}(\nu)$  & {\small{}{}$-\frac{\pi}{2}$}  & {\small{}{}$-\frac{\pi}{2}$}  & {\small{}{}$\arcsin\left(1-\tilde{\nu}\right)$ }  & {\small{}{}$-\frac{\pi}{2}$ }\tabularnewline
$\psi_{2}^{D}(\nu)$  & {\small{}{}+$\frac{\pi}{2}$}  & {\small{}{}-$\arcsin\left(1+\tilde{\nu}\right)$}  & {\small{}{}+$\frac{\pi}{2}$ }  & {\small{}{}+$\frac{\pi}{2}$ }\tabularnewline
\end{tabular}
\end{ruledtabular}
\caption{Integration limits of Eq.~(\ref{eq:G_D-3}), given by $\psi_{1}^{D}(\nu)$
and $\psi_{2}^{D}(\nu)$. The limits are not defined for $\nu=\Omega$.
(\ref{eq:G_D-3}) does not apply, and $G_{D}(\nu)=0$. }
\label{tab:T2} 
\end{table}

As $G_{S}$ and $G_{D}$ are given by Eqs. (\ref{eq:GS-final_form-1})
and (\ref{eq:Gd}), we can now return to Eq.~(\ref{eq:G-4}) and write
$G$ in its final form: 
\begin{equation}
G(\nu)=\delta\left(\nu-\Omega\right){\rm S}\left(K\right)+G_{D}(\nu),\label{eq:G-final-form}
\end{equation}
where $G_{D}(\Omega)=0$, and 
\begin{equation}
G_{D}(\nu)=\frac{1}{\pi\left|\tilde{\nu}\right|}\int_{\psi_{\nu}^{-}}^{\psi_{\nu}^{+}}h_{\nu}(\psi)\,d\psi,\label{eq:Gd}
\end{equation}
if $\nu\neq\Omega$. In Eq.~(\ref{eq:Gd}), the quantities $h_{\nu}(\psi)$,
$\chi_{\nu}$, $\psi_{\nu}^{-}$, and $\psi_{\nu}^{+}$ are defined
by
\begin{equation}
h_{\nu}(\psi)=g\left(a\sin\psi+\nu\right)U\left(\sin\psi+\tilde{\nu}\right)
\end{equation}
and 
\begin{equation}
\psi_{\nu}^{\pm}=\arcsin\left\{ -\left(\tilde{\nu}\pm2\right)\Theta\left[-\tilde{\nu}\left(\tilde{\nu}\pm2\right)\right]\pm1\right\} ,
\end{equation}
where $\Theta\left[\,.\,\right]$ denotes the Heaviside step function.
Note that $\psi_{\nu}^{-}$ and $\psi_{\nu}^{+}$ correspond to the
previously defined $\psi_{1}^{D}(\nu)$ and $\psi_{2}^{D}(\nu)$.

The singular term in (\ref{eq:G-final-form}) means that 
\begin{equation}
\lim_{\epsilon\rightarrow0^{+}}\intop_{\Omega-\epsilon}^{\Omega+\epsilon}G(\nu)d\nu={\rm S}\left(K\right),
\end{equation}
i.e. the probability of an oscillator with instantaneous frequency
in an infinitesimally small neighborhood of $\Omega$ is given by
the fraction of synchronized oscillators.

Our procedure to obtain Eq.~(\ref{eq:G-final-form}) does not depend
in any symmetry assumption related to $g$. But let us now assume
the situation where $g(\Omega+x)=g(\Omega-x)$ for any positive number
$x$. This implies that $G$ has the same property, viz. if $g(\Omega+x)=g(\Omega-x)$,
then $G(\Omega+x)=G(\Omega-x)$. To show this, it is sufficient showing
that $G_{D}$ is also symmetric. For $x>0$, we have 
\begin{equation}
G_{D}(\Omega\pm x)=\frac{a}{\pi x}\int_{\psi_{\Omega\pm x}^{-}}^{\psi_{\Omega\pm x}^{+}}h_{\Omega\pm x}(\psi)\,d\psi,\label{eq:symmetry}
\end{equation}
and $\psi_{\Omega\pm x}^{-}=-\psi_{\Omega\mp x}^{+}.$ In the formula
for $G_{D}(\Omega+x)$ (see Eq.~(\ref{eq:symmetry})), we can introduce
the following changes: first, we change $\psi_{\Omega+x}^{-}$ by
$-\psi_{\Omega-x}^{+}$ and $\psi_{\Omega+x}^{+}$ by $-\psi_{\Omega-x}^{-}$;
second, we redefine $\psi$ as $-\psi$. Then,

\begin{equation}
G_{D}(\Omega+x)=\frac{a}{\pi x}\int_{\psi_{\Omega-x}^{-}}^{\psi_{\Omega-x}^{+}}h_{\Omega+x}(-\psi)\,d\psi',\label{eq:symmetry-1}
\end{equation}
Since $g\left[\Omega-\left(a\sin\psi-x\right)\right]=g\left(a\sin\psi+\Omega-x\right)$
(from the symmetry assumption of $g$) and $U\left(-\sin\psi+\frac{x}{a}\right)=U\left(\sin\psi-\frac{x}{a}\right)$,
we have $h_{\Omega+x}(-\psi)=h_{\Omega-x}(\psi)$. Then, from Eq.
(\ref{eq:symmetry-1}), $G_{D}(\Omega+x)=G_{D}(\Omega-x)$, which
proves our initial statement.

According to Eq.~(\ref{eq:G-final-form}), $G$ consists of a delta
peak and a distribution of instantaneous frequencies for non-synchronized
oscillators ($G_{D}$). $G_{D}$ is zero at $\Omega$, where the delta
peak is located. $\overline{G}$ has a similar form: the same delta
peak at $\Omega$, and a distribution of time-averaged instantaneous
frequencies for non-synchronized oscillators ($\overline{G}_{D}$).
$\overline{G}_{D}$ is also zero at $\Omega$. Since a synchronized
oscillator's instantaneous frequency goes to $\Omega$ as time goes
to infinity, the same happens to its average in time. This explains
why the fraction of synchronized oscillators, defined by $S(K)$,
appears as a factor in the delta peaks of both $G$ and $\overline{G}$.

\section{Application to Gaussian and Beta distributions\label{sec:Applications}}

In this section we illustrate our analytical results about the instantaneous
frequency distributions on two distributions of natural frequencies, 
the normal (Gaussian) distribution and the Beta distribution. 
Both are unimodal, but one has unbounded support, whereas the other lives on a finite interval (Beta distribution).

\subsection{General features}

\begin{figure*}
\begin{centering}
\subfloat[]{\includegraphics[width=0.5\linewidth]{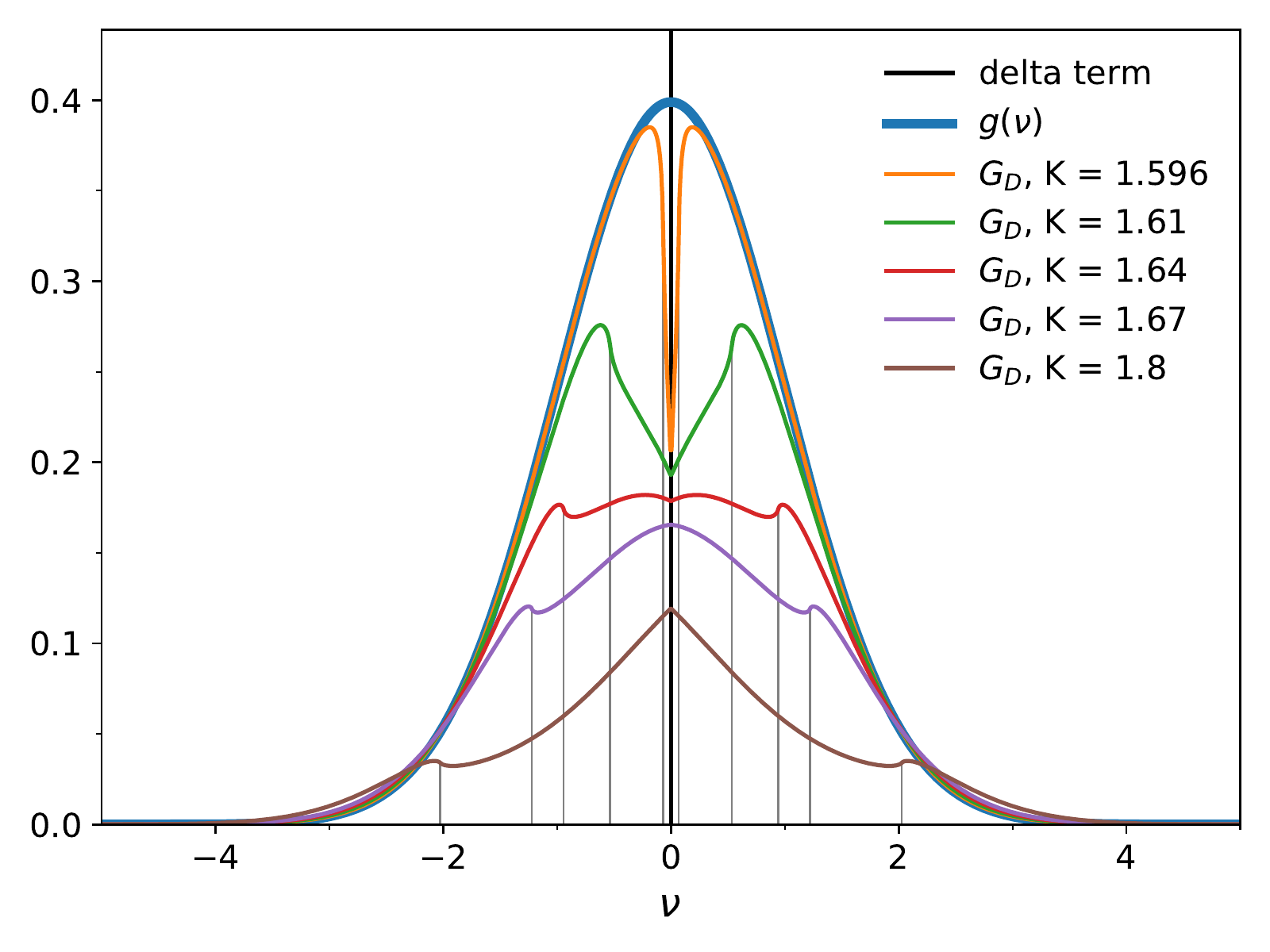} }\subfloat[]{\begin{centering}
\includegraphics[width=0.5\linewidth]{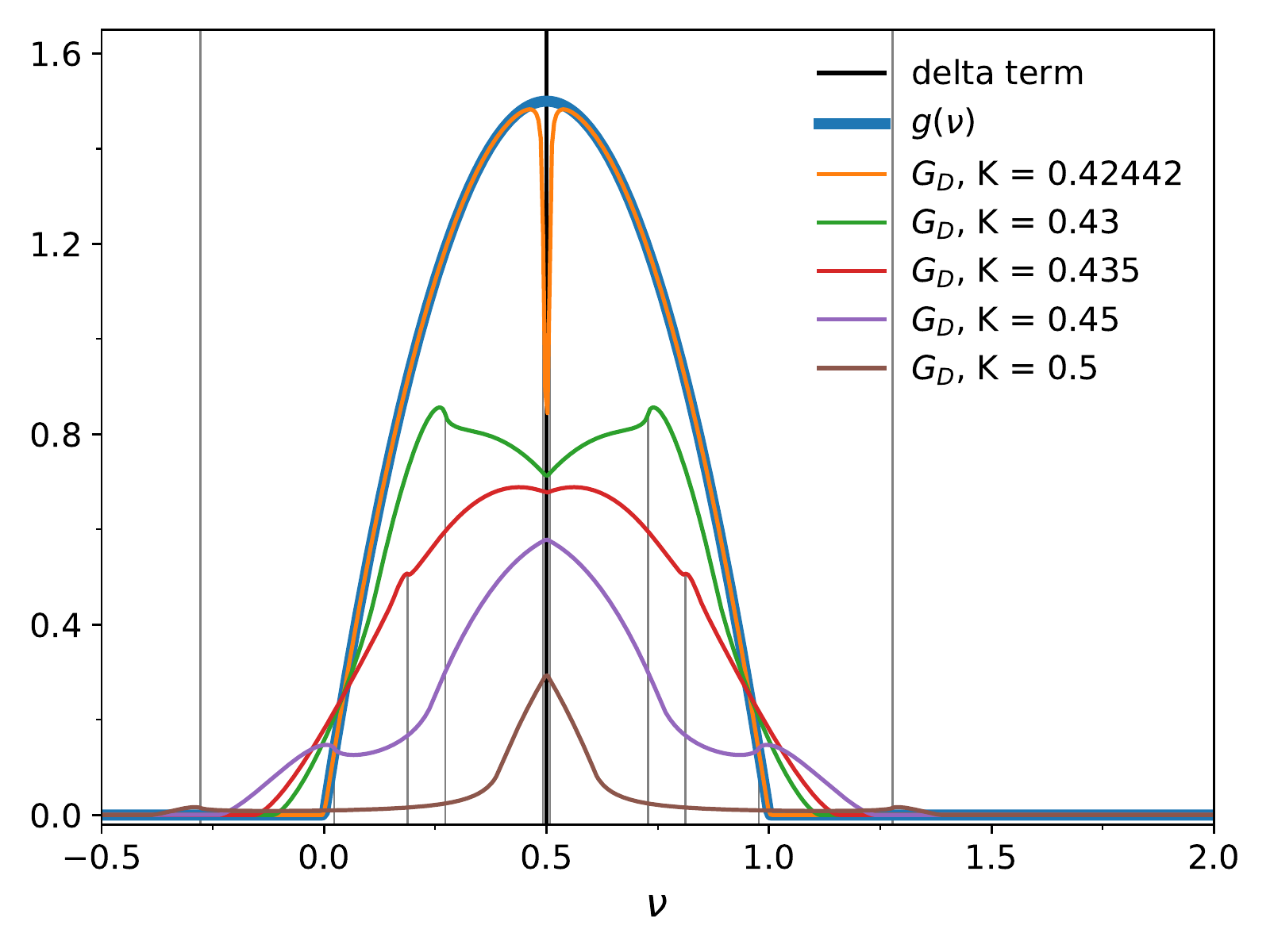}
\par\end{centering}
}
\par\end{centering}
\caption{(a) Instantaneous frequency distribution ($G$) for
natural frequencies normally distributed. The thin vertical lines
show different locations of $\Omega_{-}$ and $\Omega_{+}$, defined
by $\Omega_{\pm}=\Omega\pm2a$. Fractions of synchronized oscillators ($S(K)$): $S(1.596)=0.03$; $S(1.61)=0.21$
$S(1.64)=0.36$; $S(1.67)=0.46$; $S(1.8)=0.69$. (b) Graphs of $G$ assuming a Beta(2,2)
distribution of natural frequencies. $S(K)$: $S(0.42442)=0.01$; $S(0.43)=0.34$, 
$S(0.435)=0.45$; $S(0.45)=0.66$; $S(0.5)=0.93$.}
\label{fig:G-Gauss-Beta}
\end{figure*}

The Gaussian natural frequency distribution considered is 
\begin{equation}
g(\omega)=\frac{1}{\sqrt{2\pi}}\exp\left(-\frac{\omega^{2}}{2}\right),\label{eq:normal-dist}
\end{equation}
for which $\Omega=0$, and, according to Eq.~(\ref{eq:Kc}), $K_{c}\simeq1.5957$.

The Beta distribution considered reads  
\begin{equation}
g(\omega)=\begin{cases}
6\omega\left(1-\omega\right) & \omega\in\left[0,1\right]\\
0 & \omega\notin\left[0,1\right].
\end{cases}\label{eq:Beta(2,2)-dist}
\end{equation}
which is usually called Beta$\left(2,2\right)$.
All members of the family of Beta distributions have the support interval
$\left[0,1\right]$. So in this example oscillators have no natural
frequencies outside the interval $\left[0,1\right]$.  Given the symmetric shape,
the synchronization frequency is $\Omega=0.5$. The critical coupling strength,
given by Eq.~(\ref{eq:Kc}), is $K_{c}\simeq0.42441$.

In Fig.~\ref{fig:G-Gauss-Beta}, we show these two distributions (thick blue curves), but also 
the distribution of instantaneous frequencies for different values of the coupling strength above the critical coupling
$K_c$. (For subcritical coupling values, the instantaneous frequencies are the natural frequencies.)
For $K>K_{c}$, $G(\nu)=G_{S}(\nu)+G_{D}(\nu)$,
where $G_{S}(\nu)$ and $G_{D}(\nu)$ are defined by Eqs. (\ref{eq:GS-final_form-1})
and (\ref{eq:Gd}), respectively. Since $G_{S}(\nu)$ is a Dirac delta
term, we represent it by a black vertical line located in $\Omega$.

The thin colored curves show $G_{D}$ for different values of $K$. This continuous part of $G$ obeys the
symmetry of  $g$, the distribution of natural frequencies. The tails of $G_D$ are fatter than those of 
$g$. In particular, $G_D$ extends beyond the interval of support of $g$ in the Beta case. 
In the central region, $G_D(\nu)< g(\nu)$. For $K\gtrsim K_c$, $G_D$ is very close to $g$, but for a sharp
drop near $\Omega$, the synchronized frequency (orange curves). This drop, however, does not extend to zero:
$G_D(\nu)$ tends to a finite value when $\nu\to 0$.
Increasing $K$, $G_D$ develops a more complicated structure: the central region decreases, the tails grow,
and some special values of $\nu$ appear, marked by thin gray vertical lines on the figure.
They indicate the locations of $\Omega_{-}$
and $\Omega_{+}$, defined by $\Omega_{\pm}=\Omega\pm2a$. The quantities
$\Omega_{-}$, $\Omega$, and $\Omega_{+}$ are endpoints of intervals
which define the integration limits of $G_{D}$ (see Table \ref{tab:T2}).
Since $G_{D}$ is a piecewise function, the graph of $G_{D}$ consists
of four sub-graphs associated to the intervals $\nu\leq\Omega_{-}$,
$\Omega_{-}<\nu<\Omega$, $\Omega<\nu<\Omega_{+}$, and $\Omega_{+}\leq\nu$.

\begin{figure*}

\begin{centering}
\subfloat[$K<K_{c}\simeq1.5957$]{\centering{}\includegraphics[width=0.33333\linewidth]{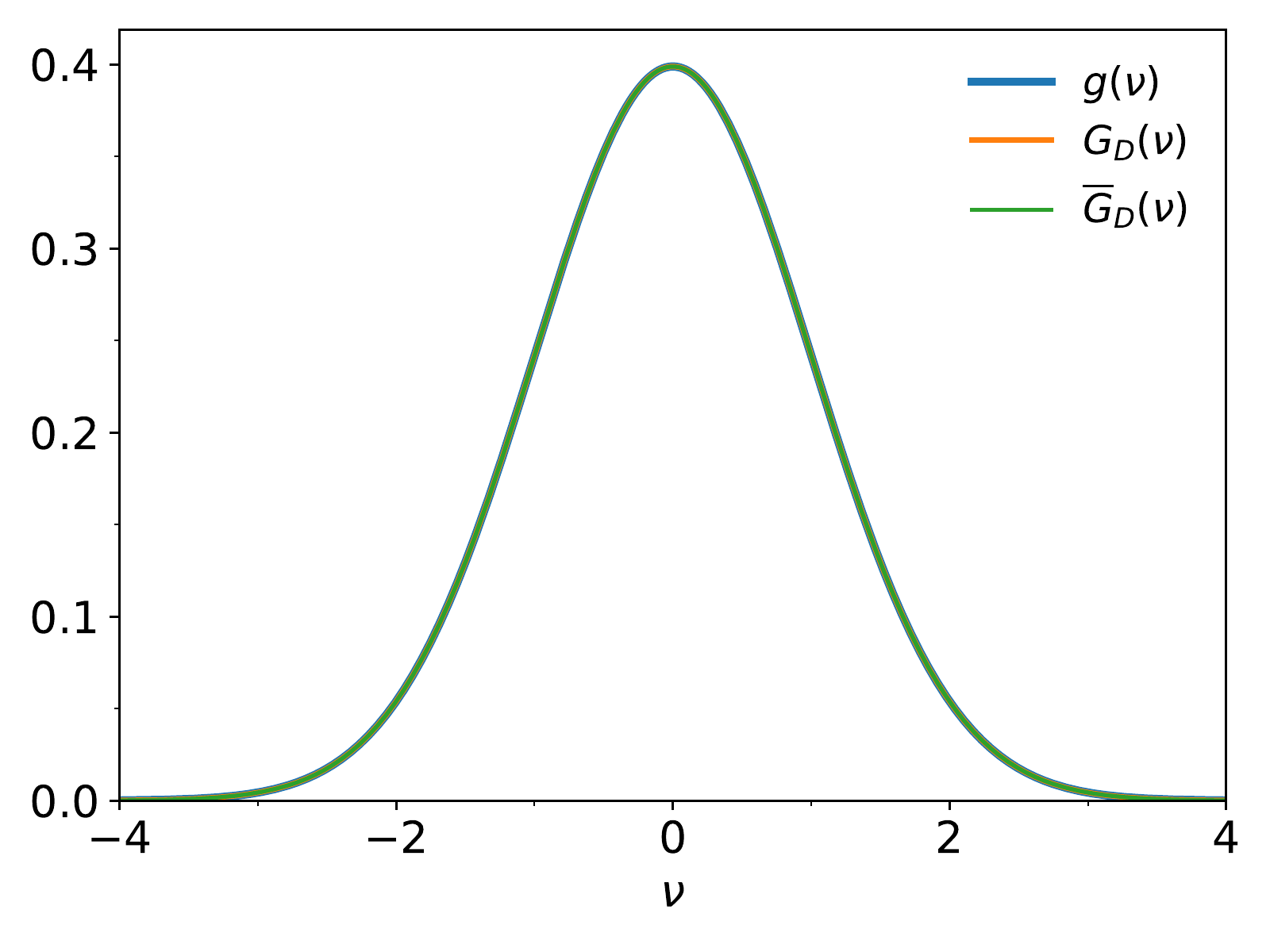}

}\subfloat[$K=1.596$]{\begin{centering}
\includegraphics[width=0.33333\linewidth]{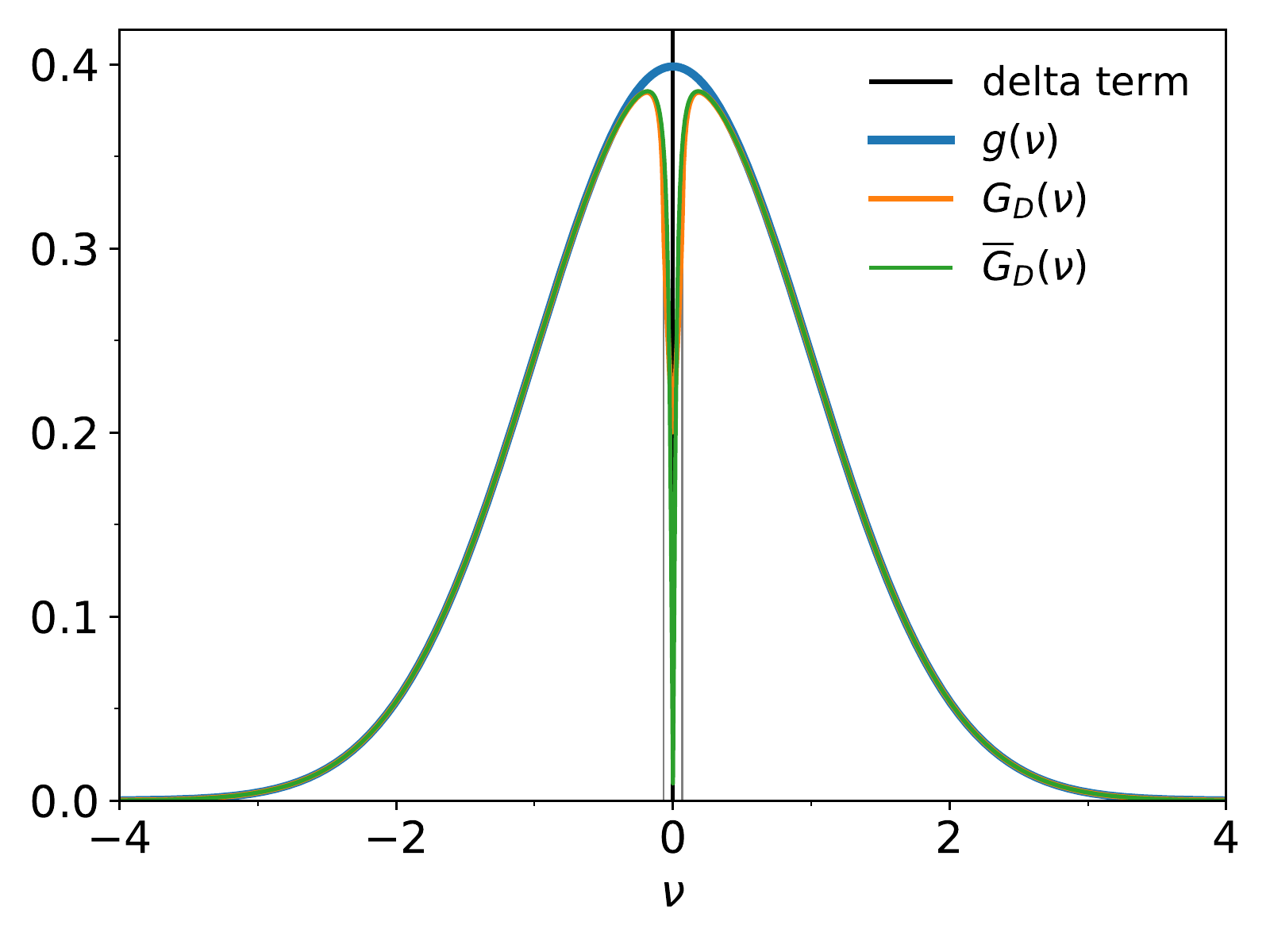}
\par\end{centering}
}\subfloat[$K=1.61$]{\begin{centering}
\includegraphics[width=0.33333\linewidth]{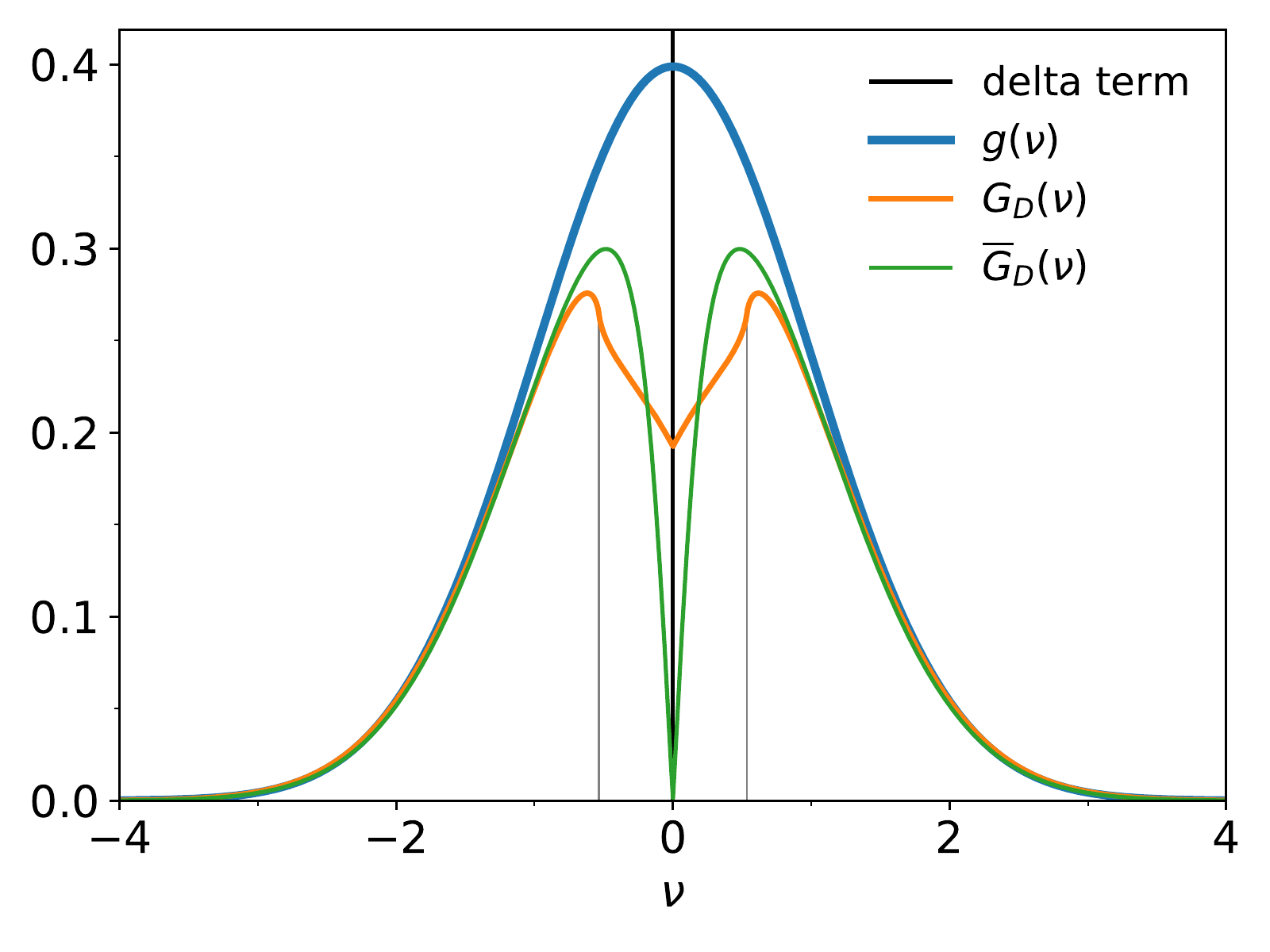}
\par\end{centering}
}
\par\end{centering}
\begin{centering}
\subfloat[$K=1.64$]{\centering{}\includegraphics[width=0.33333\linewidth]{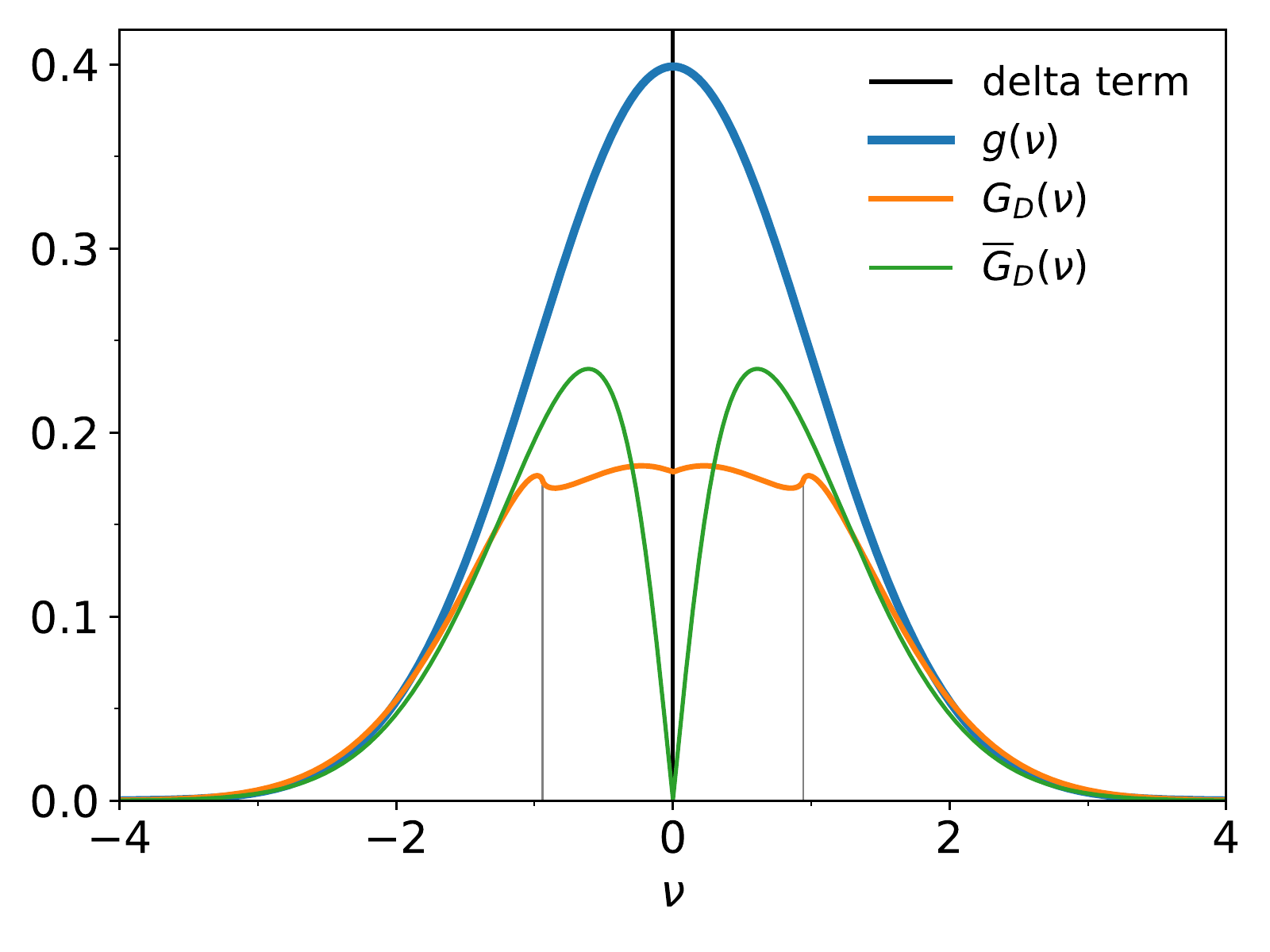}

}\subfloat[$K=1.67$]{\begin{centering}
\includegraphics[width=0.33333\linewidth]{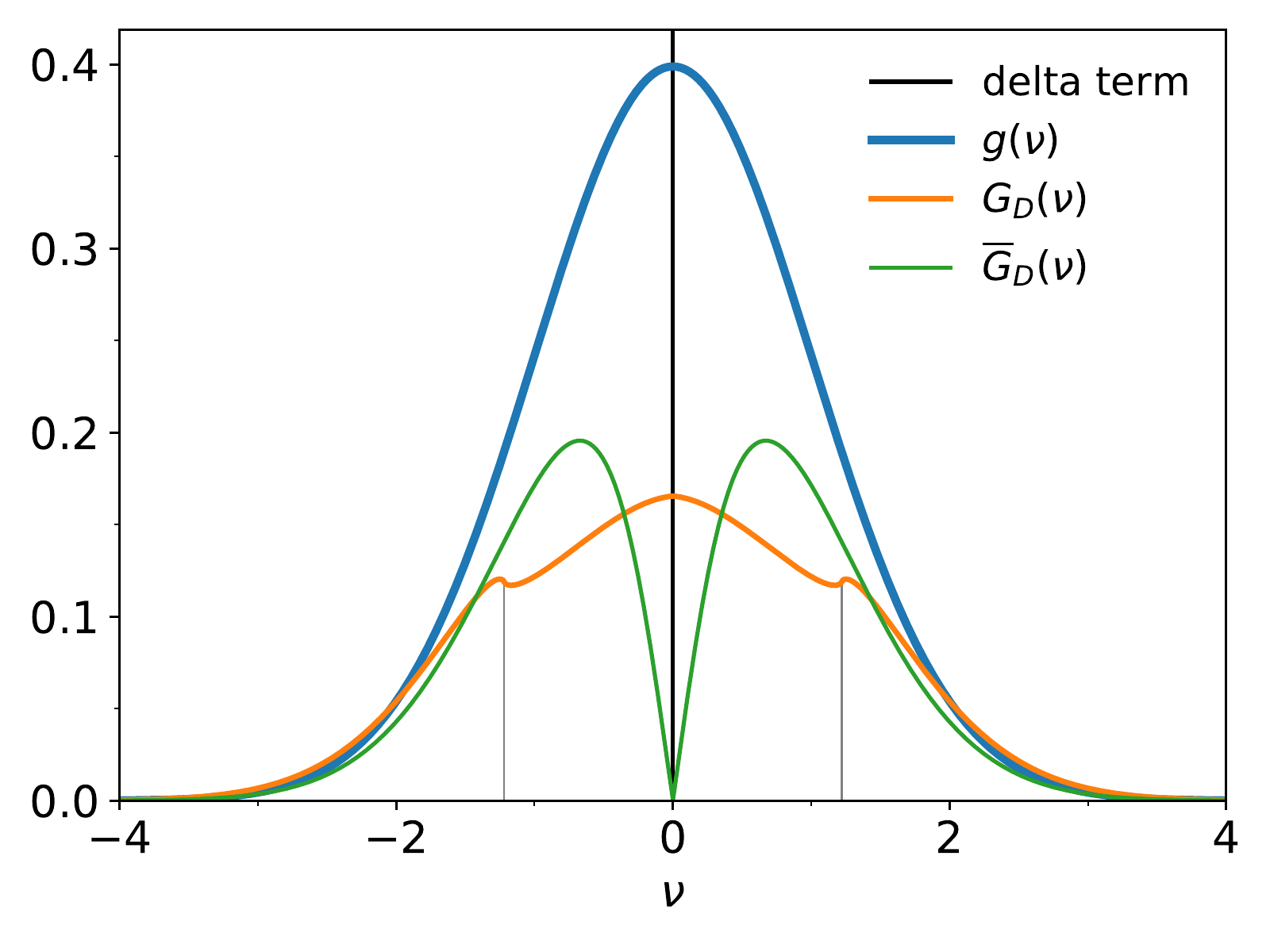}
\par\end{centering}
}\subfloat[$K=1.8$]{\centering{}\includegraphics[width=0.33333\linewidth]{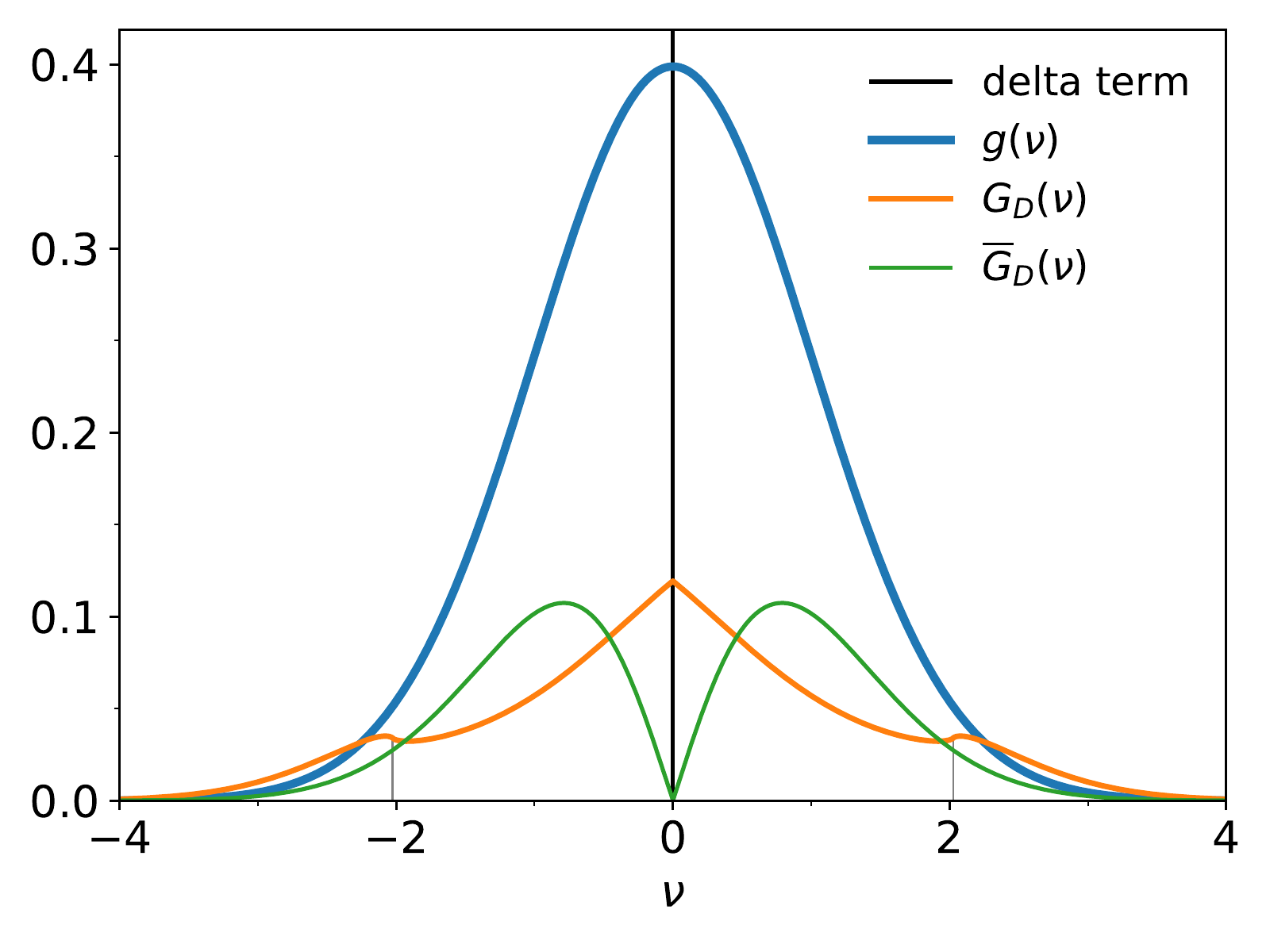}

}
\par\end{centering}

\caption{Comparison between the instantaneous frequency distribution, $G(\nu)$,
and the time-averaged frequency distribution, $\overline{G}(\nu)$.
A normal distribution of natural frequencies, $g(\nu)$, is assumed.
For $K<K_{c}$, $g(\nu)=G(\nu)=\overline{G}(\nu)$, and there is no
delta term, which means that $S(K)=0$.}
\label{fig:G-vs-Gbar-Gauss}
\end{figure*}

\begin{figure*}

\begin{centering}
\subfloat[$K<K_{c}\simeq0.42441$]{\centering{}\includegraphics[width=0.33333\linewidth]{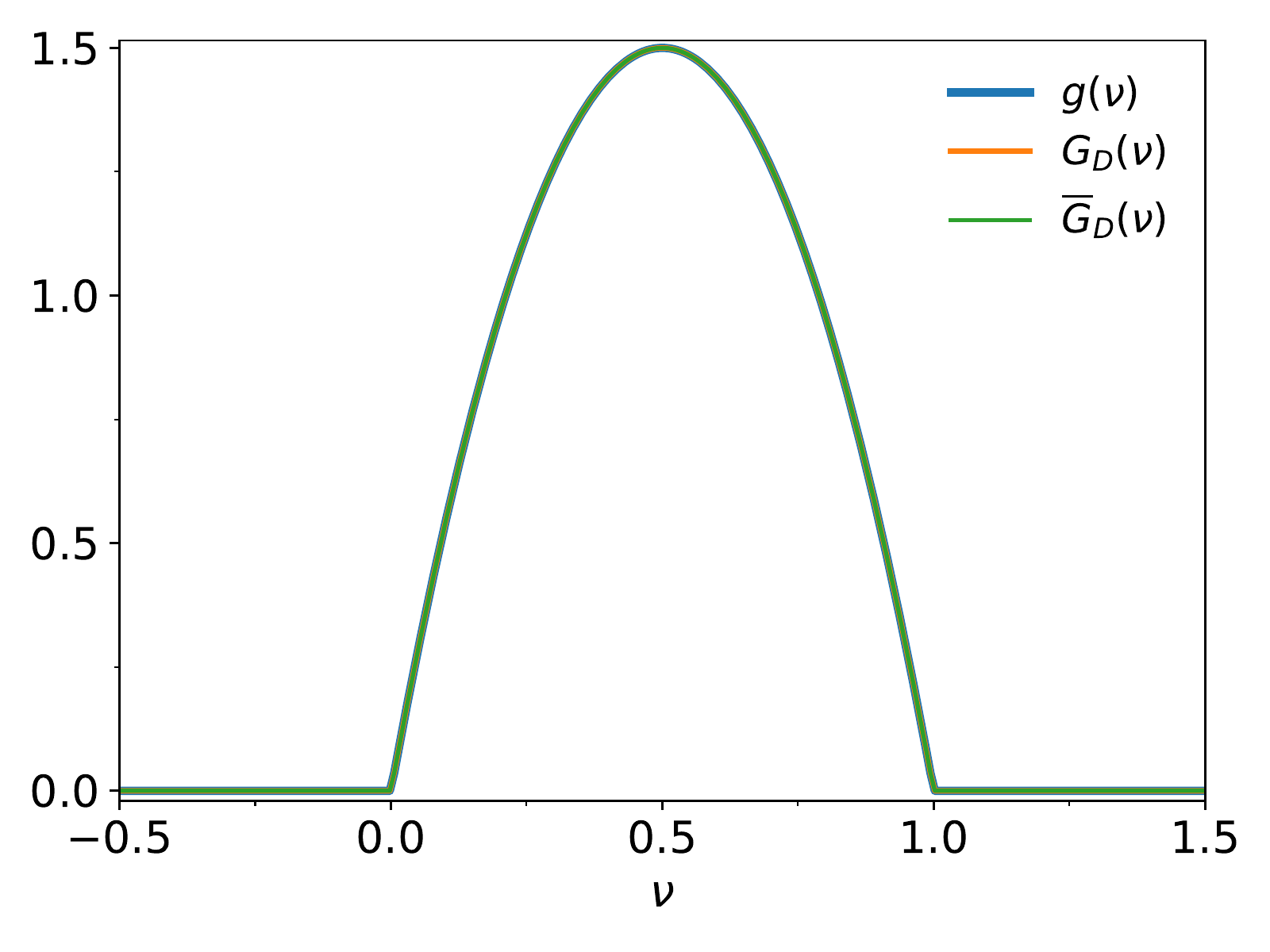}

}\subfloat[$K=0.42442$]{\centering{}\includegraphics[width=0.33333\linewidth]{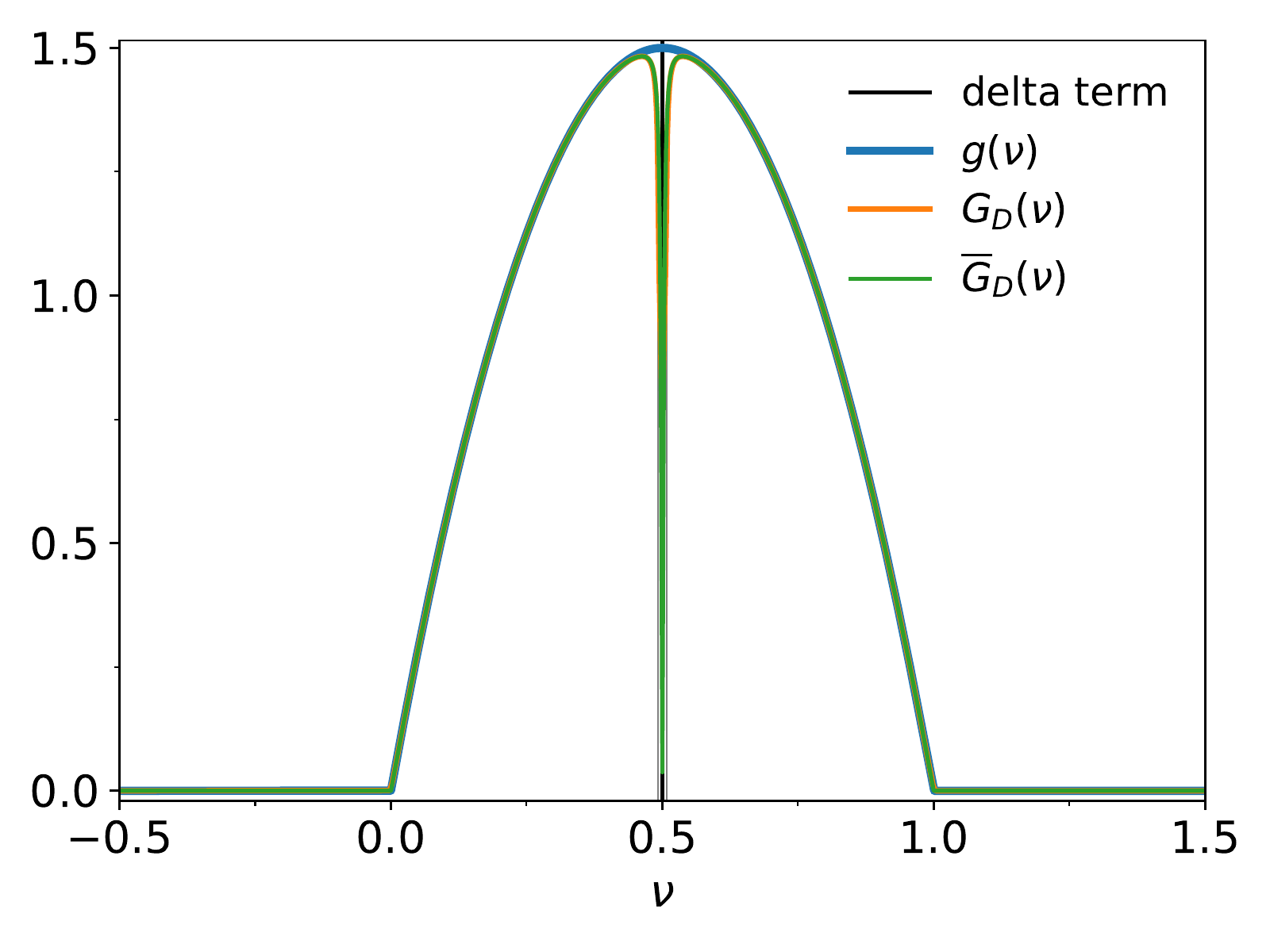}

}\subfloat[$K=0.43$]{\begin{centering}
\includegraphics[width=0.33333\linewidth]{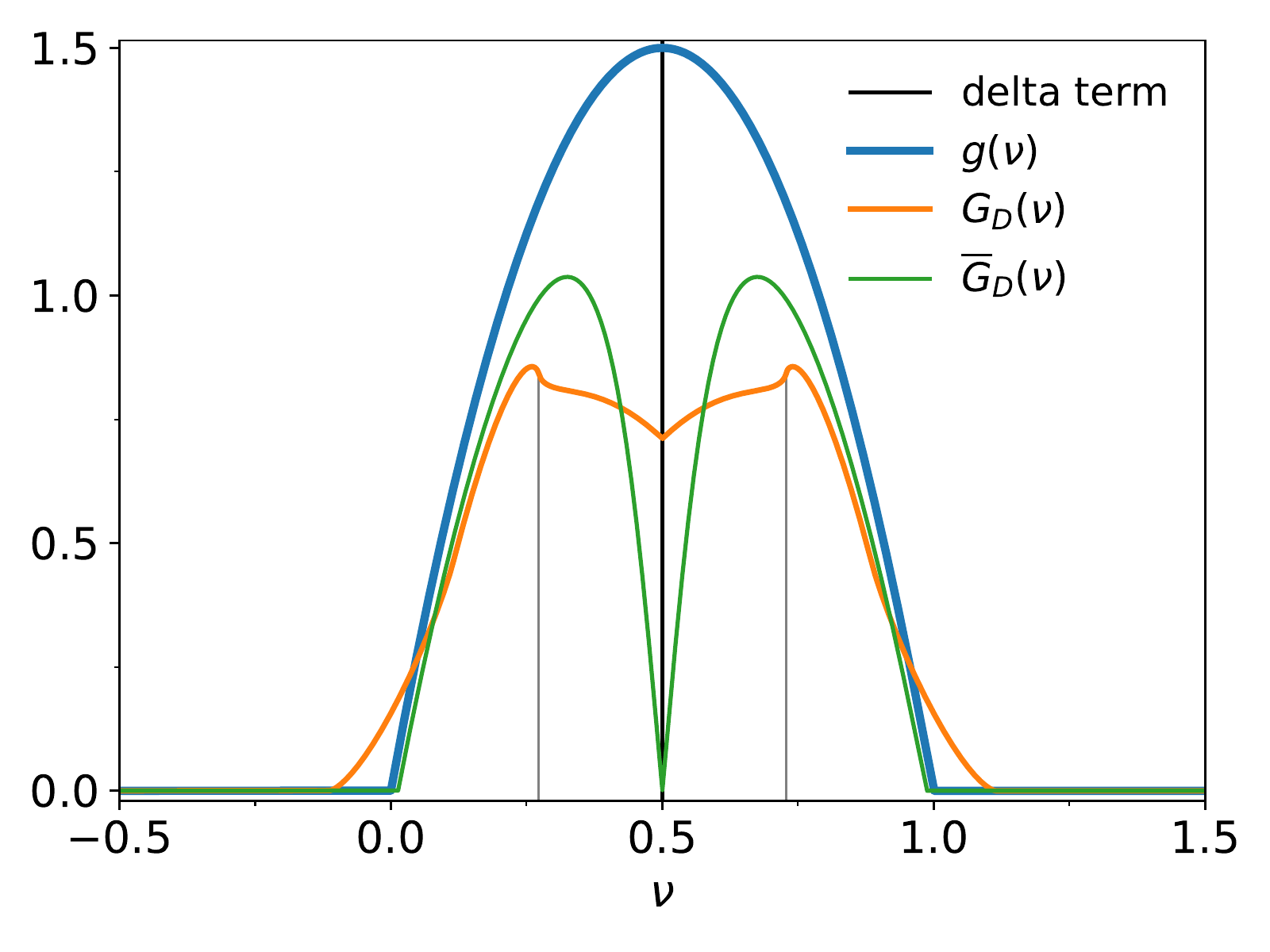}
\par\end{centering}
}
\par\end{centering}
\begin{centering}
\subfloat[$K=0.435$]{\centering{}\includegraphics[width=0.33333\linewidth]{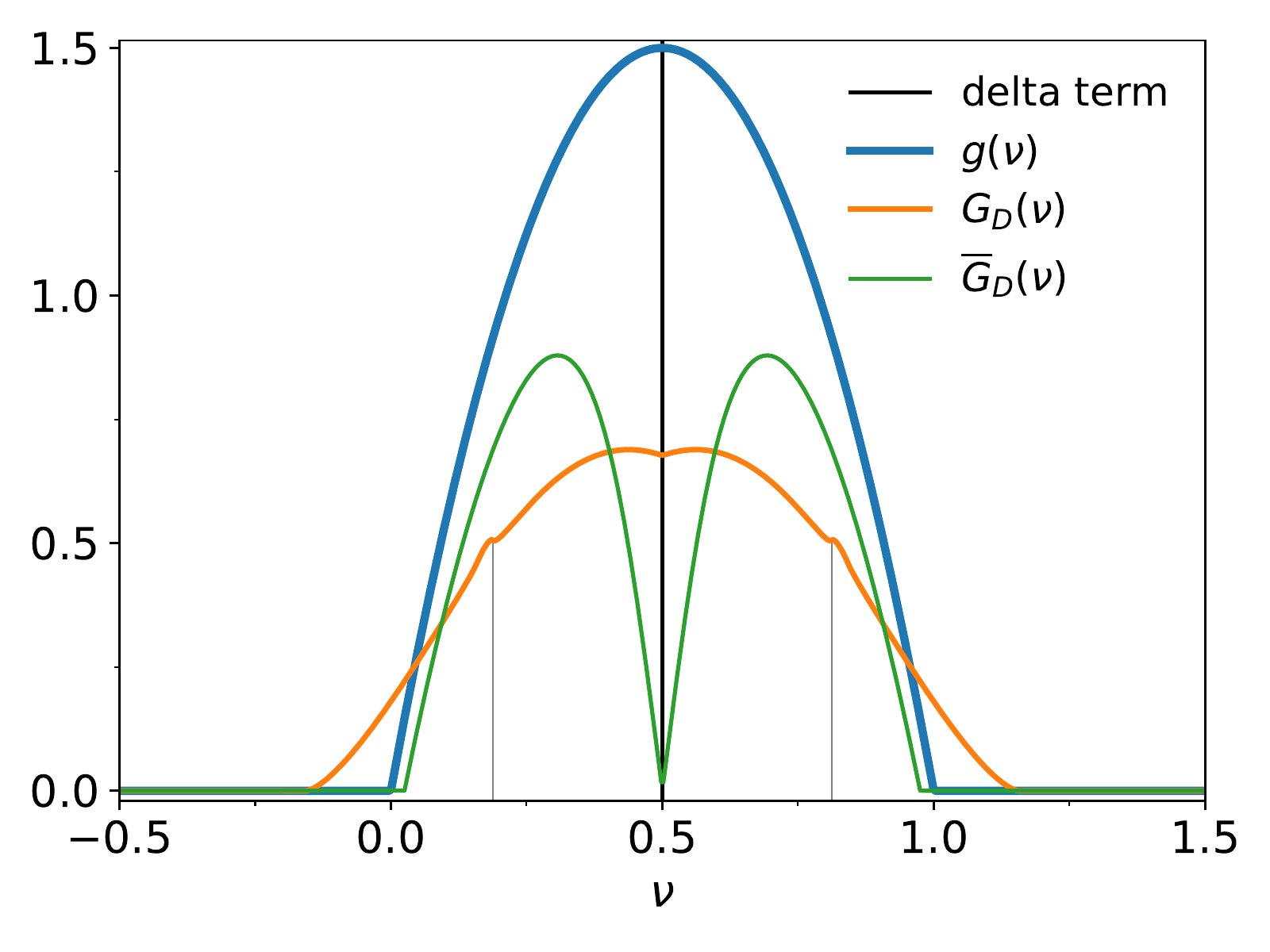}

}\subfloat[$K=0.45$]{\begin{centering}
\includegraphics[width=0.33333\linewidth]{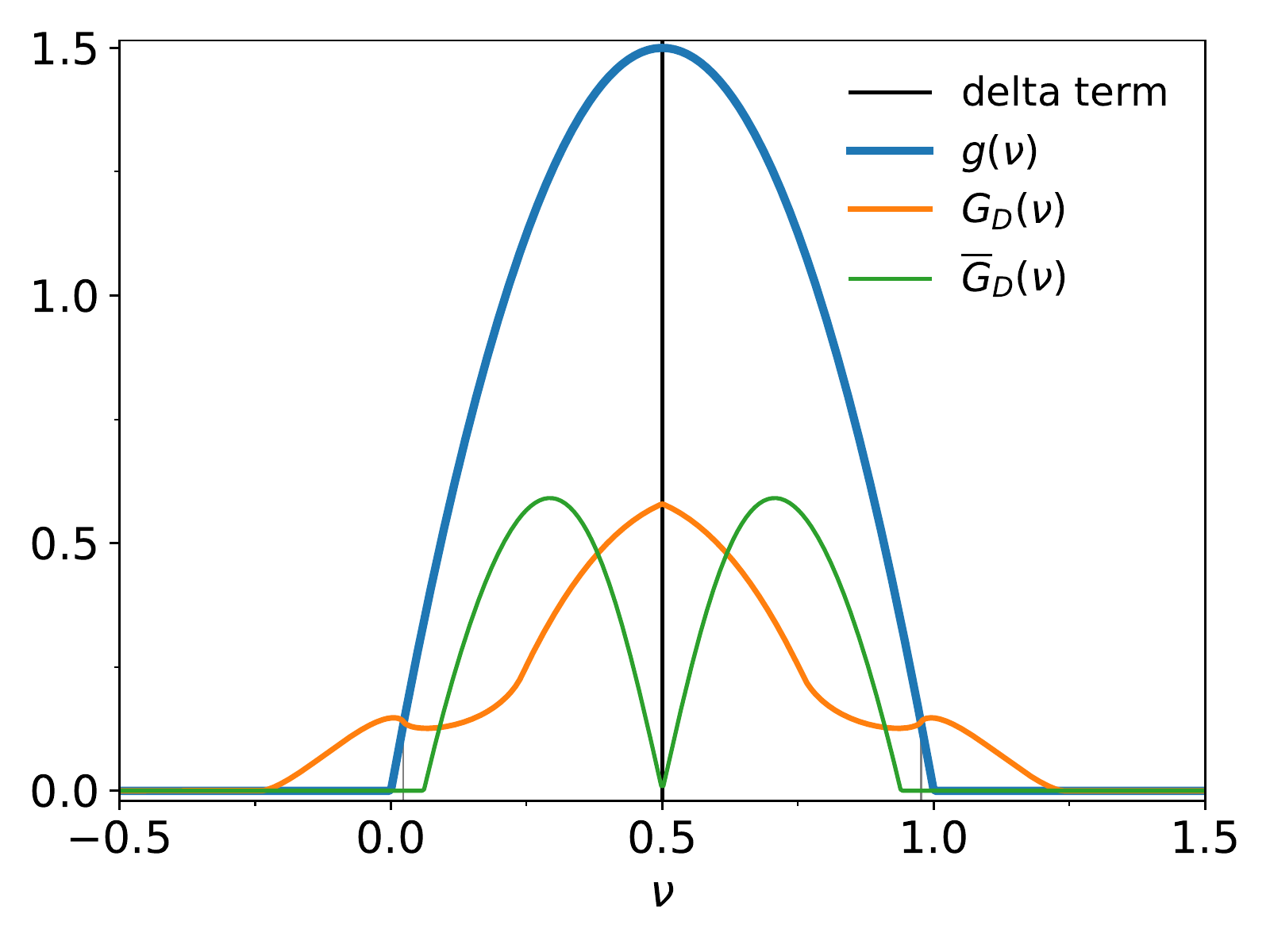}
\par\end{centering}
}\subfloat[$K=0.5$]{\centering{}\includegraphics[width=0.33333\linewidth]{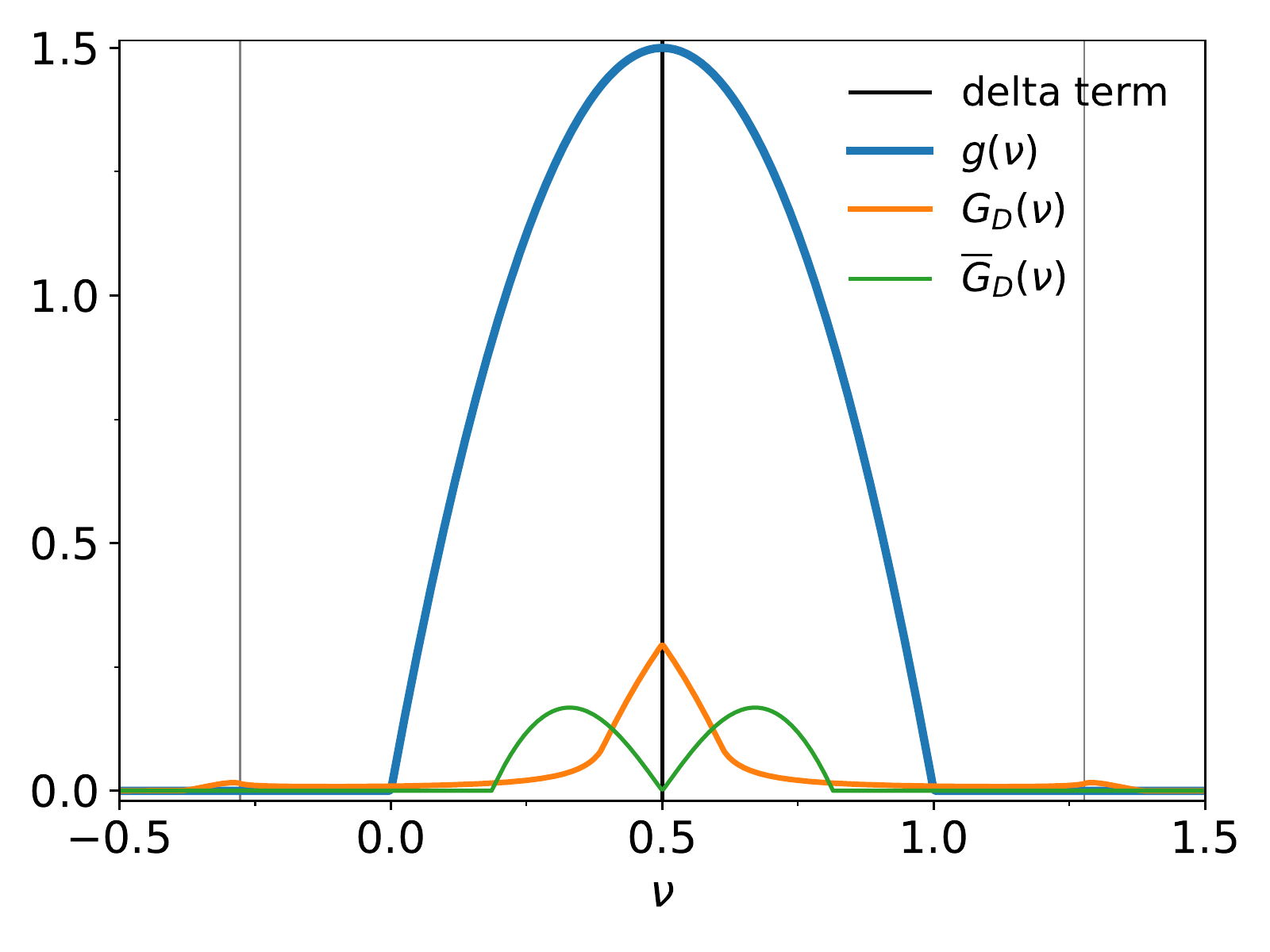}

}
\par\end{centering}

\caption{Comparison between $G$ and $\overline{G}$. $g$ is a Beta$\left(2,2\right)$
distribution. (a) For $K<K_{c}$, $g(\nu)=G(\nu)=\overline{G}(\nu)$,
and there is no delta term, which means that $S(K)=0$. (b-f) As $K$
increases, the bounding interval of $\overline{G}$ decreases its
width, while the bounding interval of $G(\nu)$ becomes larger. }
\label{fig:G-vs-Gbar-Beta}
\end{figure*}

In Figures~\ref{fig:G-vs-Gbar-Gauss} (normal distribution) and \ref{fig:G-vs-Gbar-Beta} (Beta distribution), 
we compare the distributions of instantaneous, time-averaged and natural frequencies, 
again for different values of $K$.

Except for subcritical values of $K$ or near the transition (panels (a) and (b) of each figure), the
graphs of $\overline{G}_{D}$ and $G_{D}$ are quite different from each other as
$\nu\rightarrow\Omega$. In particular while $\overline{G}_{D}(\nu)$
show a big dip to zero for $K>K_c$, $G_{D}(\nu)$ approaches non-zero values near
the synchronization frequency.

For the Beta distribution (Fig.~\ref{fig:G-vs-Gbar-Beta}) the tails of $\overline{G}_{D}$
and $G_{D}$ reach zero for large enough values of $\left|\nu\right|$.
As $K$ increases, the support interval shrinks for $\overline{G}_{D}$,
while it expands for $G_{D}$.

\begin{figure*}
\begin{centering}
\subfloat[normal.]{\includegraphics[width=0.5\linewidth]{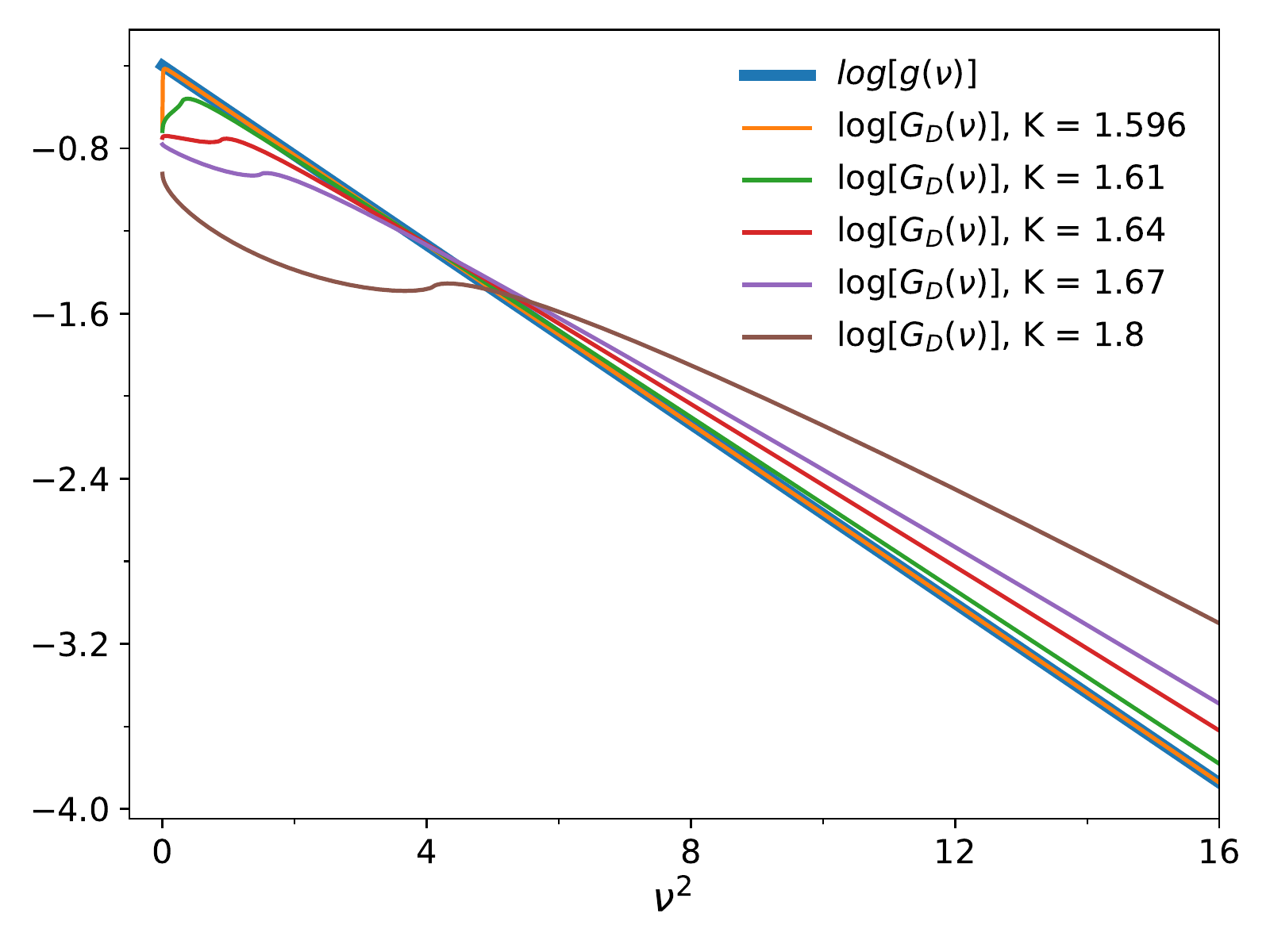}

}\subfloat[beta.]{\includegraphics[width=0.5\linewidth]{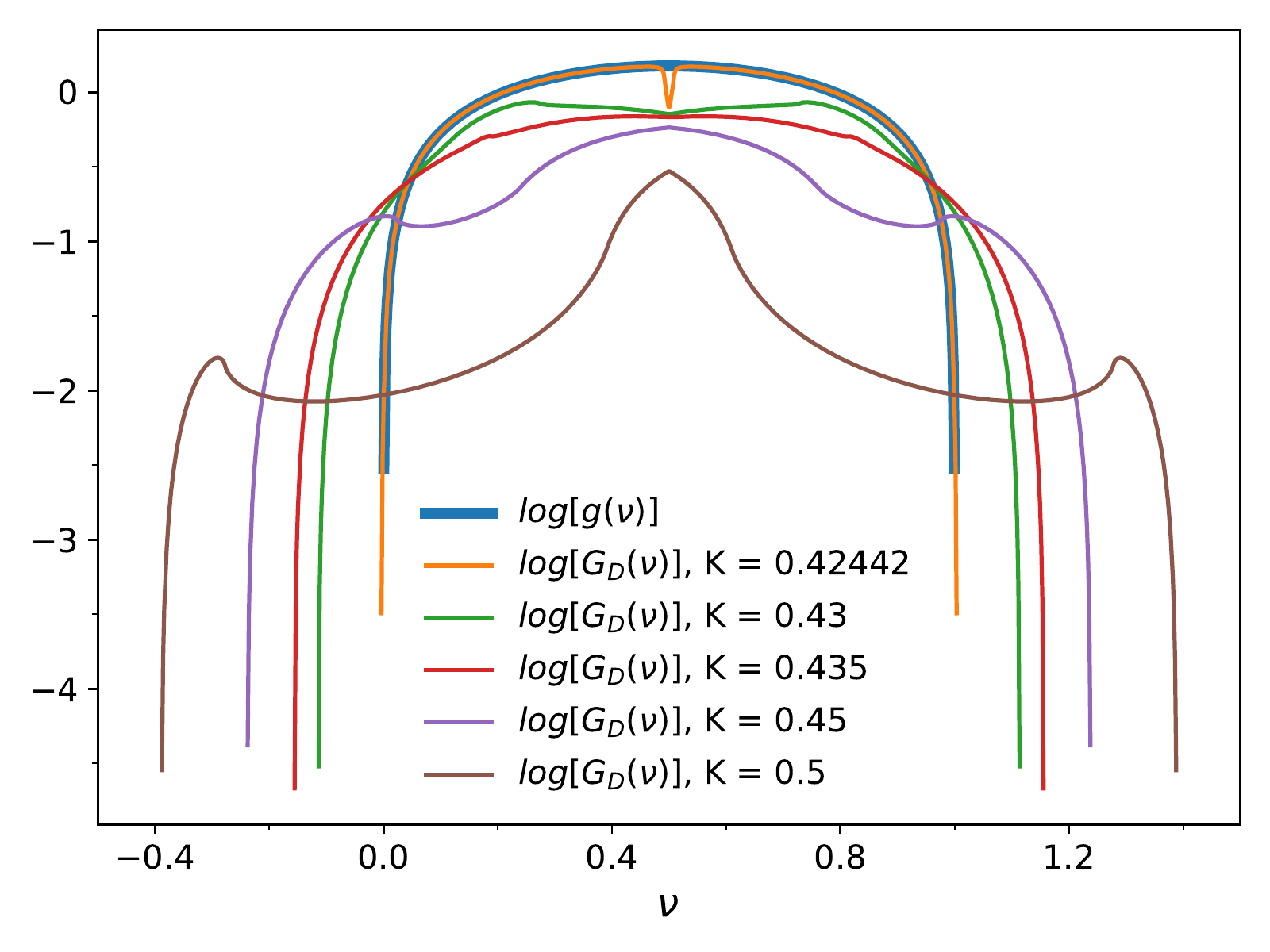}

}
\par\end{centering}
\begin{centering}
\subfloat[normal.]{\includegraphics[width=0.5\linewidth]{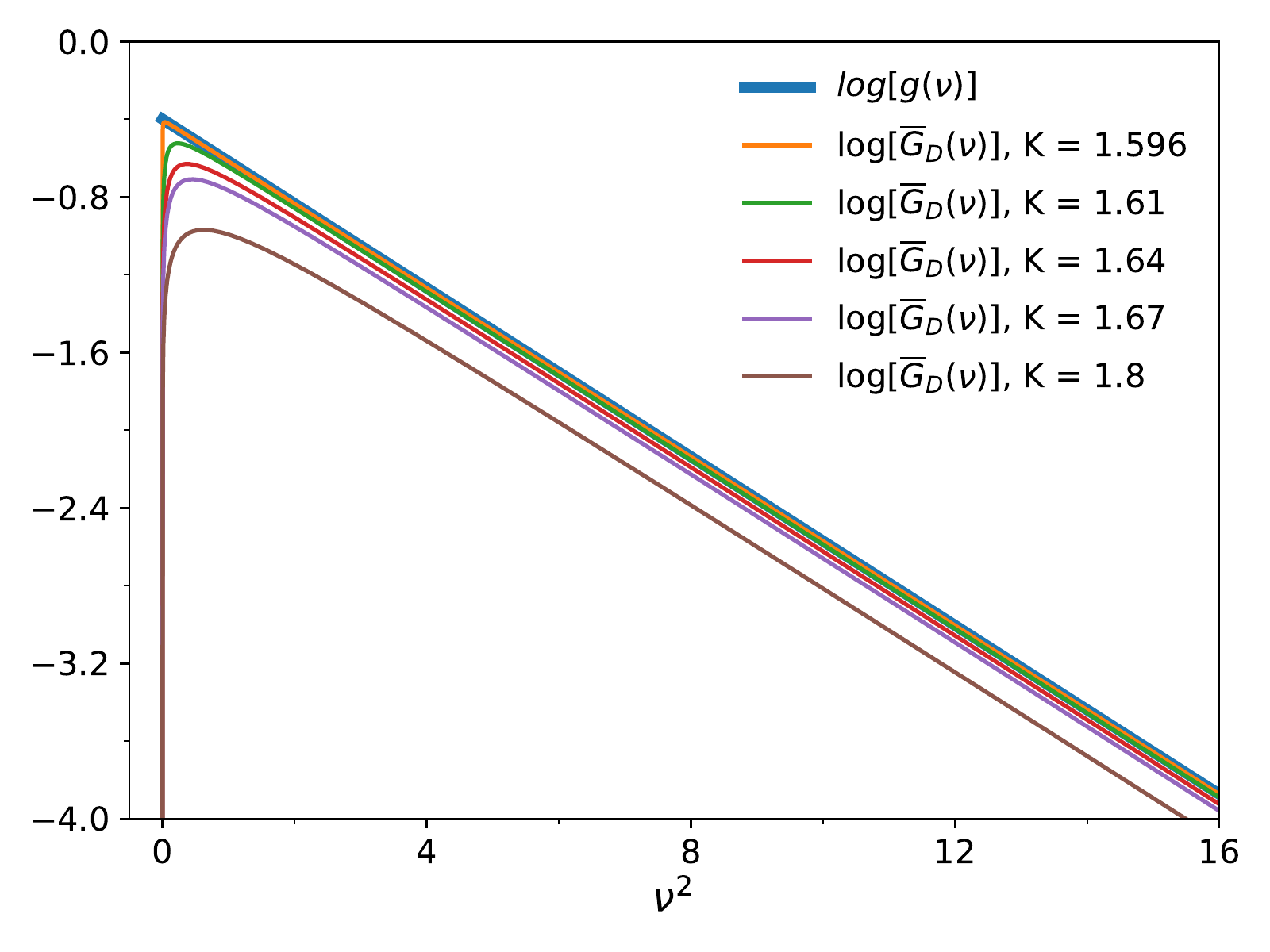}

}\subfloat[beta.]{\includegraphics[width=0.5\linewidth]{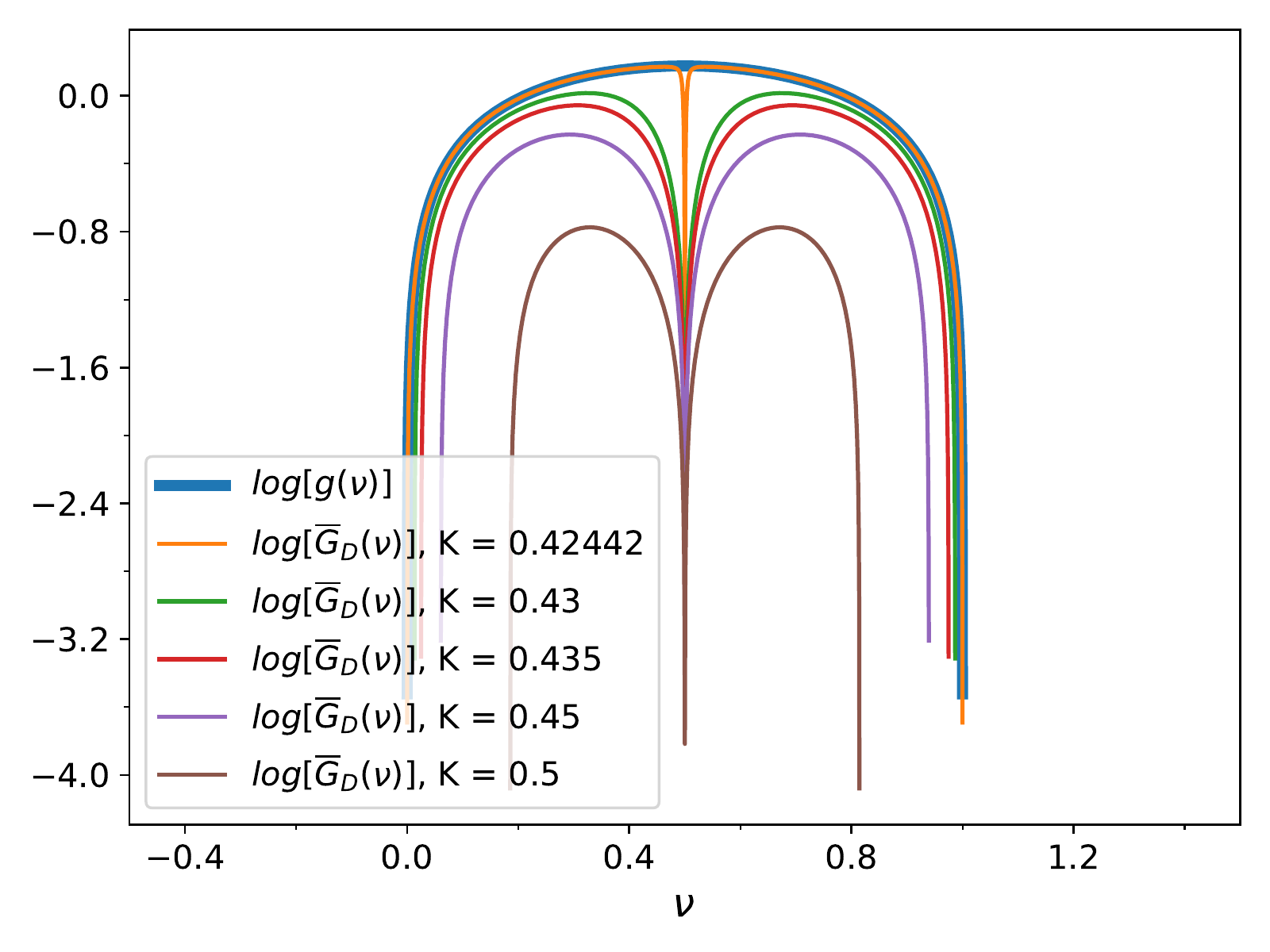}}
\par\end{centering}
\caption{Decimal logarithms of $G_{D}$ ($\log\left[G_{D}(\nu)\right]$) and
$\overline{G}_{D}(\nu)$ ($\log\left[\overline{G}_{D}(\nu)\right]$)
assuming that $g$ is a normal and a Beta$\left(2,2\right)$ PDF.}

\label{fig:log-G-Gbar-Gauss-Beta}
\end{figure*}

\subsection{A focus on tails}

The tails of $G$ describe rare events, viz. large instantaneous
frequencies with small occurrence probabilities. Here we analyze these
events for the Gaussian and Beta cases examined above. 

Figure~\ref{fig:log-G-Gbar-Gauss-Beta} shows the decimal logarithms
of $G_{D}$ (panels (a) and (c)) and $\overline{G}_D$ (panels (b) and (d)) for the
normal and Beta distributions studied above,
using the same set of $K$ values as in previous figures.

For clarity, in the Gaussian case, the distributions are plotted 
as functions of $\nu^{2}$, so that Gaussian tails appear as straight lines.
For large values of $\left|\nu\right|$,
the tails of $G_{D}$ stay above the tails of $g$, and the difference
between $\log\left[G_{D}(\nu)\right]$ and $\log\left[g(\nu)\right]$
increases with $K$. Nevertheless, all $G_{D}$ distributions keep the same asymptotic tail as $g$, rescaled by a 
$K$-dependent factor (Fig.~\ref{fig:log-G-Gbar-Gauss-Beta}(a)). 
The tails of $\overline{G}_D$ are also Gaussian, and asymptotically identical but below those of $g$, rescaled by a $K$-dependent factor that decreases with $K$  (Fig.~\ref{fig:log-G-Gbar-Gauss-Beta}(c)).

In the case of the Beta distribution, both $G_D$ and $\overline{G}_D$ have a bounded support, and 
they behave in a qualitatively-similar manner to $g$ near the limit values of their support intervals.
The tails of $G_D$ extends beyond the support interval of $g$, and all the more so as $K$ increases  (Fig.~\ref{fig:log-G-Gbar-Gauss-Beta}(b)), while the tails of $\overline{G}_D$ show the opposite tendency
 (Fig.~\ref{fig:log-G-Gbar-Gauss-Beta}(d)).

\begin{figure*}
\begin{centering}
\subfloat[normal]{\begin{centering}
\includegraphics[width=0.5\linewidth]{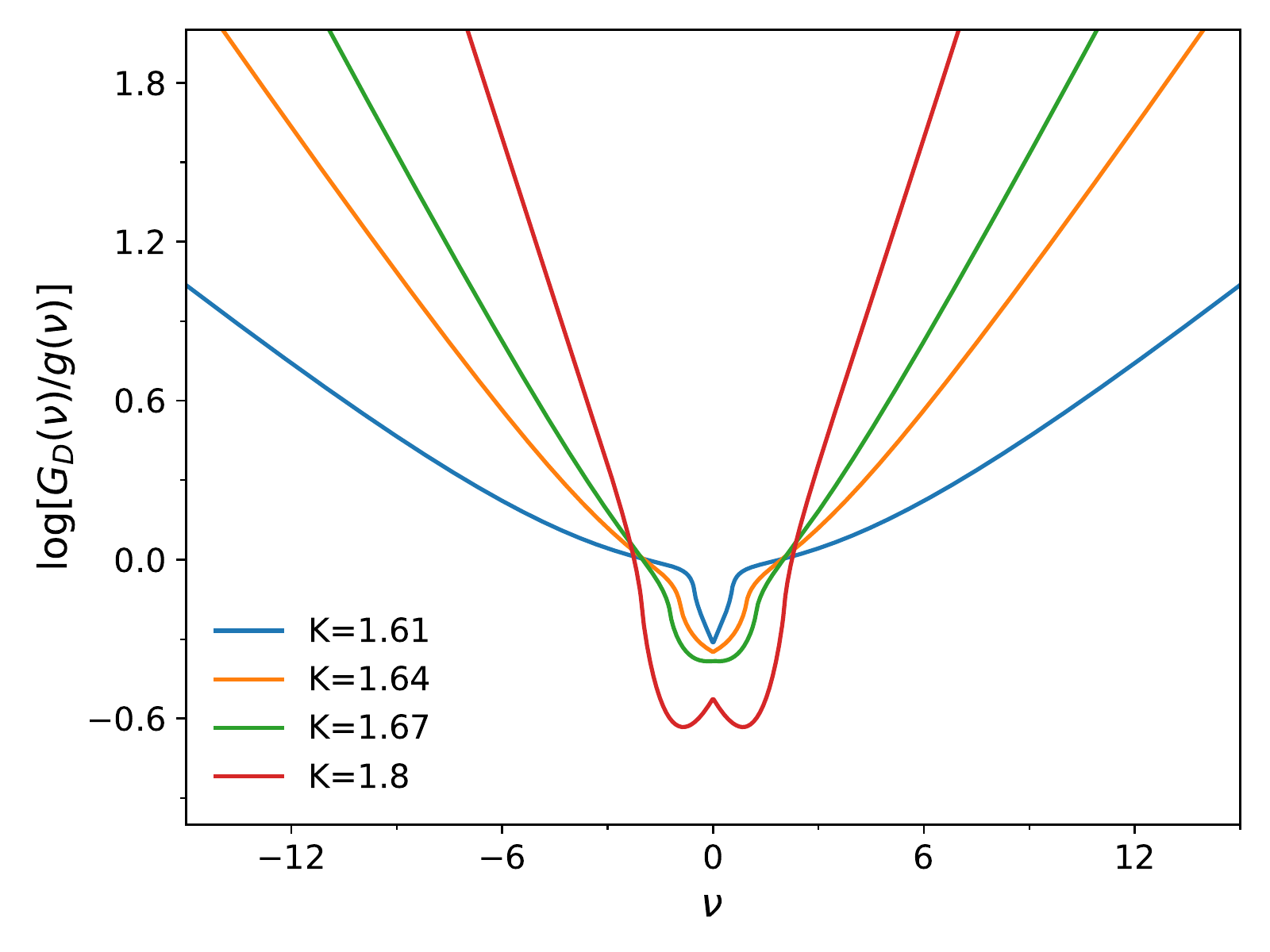}
\par\end{centering}
}\subfloat[normal]{\includegraphics[width=0.5\linewidth]{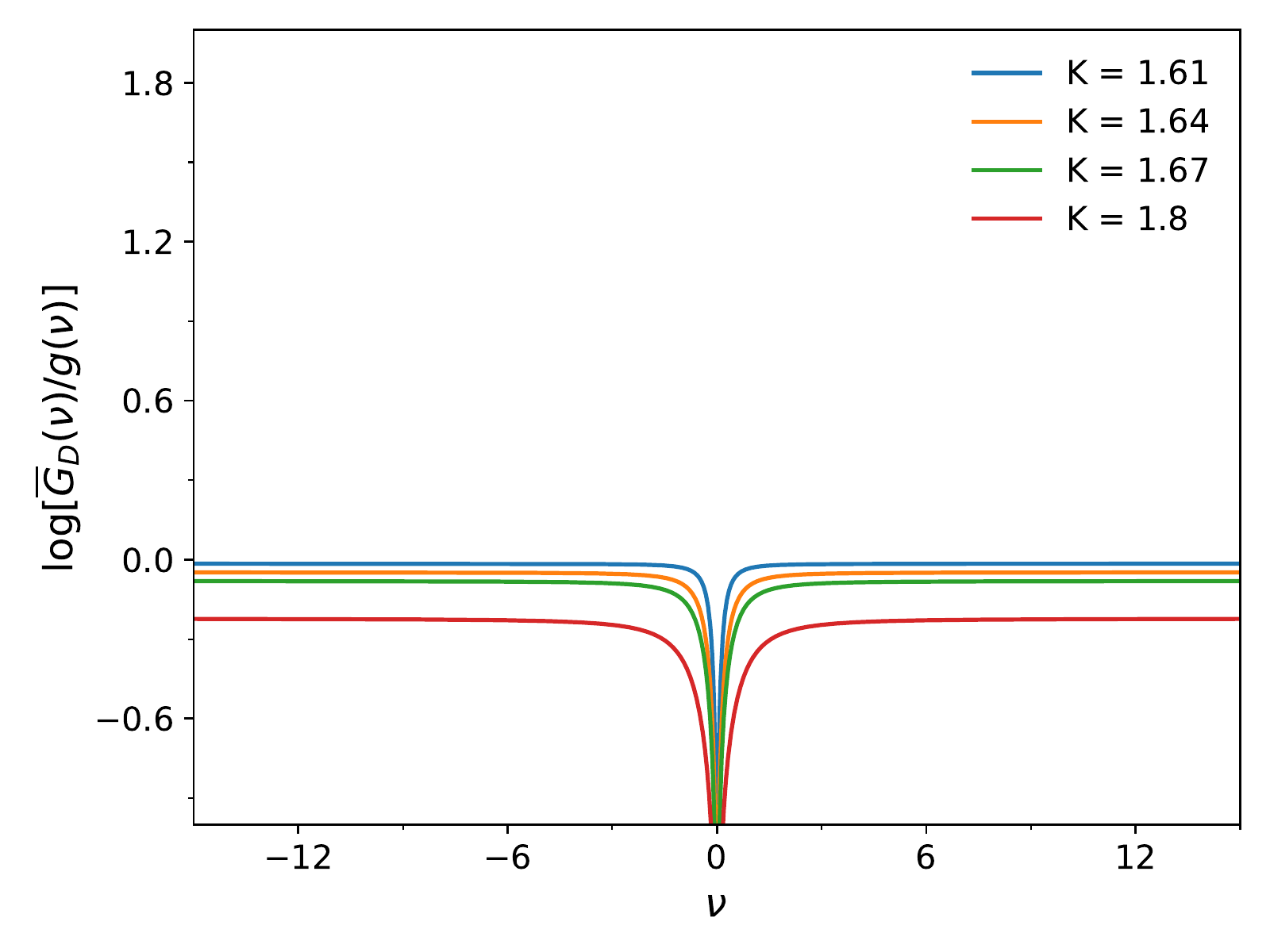}}
\par\end{centering}
\subfloat[beta]{\includegraphics[width=0.5\linewidth]{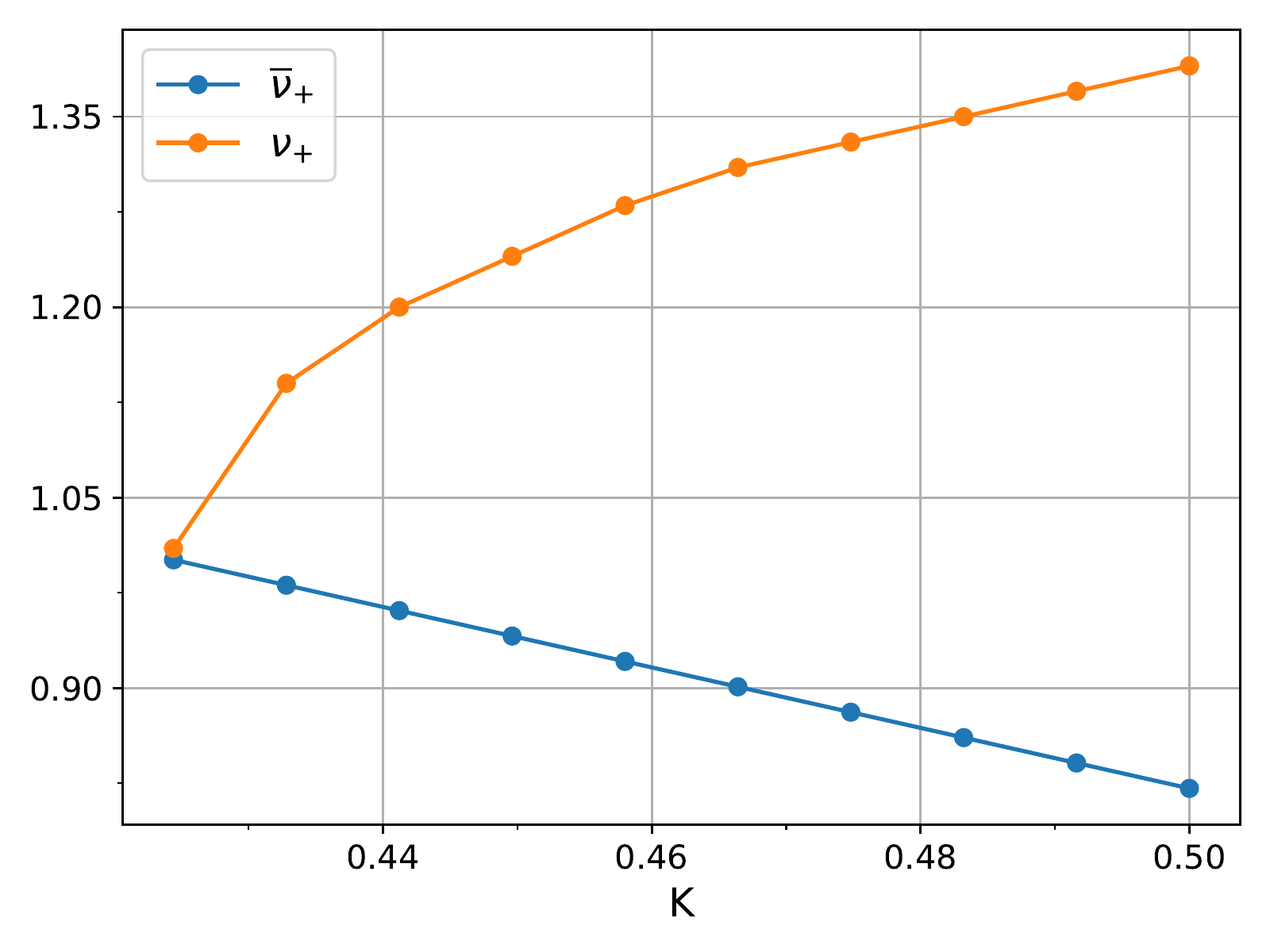}}\subfloat[beta]{\includegraphics[width=0.5\linewidth]{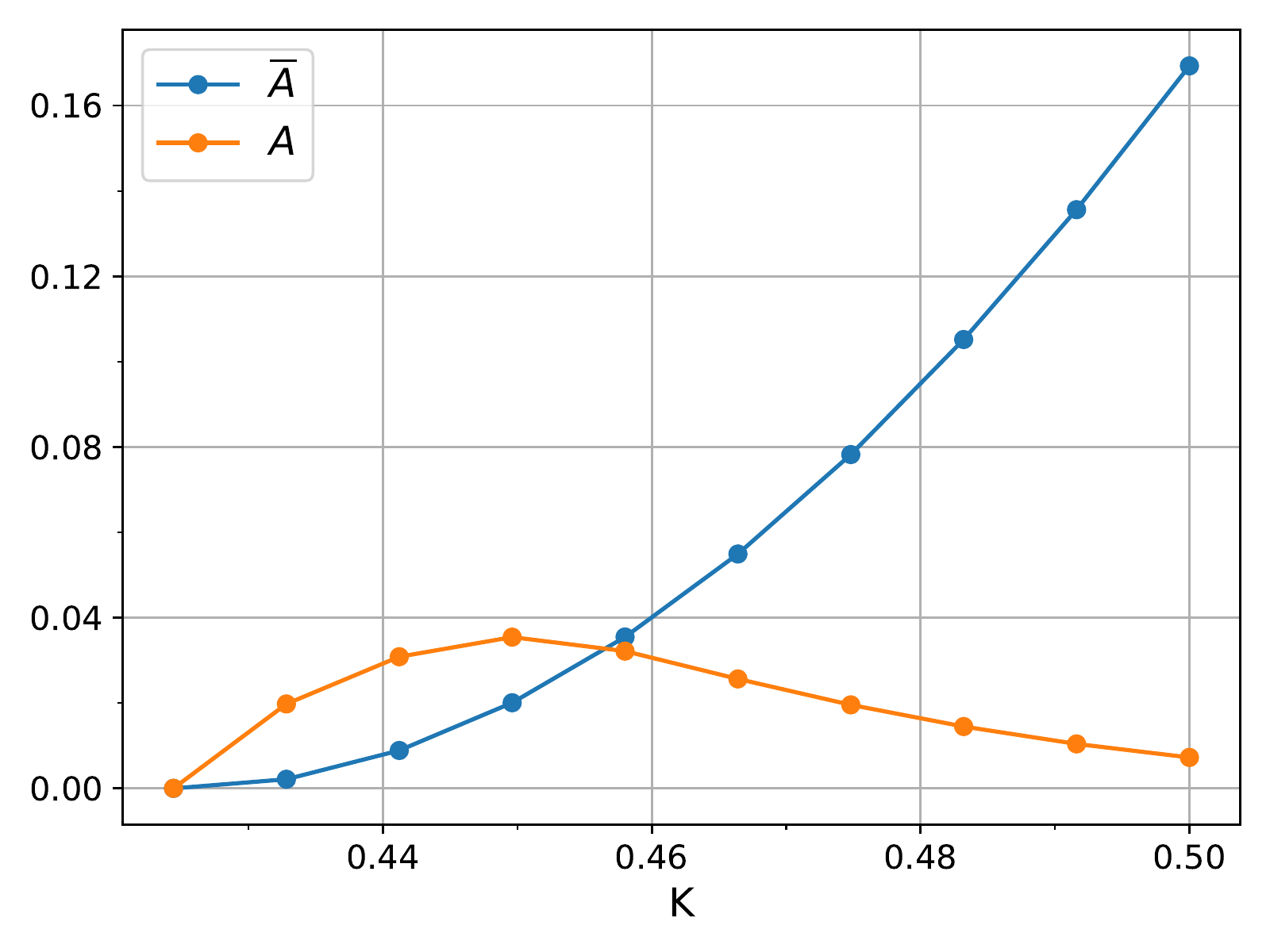}

}

\caption{(a)Decimal logarithms of the ration $\frac{G(\nu)}{g(\nu)}$: as $\nu$
increases, $\frac{G(\nu)}{g(\nu)}$ diverges. (b)Decimal logarithms
of $\frac{\overline{G}(\nu)}{g(\nu)}$ remains constant as $\nu$
increases. (c)Positive endpoints of the support intervals of $G$
and $\overline{G}$, denoted by $\nu_{+}$ and $\overline{\nu}_{+}$.
(d) Area of $G$ outside the support interval of $g$ ($A$), and
area of $g$ outside the support interval of $\overline{G}$ ($\overline{A}$).}
\label{fig:G-vs-Gbar-Gauss-Tails}
\end{figure*}

Further information about the above results is presented in Figure~\ref{fig:G-vs-Gbar-Gauss-Tails}:
In panels (a) and (b), we show the decimal logarithms of the ratios $G_{D}/g$
and $\overline{G}_{D}/{g}$, which shows clearly that $G_D$ ``goes away''
from $g$ as $K$ increases: in the central region $G_D$ becomes smaller and smaller than $g$; in the tails,
the difference grows. In comparison, the behavior of $\overline{G}_D$ is much more ``gentle''.
In the Beta case, we show how the bounds of the support interval varies with $K$ (Fig~\ref{fig:G-vs-Gbar-Gauss-Tails}(c)). 
For the average frequencies ($\overline{G}_D$), the support shrinks almost linearly with $K$, while it grows more slowly than linearly for $G_D$. 
Finally, in Fig.~\ref{fig:G-vs-Gbar-Gauss-Tails}(d) we show $A$, the area of $G_D$ beyond the support interval of $g$. 
This quantity measures the overall likelihood to observe
instantaneous frequencies beyond the possible nominal frequencies. 
Interestingly, $A$ first grows with $K$, then decreases, in spite of the monotonous increase of the support of 
$G_D$. (In Fig.~\ref{fig:log-G-Gbar-Gauss-Beta}(b), one can understand that this comes from the increasingly trimodal nature of $G_D$.)
We also plot $\overline{A}$, the area
of $g$ outside the support interval of $\overline{G}_D$, which indicates the overall weight of natural frequencies
unobservable as time-averaged frequencies. It increases monotonously with $K$. 

\section{Rare events and power-law tails\label{sec:Rare-events}}

In this section we  consider natural frequency distributions with power-law tails and 
develop a power series expansion of Eq.~(\ref{eq:Gd})
in order to deepen our understanding of rare events. 
By rare events we mean occurrences of large instantaneous frequency values such
that $\nu\ll\Omega_{-}$ or $\nu\gg\Omega_{+}$. We assume that natural
frequency distributions are smooth and have unimodal and symmetric
profiles centered at $\Omega=0$.

According to Table \ref{tab:T2}, the instantaneous frequency distribution 
for $\Omega=0$ and $\left|\nu\right|>2a$ can be written
\begin{equation}
G(\nu)=\frac{1}{\pi\left|\nu\right|}\int_{-\frac{\pi}{2}}^{+\frac{\pi}{2}}h\left(\nu+a\sin\psi\right)d\psi.\label{eq:Gd-large-1}
\end{equation}
with
\begin{equation}
h(x)=u_{+}(x)g(x)u_{-}(x) \;\;{\rm where}\;\; u_{\pm}(x)=\sqrt{x\pm a} \;.
\label{eq:h}
\end{equation}

Expanding $h(\nu+a\sin\psi)$ as a Taylor series, we have
\begin{equation}
h\left(\nu+a\sin\psi\right)=h\left(\nu\right)+\sum_{m=1}^{\infty}\frac{a^{m}}{m!}h^{(m)}\left(\nu\right)\sin^{m}\psi,\label{eq:h-taylor}
\end{equation}
where $h^{(m)}$ denotes the $m$th-order derivative of $h$.
Substituting Eq.~(\ref{eq:h-taylor}) in Eq.~(\ref{eq:Gd-large-1}),
we obtain 
\begin{equation}
G(\nu)\!=\!\frac{h(\nu)}{|\nu|}\!+\!\frac{1}{\pi|\nu|}\sum_{m=1}^{\infty}\frac{a^{m}}{m!}h^{(m)}\!(\nu)\!\int_{-\frac{\pi}{2}}^{+\frac{\pi}{2}}\!\! \sin^m\psi\,d\psi\,. \label{eq:eq:Gd-large-2}
\end{equation}
For any integer $n>0$, 
$\int_{-\frac{\pi}{2}}^{+\frac{\pi}{2}}\sin^{2n}\!\psi\,d\psi \!=\! 2\!\int_{_{0}}^{+\frac{\pi}{2}}\sin^{2n}\!\psi\,d\psi$
and $\int_{-\frac{\pi}{2}}^{+\frac{\pi}{2}}\sin^{2n-1}\!\psi\,d\psi=0$.
So only even order terms are present in Eq.~(\ref{eq:eq:Gd-large-2}).
According to Eq.~(3.621-3) in Ref.\citep{Gradshteyn}, $2\int_{_{0}}^{+\frac{\pi}{2}}\sin^{2n}\psi\,d\psi=\pi\tfrac{\left(2n-1\right)!!}{(2n)!!}$,
whence 
\begin{equation}
G(\nu)=\frac{h\left(\nu\right)}{\left|\nu\right|}+\frac{1}{\left|\nu\right|}\sum_{n=1}^{\infty}\frac{a^{2n}}{\left(2n\right)!}\frac{\left(2n-1\right)!!}{(2n)!!}\,h^{(2n)}(\nu).\label{eq:eq:Gd-large-3}
\end{equation}
The Leibniz derivative rule allows us to write $h^{(2n)}$ as 
\begin{align}
\nonumber h^{(2n)}\!(\nu)= 
 \sum_{k_{1}+k_{2}+k_{3}=2n} &\frac{(2n)!}{k_{1}!k_{2}!k_{3}!} \\
& \times u_{+}^{(k_{1})}\!(\nu)\,g^{(k_{2})}\!(\nu)\, u_{-}^{(k_{3})}\!(\nu),\label{eq:h-2n}
\end{align}

where summation is taken over all partitions $\left(k_{1},k_{2},k_{3}\right)$
of $2n$ into non-negative integers, $g^{(k)}$ and $u_{\pm}^{(k)}$
denote the $k$th-order derivatives of $g$. The latter are given by 
\begin{equation}
u_{\pm}^{(k)}\left(\nu\right)=p_{k}\left(\frac{1}{2}\right)\left(\nu\pm a\right)^{-k}u_{\pm}\left(\nu\right),\label{eq:diff-u}
\end{equation}
where 
\begin{equation}
p_{k}\left(q\right)=\prod_{l=0}^{k-1}\left(q-l\right).\label{eq:pk}
\end{equation}
From Eqs. (\ref{eq:h-2n}) and (\ref{eq:diff-u}), it follows that
\begin{align}
\nonumber
h^{(2n)}(\nu)= &  u_{+}(\nu) u_{-}(\nu) \!\!\! \! \sum_{k_{1}+k_{2}+k_{3}=2n} \!\!\!\!
\frac{p_{k_{1}}\!(\tfrac{1}{2})\left(2n\right)! \,p_{k_{3}}\!(\tfrac{1}{2})}{k_{1}!\,k_{2}!\,k_{3}!}\\
&\times \frac{g^{(k_{2})}(\nu)}{\left(\nu\!+\!a\right)^{k_{1}}\left(\nu\!-\!a\right)^{k_{3}}}.\label{eq:h-2n-1}
\end{align}
We can now use Eqs. (\ref{eq:h-2n-1}), and (\ref{eq:h})
in Eq.~(\ref{eq:eq:Gd-large-3}) to obtain the ratio $G(\nu)/g(\nu)$, which is given by
\begin{equation}
\frac{G}{g}(\nu)=\sqrt{1-\left(\frac{a}{\nu}\right)^{2}}\left[1+\Lambda(\nu)\right],\label{eq:G-g-ration}
\end{equation}
where 
\begin{equation}
\Lambda(\nu)=\frac{1}{g(\nu)}\sum_{n=1}^{\infty}c_{n}\left(\nu\right)\left(\frac{a}{\nu}\right)^{2n}\label{eq:Lambda}
\end{equation}
and
\begin{align}
\nonumber
c_{n}\left(\nu\right)=& 
\frac{\left(2n-1\right)!!}{\left(2n\right)!!}\sum_{k_{1}+k_{2}+k_{3}=2n}
\!\! \frac{p_{k_{1}}(\tfrac{1}{2})p_{k_{3}}(\tfrac{1}{2})}{k_{1}!k_{2}!k_{3}!} \\
& \times \frac{\nu^{k_{2}}g^{(k_{2})}(\nu)}{\left(1+\frac{a}{\nu}\right)^{k_{1}}\left(1-\frac{a}{\nu}\right)^{k_{3}}}.\label{eq:cn}
\end{align}
(If $g$ is centered at a non-zero synchronization frequency $\Omega$,
more general formulas than Eqs. (\ref{eq:G-g-ration}), (\ref{eq:Lambda}),
and (\ref{eq:cn}) can be obtained by changing in them the terms $\frac{a}{\nu}$
by $\frac{a}{\left(\nu-\Omega\right)}$.)

As an application of the result given by (\ref{eq:G-g-ration}), let
us now consider a class of natural frequency distributions of the
form 
\begin{equation}
g\left(\nu\right)\sim C\nu^{-2\mu}\qquad\left(\left|\nu\right|\longrightarrow\infty\right),\label{eq:power-law}
\end{equation}
where $\mu$ is a positive integer, and $C$ a real constant. 

Its derivatives read
\begin{equation}
g^{(k)}\left(\nu\right)\sim p_{k}\left(-2\mu\right)\nu^{-k}g\left(\nu\right)\qquad\left(\left|\nu\right|\longrightarrow\infty\right),\label{eq:power-law-1}
\end{equation}
where $p_{k}$ is given by Eq.~(\ref{eq:pk}), and $k=0,...,2n$.
Since $\left(1\pm\frac{a}{\nu}\right)^{k}\sim1$ as $\left|\frac{a}{\nu}\right|\longrightarrow0$,
it follows from (\ref{eq:cn}) and (\ref{eq:power-law-1}) that 
\begin{equation}
c_{n}\left(\nu\right)\sim\overline{c}_{n}g\left(\nu\right)\qquad\left(\left|\frac{a}{\nu}\right|\longrightarrow0\right),\label{eq:cn-2}
\end{equation}
where the constant coefficient $\overline{c}_{n}$ is defined by
\begin{equation}
\overline{c}_{n}=\frac{\left(2n\!-\!1\right)!!}{\left(2n\right)!!}\!\!\!\sum_{k_{1}+k_{2}+k_{3}=2n}
\!\!\!\frac{p_{k_{1}}\!\left(\tfrac{1}{2}\right) \, p_{k_{2}}\!(-2\mu) \, p_{k_{3}}\!\left(\tfrac{1}{2}\right)}{k_{1}!k_{2}!k_{3}!}.\label{eq:cn-1}
\end{equation}
Therefore, 
\begin{equation}
\Lambda(\nu) \underset{\left|\frac{a}{\nu}\right|\to 0}{\sim} \sum_{n=1}^{\infty}\overline{c}_{n}\left(\frac{a}{\nu}\right)^{2n}
,\label{eq:Lambda-1}
\end{equation}
and
\begin{equation}
\frac{G}{g}(\nu)\underset{\left|\frac{a}{\nu}\right|\to 0}{\sim}\sqrt{1\!-\!\left(\tfrac{a}{\nu}\right)^{2}}
\left\{ 1\!+\!\overline{c}_{1}\left(\tfrac{a}{\nu}\right)^{2}\!+\!O\left[\left(\tfrac{a}{\nu}\right)^{4}\right]\right\}
 .\label{eq:asym-G-g}
\end{equation}
This means: assuming that $g(\nu)$ has power-law tails
of form (\ref{eq:power-law}), $G\left(\nu\right)$ approaches asymptotically
$g(\nu)$ for large instantaneous frequencies (compared to $a$) or
small order parameter values (compared to $\left|\frac{K}{\nu}\right|$).

\begin{figure}
\begin{centering}
\includegraphics[width=1\linewidth]{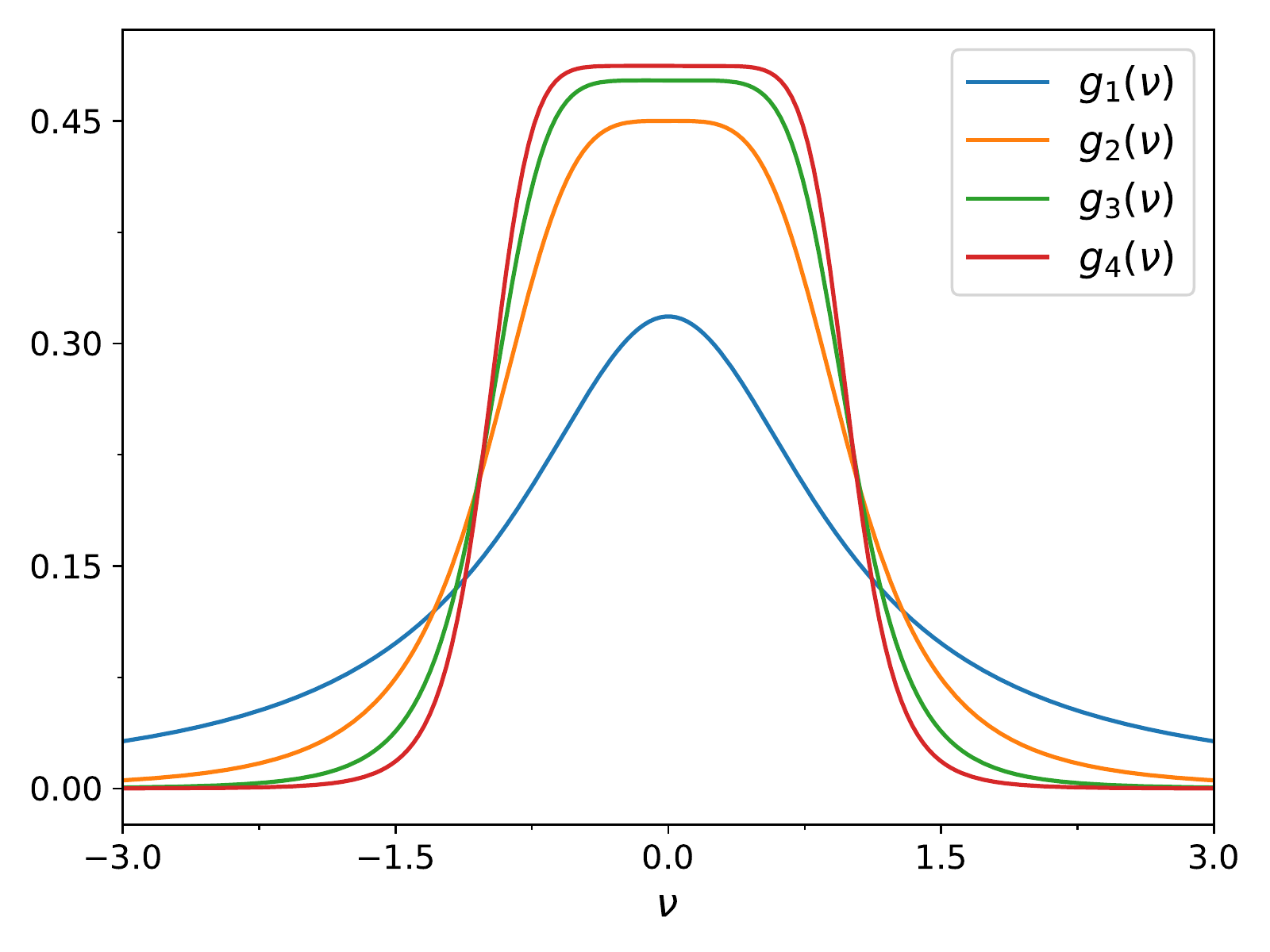}
\par\end{centering}
\caption{Graphs of $g_{\mu}$, defined by (\ref{eq:gen-Lorentz}), for $\mu=1,2,3,4$. }
\label{fig:g-mu}
\end{figure}

\begin{figure*}
\begin{centering}
\subfloat[$\mu=1$]{\includegraphics[width=0.5\linewidth]{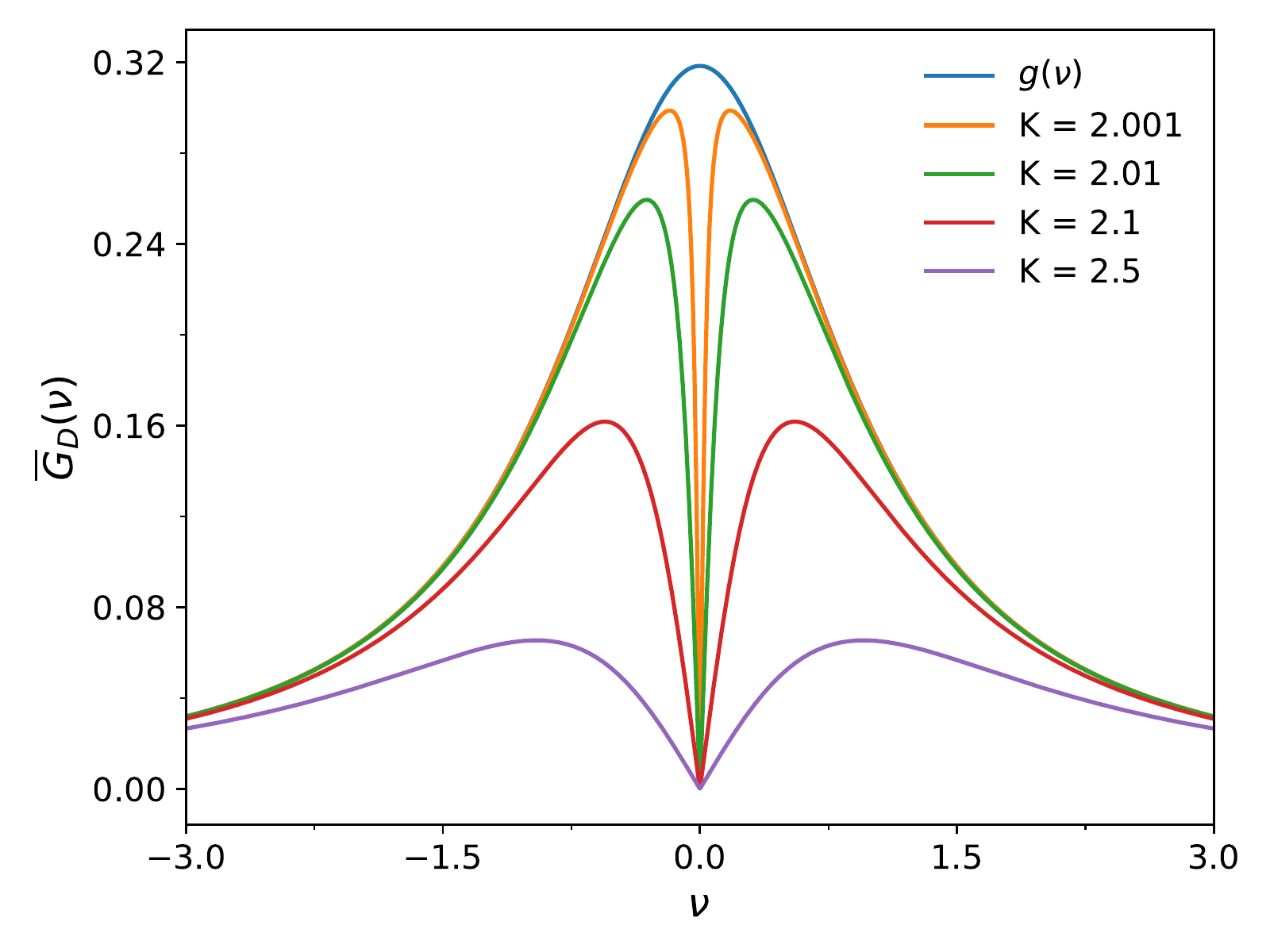}

}\subfloat[$\mu=1$]{\includegraphics[width=0.5\linewidth]{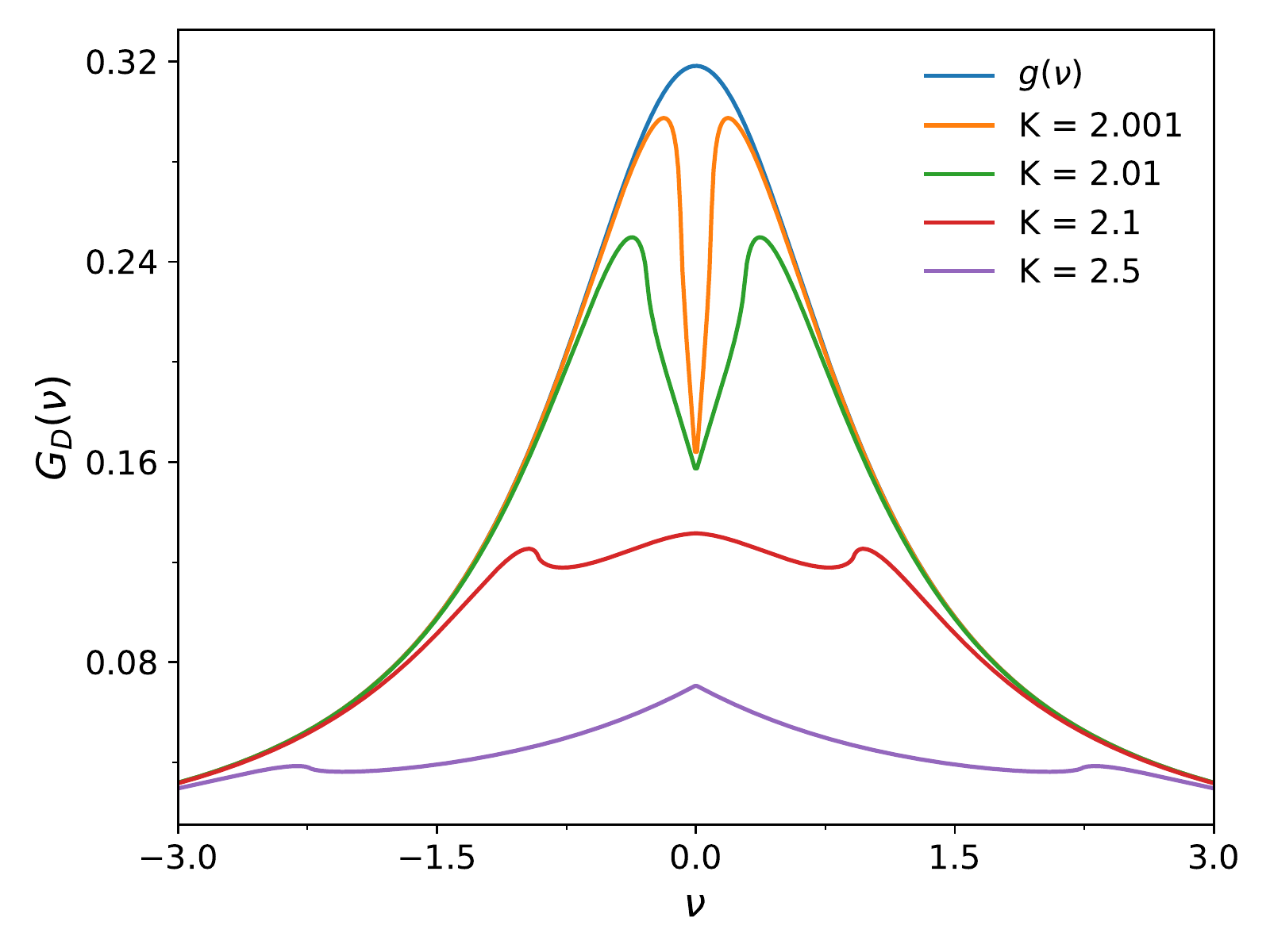}

}
\par\end{centering}
\begin{centering}
\subfloat[$\text{\ensuremath{\mu}}=3$]{\includegraphics[width=0.5\linewidth]{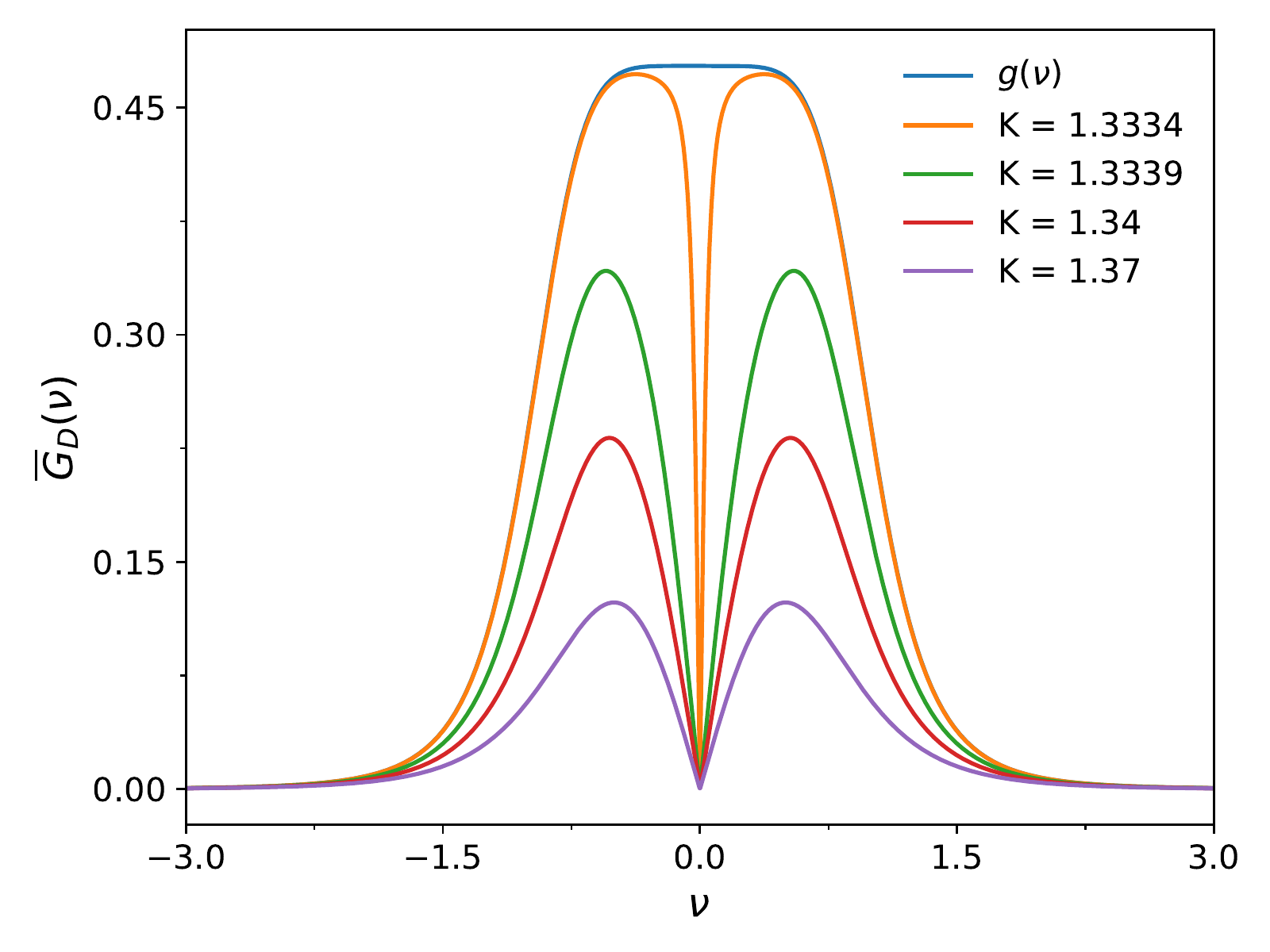}

}\subfloat[$\mu=3$]{\includegraphics[width=0.5\linewidth]{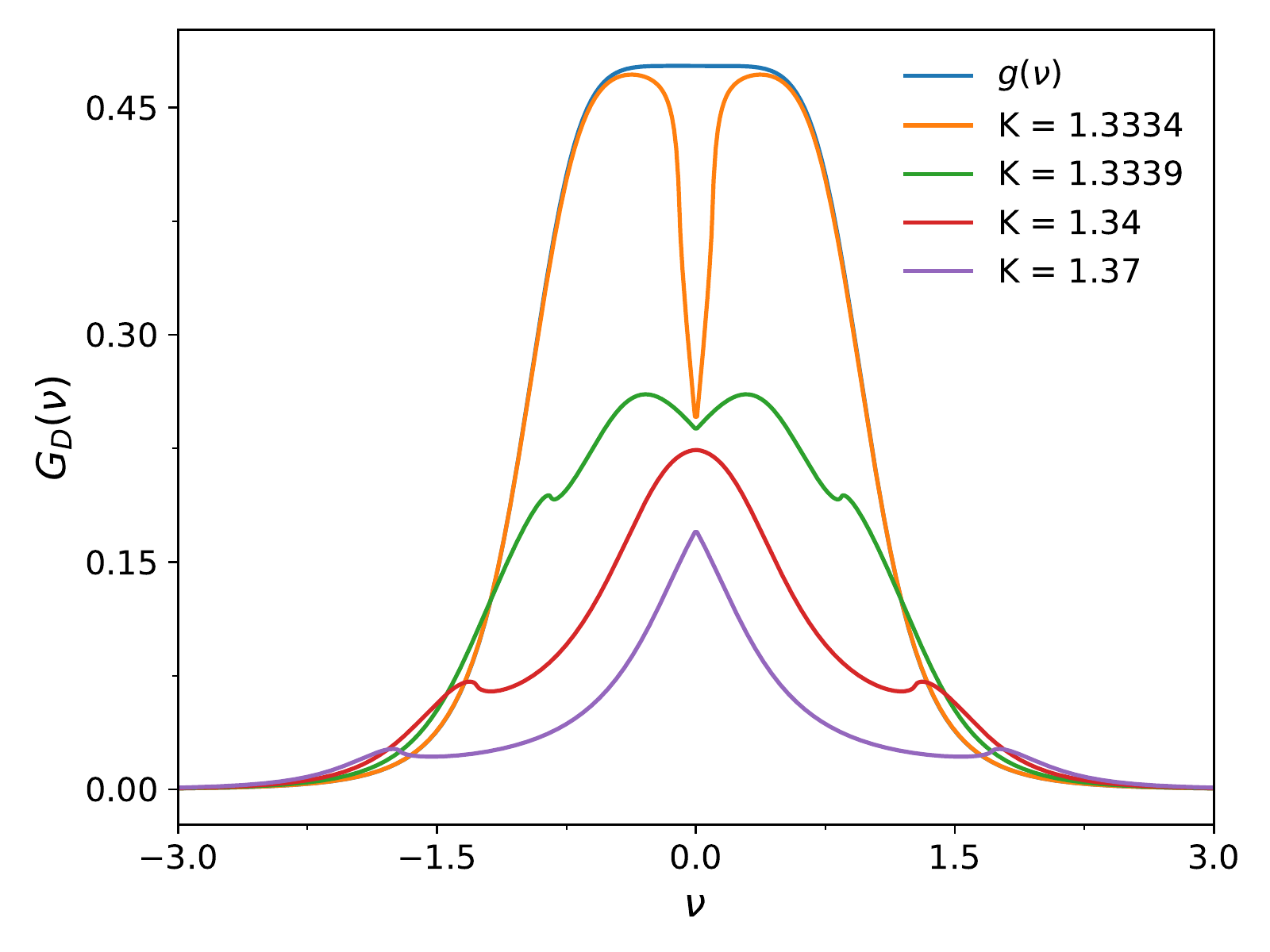}

}
\par\end{centering}
\caption{$\overline{G}_{D}$ and $G$ for $g=g_{\mu}$, defined by (\ref{eq:gen-Lorentz}). }
\label{fig:g-Gbar-G-Lorentz}
\end{figure*}

To illustrate this point, we consider the family of natural frequency
distributions 
\begin{equation}
g_{\mu}\left(\nu\right)=\frac{\mu}{\pi(1+\nu^{2\mu})}\sin\left(\frac{\pi}{2\mu}\right),\label{eq:gen-Lorentz}
\end{equation}
where $\mu$ is a positive integer. Formula (\ref{eq:gen-Lorentz})
generalizes the standard Cauchy-Lorentz distribution, which corresponds
to the particular case $g_{1}$. Graphs of $g_{\mu}\left(\nu\right)$
are shown in Fig.~\ref{fig:g-mu} for $\mu=1,2,3,4$.
By increasing $\mu$, the tails of $g_{\mu}\left(\nu\right)$ gets
thinner, and high natural frequencies have lower occurrence probabilities.
In Figs. \ref{fig:g-Gbar-G-Lorentz}(a)-(d), we show graphs of $\overline{G}_{D}$
and $G$ for different values of $K$ considering the cases $g_{1}$
and $g_{3}$.

\begin{figure*}
\begin{centering}
\subfloat[$\mu=1$]{\begin{centering}
\includegraphics[width=0.5\linewidth]{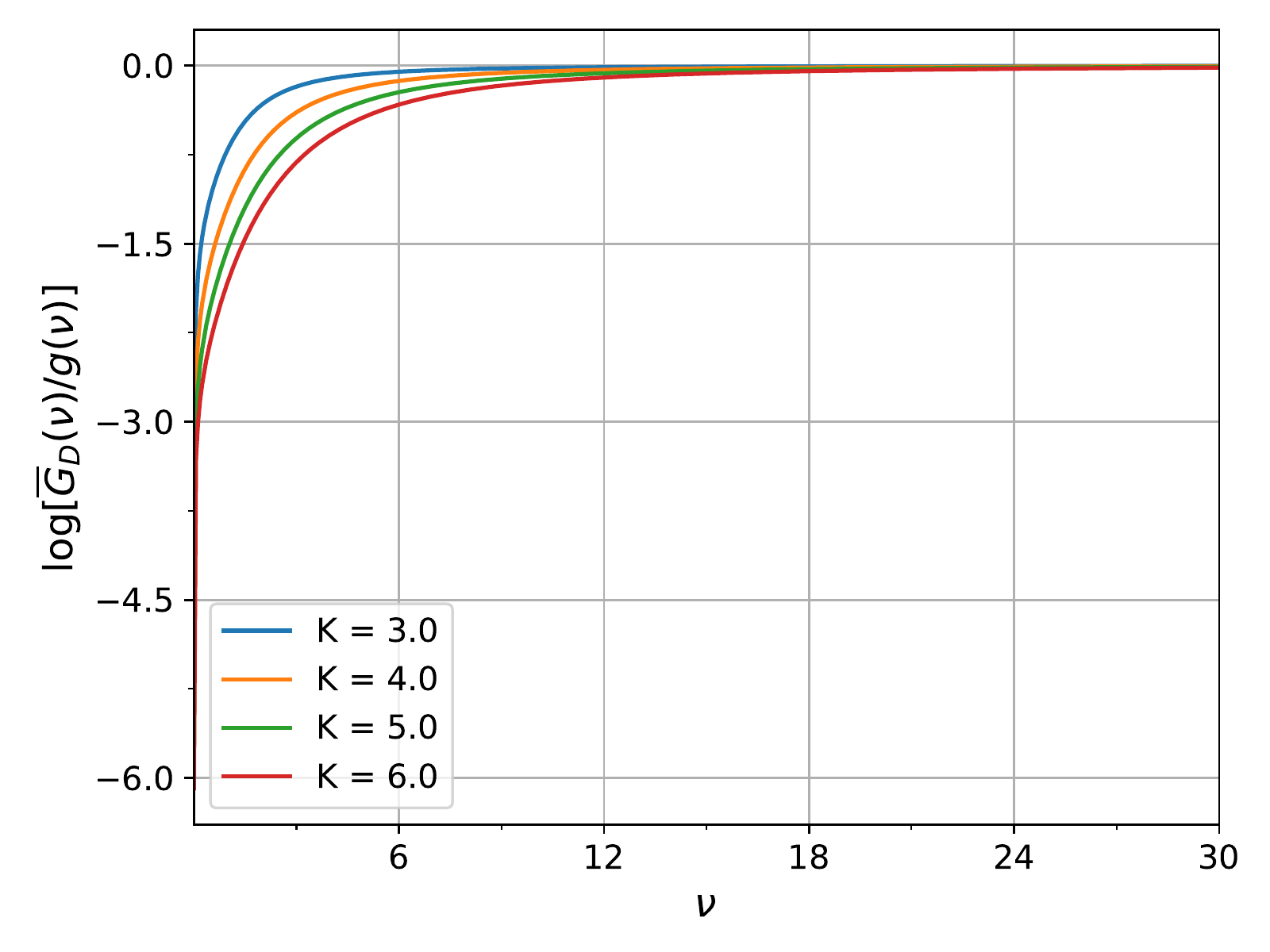}
\par\end{centering}
}\subfloat[$\mu=1$]{\begin{centering}
\includegraphics[width=0.5\linewidth]{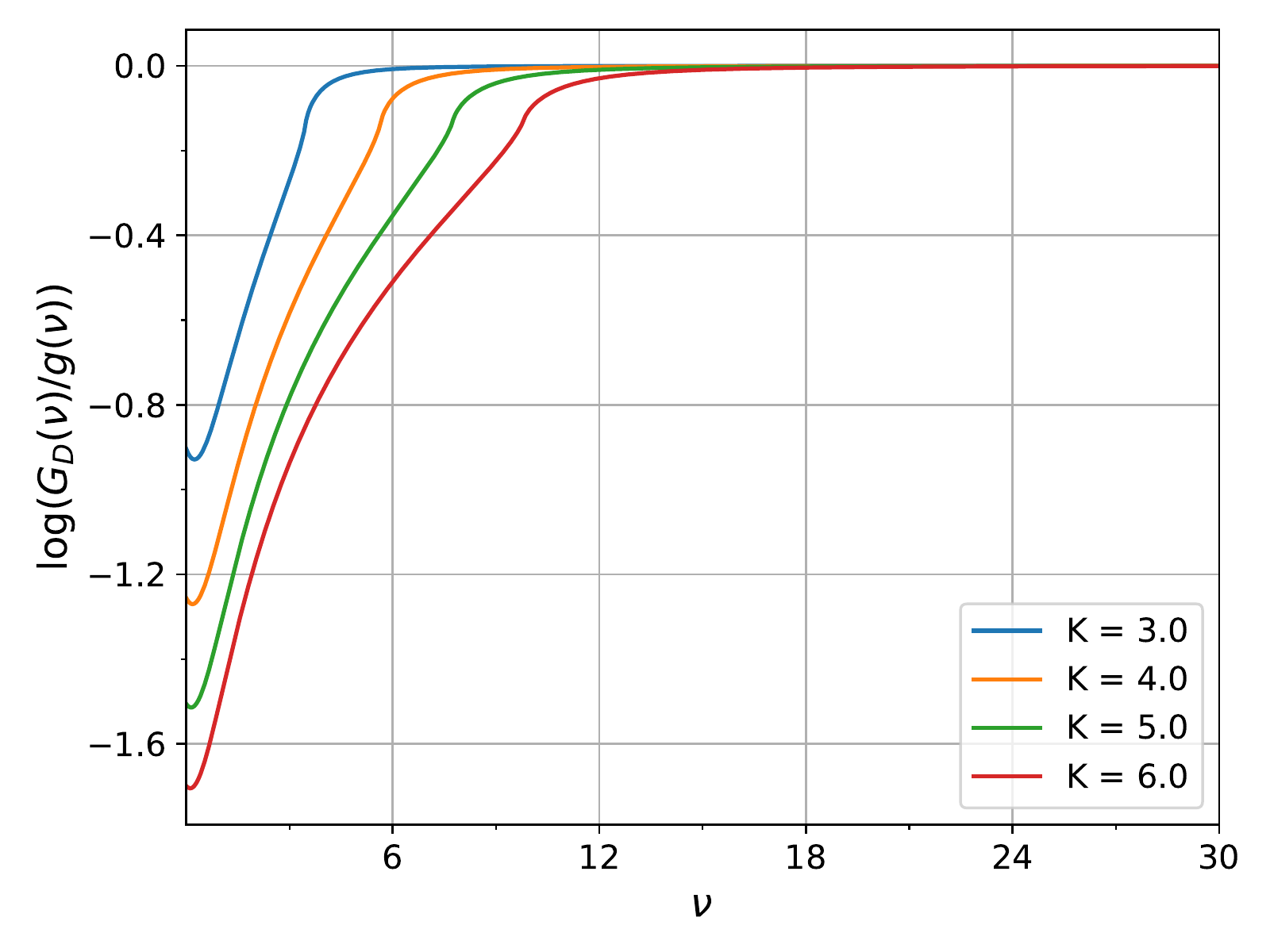}
\par\end{centering}
}
\par\end{centering}
\begin{centering}
\subfloat[$\mu=3$]{\begin{centering}
\includegraphics[width=0.5\linewidth]{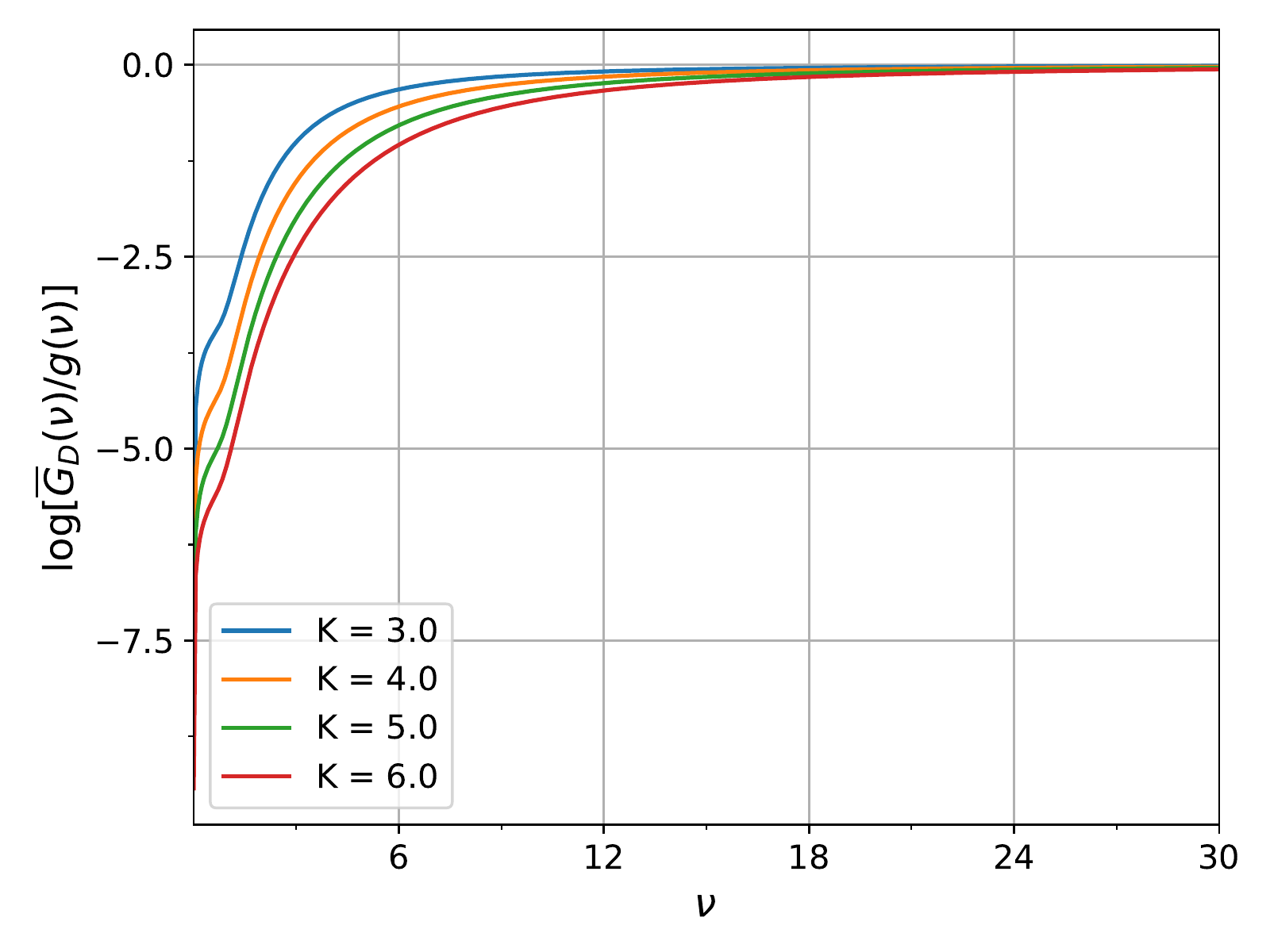}
\par\end{centering}
}\subfloat[$\mu=3$]{\begin{centering}
\includegraphics[width=0.5\linewidth]{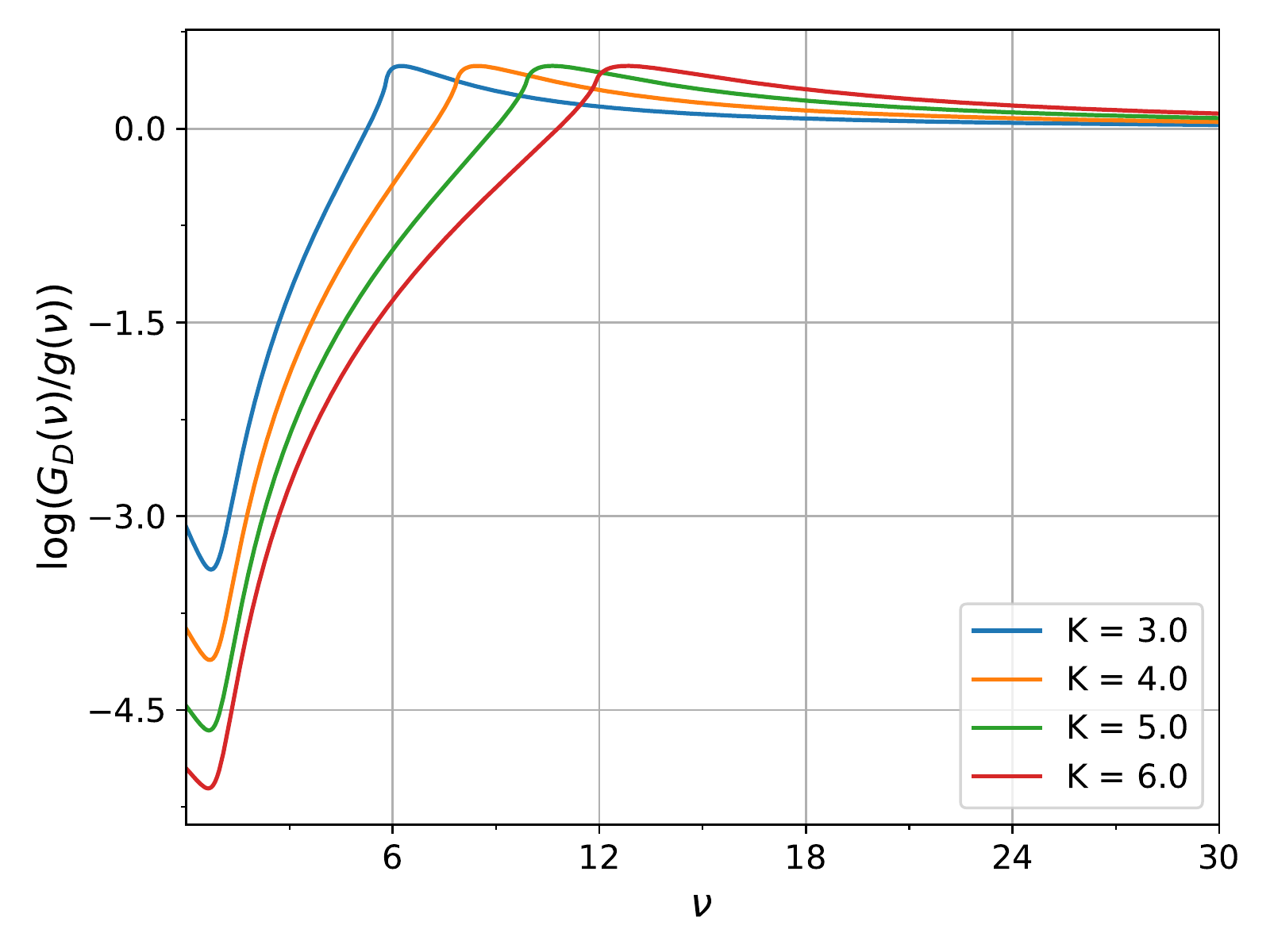}
\par\end{centering}
}
\par\end{centering}
\caption{Decimal logarithms of the ratios $\frac{\overline{G}_{D}}{g}\left(\nu\right)$
and $\frac{G}{g}\left(\nu\right)$. $g(\nu)=g_{\mu}(\nu)$, defined
by Eq.~(\ref{eq:gen-Lorentz}).}
\label{fig:Gbar-g-GenLorentz-Tails}
\end{figure*}

Figures \ref{fig:Gbar-g-GenLorentz-Tails}(a) and (c) show decimal
logarithms of the ratio $\overline{G}_{D}/{g}$, where
$\overline{G}_{D}(\nu)$ is computed using Eq.~(\ref{eq:GbarD}) with
$g(\nu)=g_{\mu}(\nu)$. Like in the Gaussian and Beta examples
of Sec. \ref{sec:Applications}, the graphs of $\overline{G}_{D}$
are more and more below the graph of $g$ as increasing $K$ increases.
Yet, for $|\nu|\to\infty$, $\overline{G}_{D}(\nu)$
and $g(\nu)$ show the same asymptotic behavior.

In Figs. \ref{fig:Gbar-g-GenLorentz-Tails}(b) and (d), we show decimal
logarithms of the ratio ${G}/{g}$ for the same
$K$ values used in Figs. \ref{fig:Gbar-g-GenLorentz-Tails}(a) and (c). 
All graphs show that $\log[\tfrac{G}{g}(\nu)]\to 0$
as $\left|\tfrac{a}{\nu}\right|\to 0$, which is in agreement
with formula (\ref{eq:asym-G-g}). So $G(\nu)$ converges to $g(\nu)$
as $\nu$ increases, albeit this convergence is restrained by increasing
$K$. 

Another somewhat counterintuitive effect is related to the tail thickness of $g_{\mu}$. Compared
to the other distributions $g_{\mu>1}$, $g_{1}$ decays more slowly
as $\nu$ increases, and $\log[\tfrac{G}{g_{1}}(\nu)]$
decays more easily to zero. When $\mu$ increases , $g_{\mu}$'s tails
become thinner, and convergence of $\log[\tfrac{G}{g_{\mu}}(\nu)]$
to zero requires larger values of $\nu$.
A simple explanation to this tail thickness effect is related to the
critical order parameter, which is defined by $K_{c}^{(\mu)}=\frac{2}{\pi g_{\mu}\left(0\right)},$
if $g=g_{\mu}$ (see Eq.~(\ref{eq:Kc})). 
Since the normalization condition remains valid
for any $\mu$, thinner tails result in higher $g(0)$. If $g_{\mu}(0)>\ldots>g_{2}(0)>g_{1}(0)$,
then $K_{c}^{(\mu)}<\ldots<K_{c}^{(2)}<K_{c}^{(1)}$. So, for $K$ fixed,
the difference $K-K_{c}^{\mu}$ decreases with decreasing $\mu$,
$R$ (and $a$) diminishes, and $G$ resembles more closely $g_{\mu}$.
When $R$ is small, $G(\nu)$ converges more easily $g_{\mu}(\nu)$
as $\left|\tfrac{a}{\nu}\right|\to0$. This is shown by
Eq.~(\ref{eq:asym-G-g}) and Figs. \ref{fig:Gbar-g-GenLorentz-Tails}(b)
and (d). 

\section{Conclusion\label{sec:Conclusion}}

Based on Kuramoto theory, we have obtained an analytical formulation of
the instantaneous frequency distribution in the Kuramoto model. Numerical data
show excellent agreement with our formula, provided they are obtained on very large 
collections of oscillators studied in their steady state (see Appendix), i.e. 
in the limits where our results are expected to be valid. 
Access to the distribution of instantaneous frequencies $G$ extends Kuramoto theory,
which was limited heretofore to the knowledge of $\overline{G}$, the distribution of time-averaged,
or ``coupling-modified'', frequencies.

Distributions $G$ and $\overline{G}$ are functionals of the natural frequency distribution $g$.
Irrespective of $g$, the synchronization scenario keeps the same basic features: 
beyond the critical coupling strength value $K_c=\tfrac{2}{\pi g(\Omega)}$, 
a subset of oscillators synchronize, so that both $\overline{G}$ and $G$ comprise a delta peak 
at the frequency $\Omega$. This delta peak is identical for both $G$ and $\overline{G}$ and 
represents the fraction of synchronized oscillators, which grows monotonously with $K$
(at least for the $g$ distributions considered here, see e.g. Fig. \ref{fig:G-Gauss-Beta}). 
As soon as $K>K_c$, the continuous part of both $G$ and $\overline{G}$ departs from $g$.
Whereas $\overline{G}$ remains below $g$ everywhere, and displays a characteristic dip near the 
synchronized frequency $\Omega$, the continuous part of $G$ has tails that pass over $g$ and shows a maximum
at $\Omega$ for $K$ large enough. 
This last fact reflects the relaxation-oscillation-like dynamics of oscillators with 
natural frequency outside but close to the synchronization band. 

The distribution of instantaneous frequencies $G$ typically displays a rather complicated structure.
It is in fact trimodal for $K$ large enough, even as the natural frequency distributions considered here are unimodal. 
Nevertheless, from the 3 qualitatively-different examples of $g$ studied here, 
we have shown that $G$, like $\overline{G}$, displays the same tails as $g$: a normal $g$ yields Gaussian tails for $G$ and $\overline{G}$ which are just rescaled versions of those of $g$. For a Beta $g$ with a bounded support interval, $G$ and $\overline{G}$ both have bounded supports, respectively wider and narrower than that of $g$.
 
Due to the difficulty in extracting explicit expressions from our results, 
most of the information about the distribution of instantaneous frequencies presented here has been
obtained by numerical analysis of our formula.
Yet, when dealing with rare events in the example of powerlaw tailed distribution of natural frequencies,
a power-series analysis of the distribution of instantaneous frequencies has allowed us 
to obtain an asymptotic expansion in frequency. This has confirmed that $g$, $G$, and $\overline{G}$
are asymptotically equivalent in the limit of large frequencies. 

Beyond their intrinsic interest for a deeper understanding of synchronization, our results 
are useful when it comes to choosing a numerical scheme and resolution to simulate coupled oscillators: indeed
a faithful simulation must account properly for the largest instantaneous frequencies displayed by the system. 
As seen and quantified here, these are larger than the largest natural frequency present, which implies, e.g., to choose higher-order integration schemes and/or smaller timesteps than naively suggested by the natural frequencies
at play.

The approach followed here can easily be extended to non-symmetric and/or non-unimodal distributions of natural
frequencies. We also believe that important variants of the Kuramoto model, such as the Kuramoto-Sakaguchi
model \citep{Sakaguchi} are amenable to the same type of analysis as developed here. 
More generally, we hope that this work opens new perspectives on synchronization
phenomena beyond the usual order-parameter analysis. 

\begin{acknowledgments}
This work was made possible through financial support from Brazilian
research agency FAPESP (grant n. 2019/12930-9 ). JDF warmly thanks
Prof. Joao Peres for valuable discussions. EDL thanks support from
Brazilian agencies CNPq (301318/2019-0) and FAPESP (2019/14038-6).
\end{acknowledgments}

\appendix*
\section{Numerical checks}
\label{num}

In this section we validate the formula of $G_{D}$ against numerical
results from simulations of the Kuramoto model. Our check is restricted
to the Gaussian and Beta examples discussed of Sec. \ref{sec:Applications}.
Simulations were performed with the numerical library ODEPACK \citep{Hindmarsh}.
The ODEPACK's solver used in all simulations was \emph{LSODA}, an
hybrid implementation of Adams and BDF methods \citep{Petzold}. 

\subsection{Gaussian case}

\begin{figure*}

\begin{centering}
\subfloat[$K=0.8$]{\begin{centering}
\includegraphics[width=0.33333\linewidth]{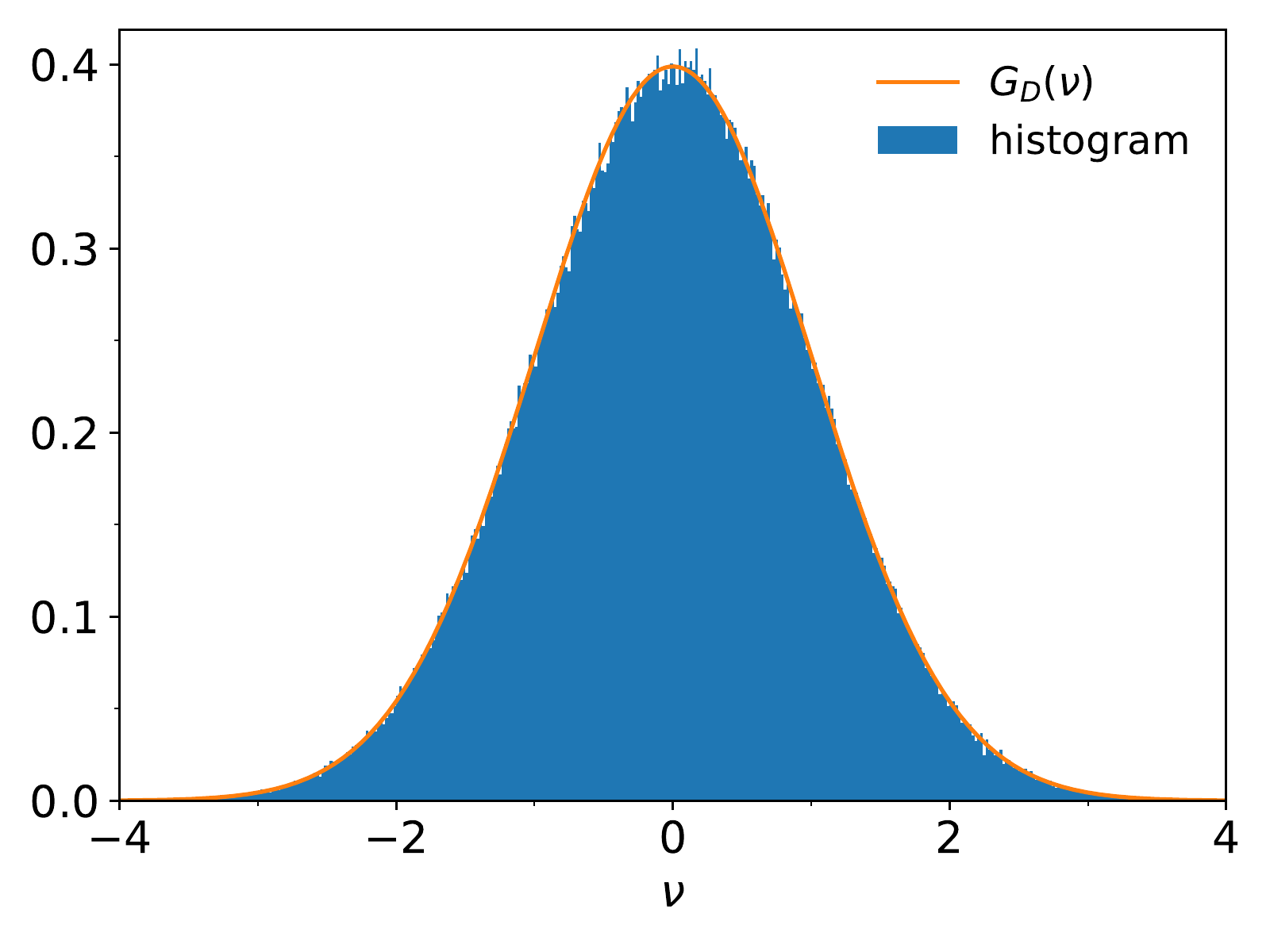}
\par\end{centering}
}\subfloat[$K=1.596$]{\begin{centering}
\includegraphics[width=0.33333\linewidth]{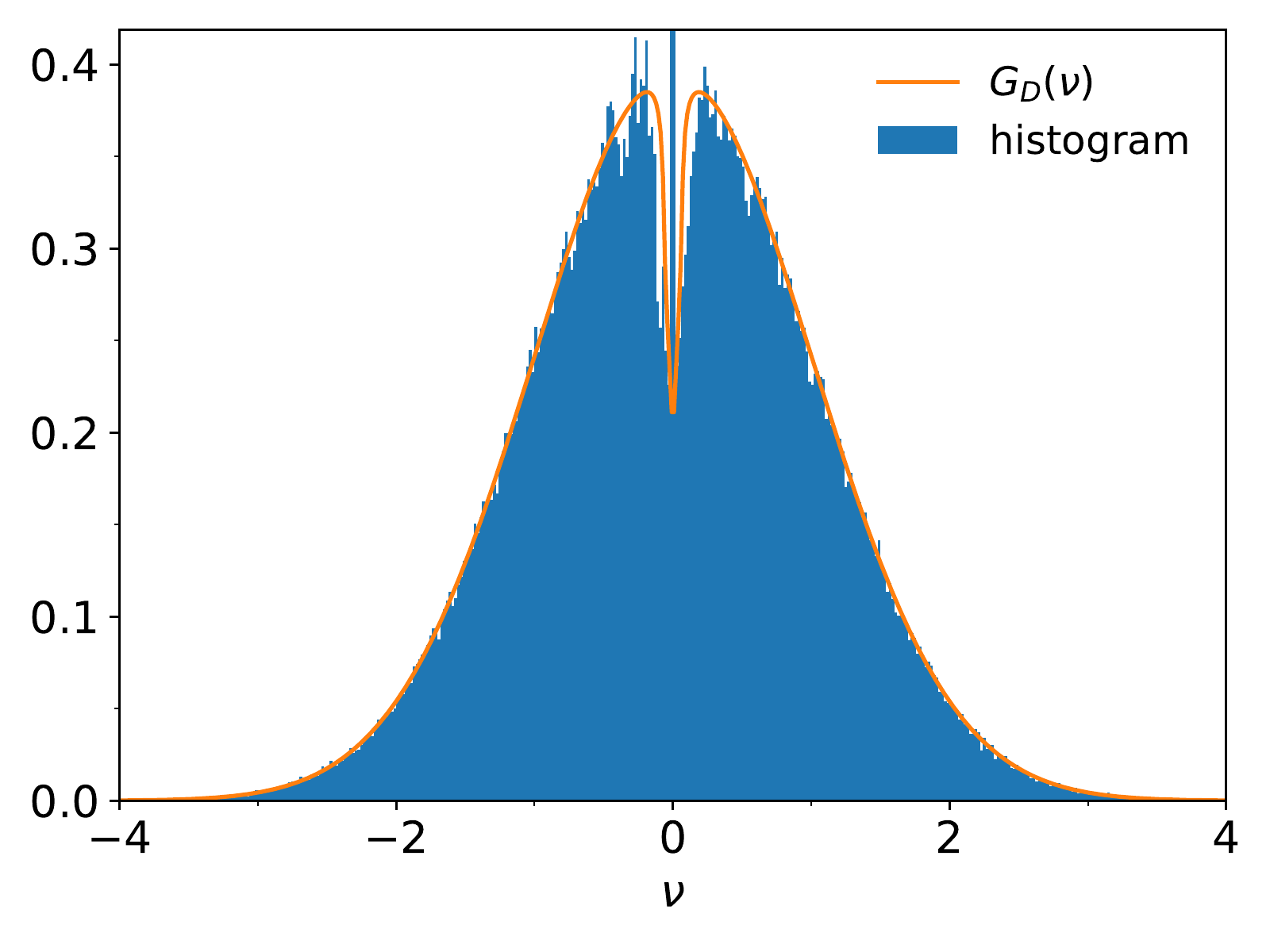}
\par\end{centering}
}\subfloat[$K=1.61$]{\begin{centering}
\includegraphics[width=0.33333\linewidth]{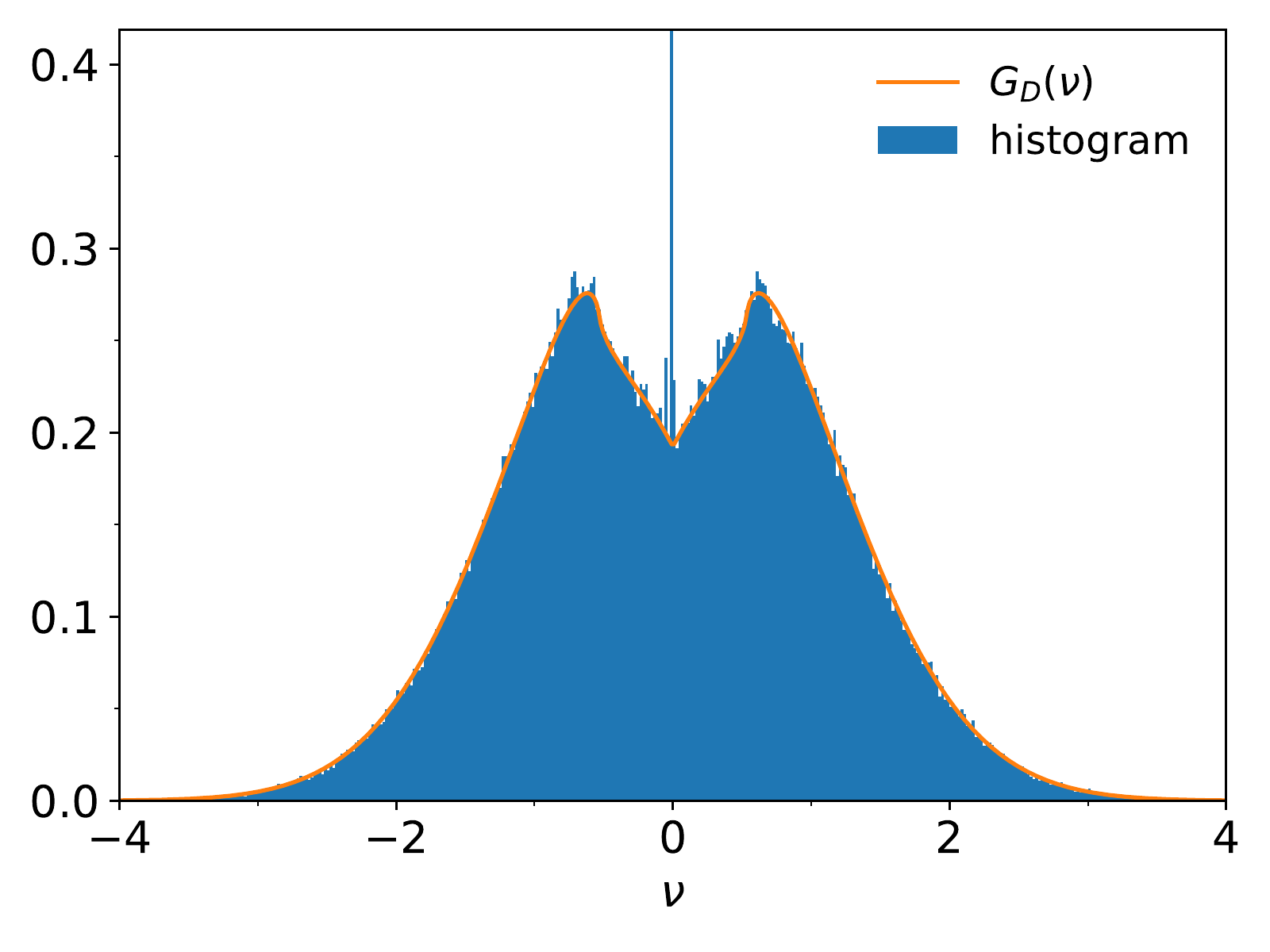}
\par\end{centering}
}
\par\end{centering}
\begin{centering}
\subfloat[$K=1.64$]{\centering{}\includegraphics[width=0.33333\linewidth]{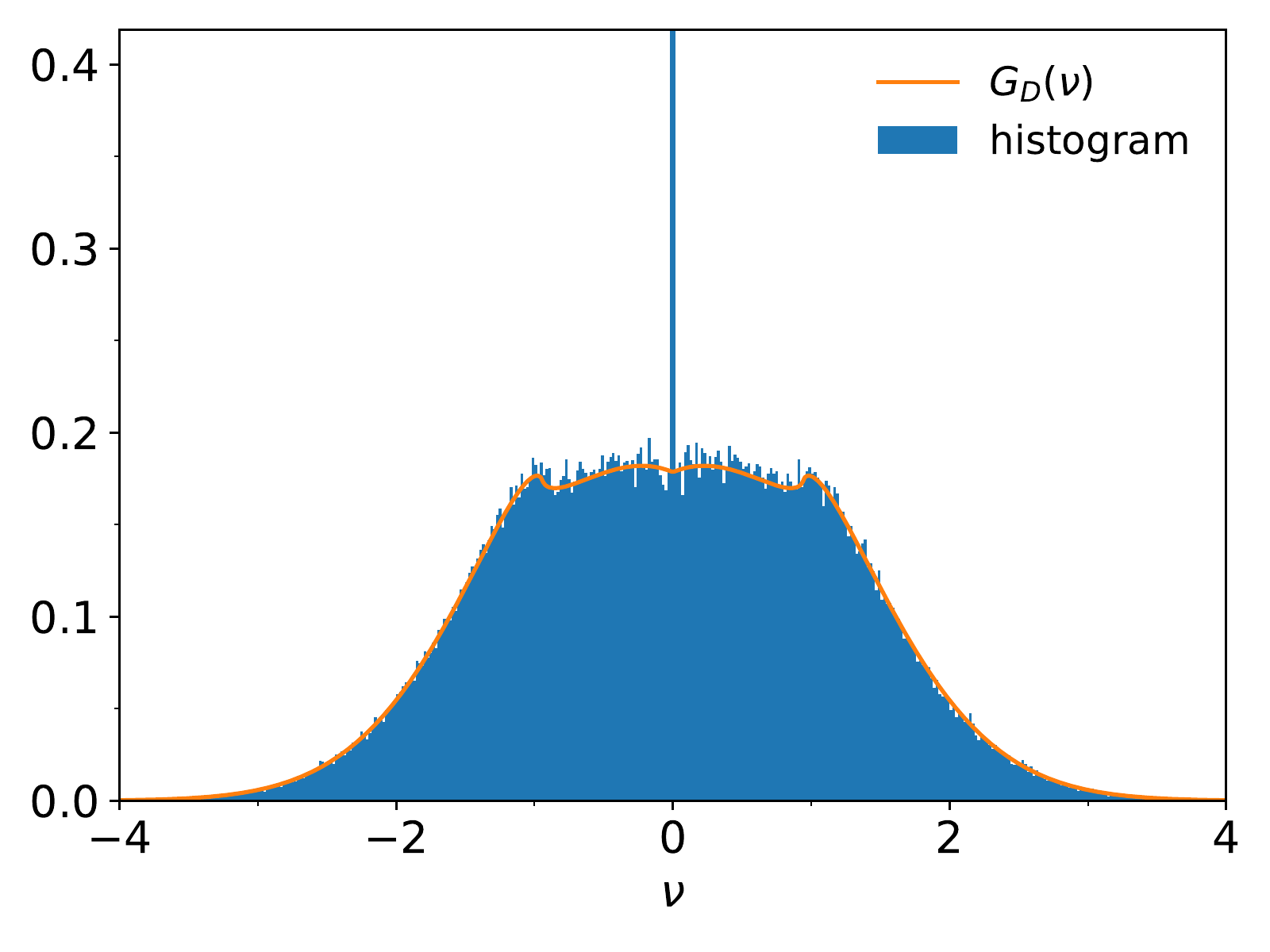}

}\subfloat[$K=1.67$]{\begin{centering}
\includegraphics[width=0.33333\linewidth]{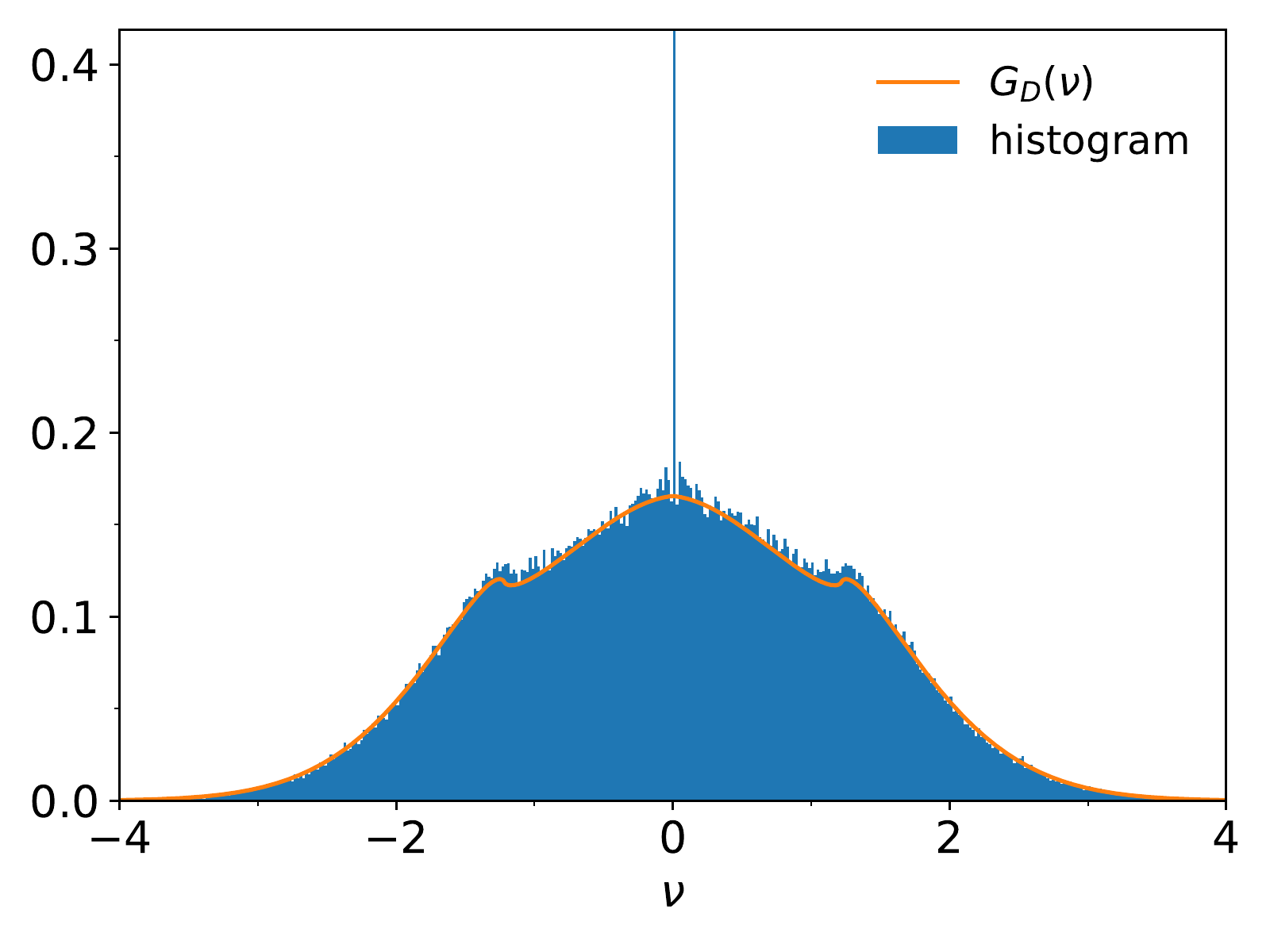}
\par\end{centering}
}\subfloat[$K=1.8$]{\centering{}\includegraphics[width=0.33333\linewidth]{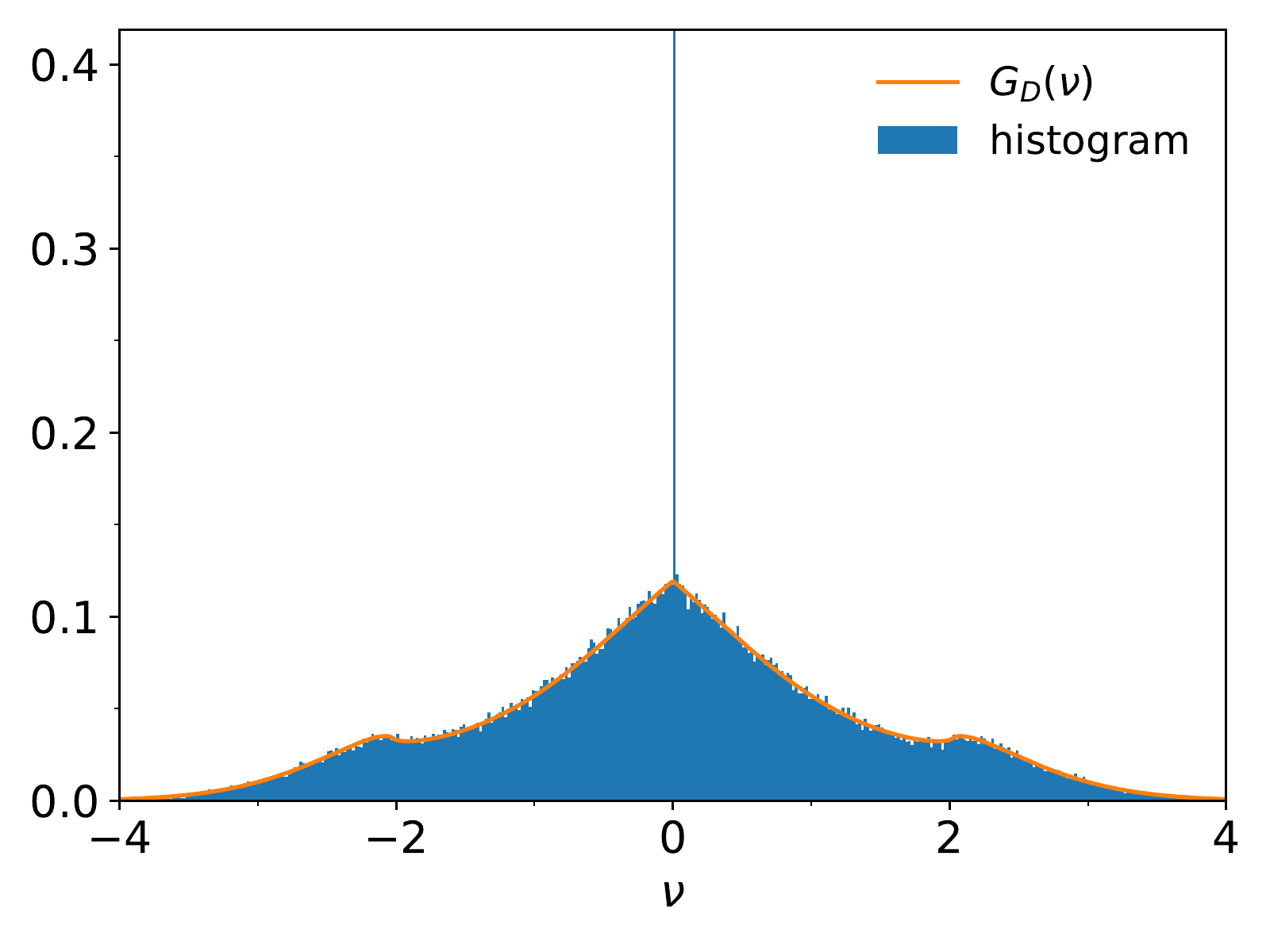}

}
\par\end{centering}

\caption{Comparison between normalized histograms of instantaneous frequencies
(in blue), obtained from numerical simulations of the Kuramoto model,
and the curve of $G_{D}$ (in red). In all simulation, $N=5\times10^{5}$.}
\label{fig:ThSim-Gauss}
\end{figure*}

In Figs. \ref{fig:ThSim-Gauss}(a)-(f), we compare graphs of $G_{D}$,
the same shown in Fig.~\ref{fig:G-Gauss-Beta}, to numerically-obtained normalized histograms
of the instantaneous frequencies. 

The histograms were obtained from numerical simulations of the Kuramoto
model with $N=5\times10^{5}$. In all simulations, the Kuramoto system
of equations is numerically integrated from a initial time 
$t_{0}=0$ to a final time  $t_{f}=5\times10^{2}$. Each histogram
is created from the set of instantaneous frequencies $\{\dot{\theta_{i}}(t_{f})\}_{i=1}^{N}$.
Simulations are performed considering random samples of
natural frequencies and initial phases. Initial phases are uniformly
sampled: a sample $\{\theta_{i}(t_{0})\}_{i=1}^{N}$ is generated
according to a uniform distribution in the interval between $0$ and
$2\pi$.

\begin{figure}
\begin{centering}
\includegraphics[width=\columnwidth]{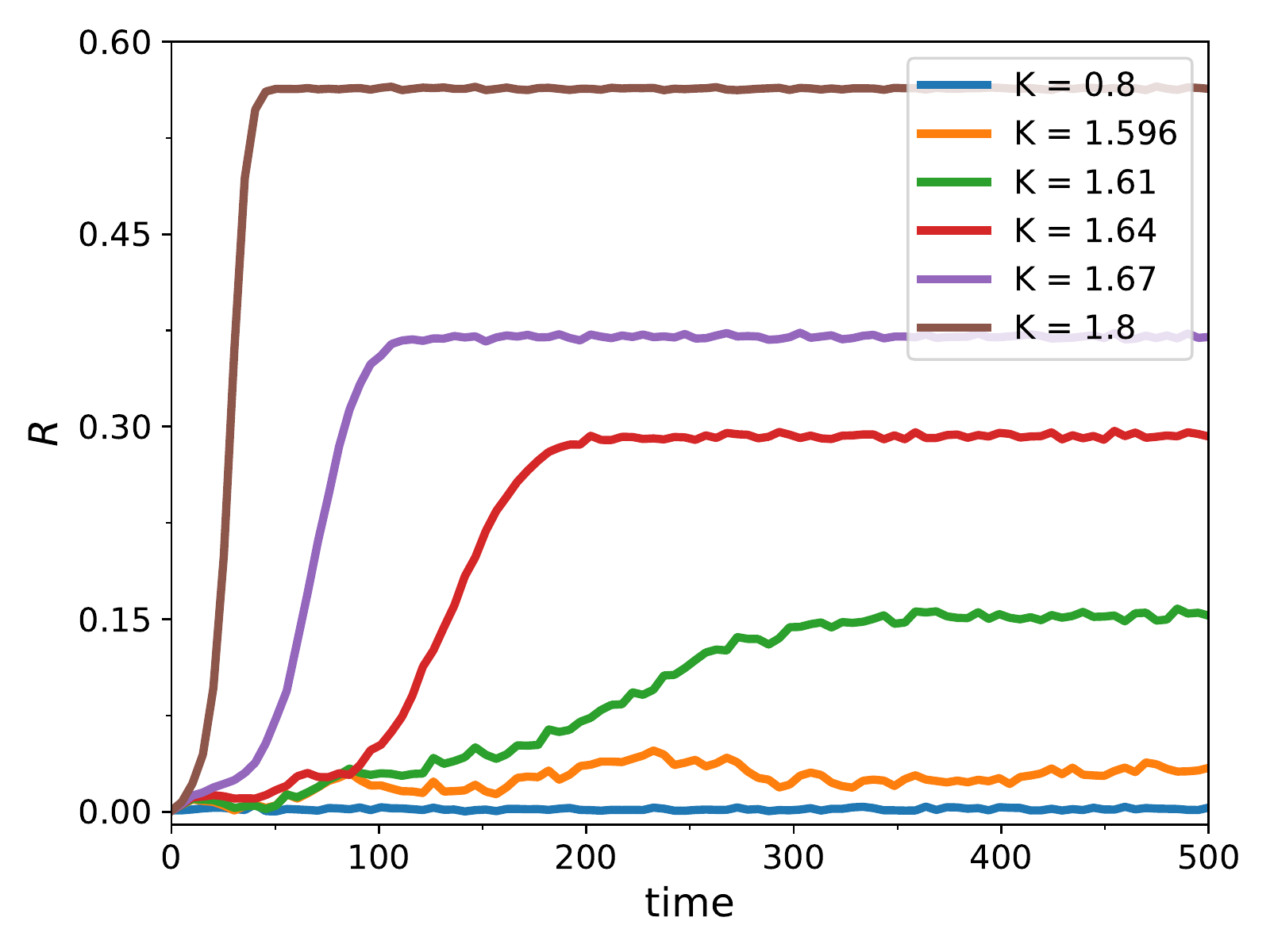}
\par\end{centering}
\caption{Time-dependence of the numerical order parameter. Except near the
transition, where $K=1.596$, low fluctuations are observed after
a sufficiently long time. Created with the same simulation near the
transition, histogram of Fig.~\ref{fig:ThSim-Gauss}(b) does not fit
properly the theoretical curve.}
\label{fig:sigma-t-gauss}
\end{figure}

Figure \ref{fig:sigma-t-gauss} shows the typical evolution of the 
 numerical order parameter. 
 When the order parameter exhibits small
fluctuations after a sufficiently long time, the corresponding histograms
are in good agreement with the analytical curves. However, stronger
order parameter fluctuations are observed near the transition ($K=1.596$),
and the histogram shown in Fig.~\ref{fig:ThSim-Gauss}(b), obtained
with the same value of $K$, does not fit properly the curve.

\begin{figure}
\begin{centering}
\subfloat[$N=5\times10^{5}$; $K=1.64$]{\begin{centering}
\includegraphics[width=1\linewidth]{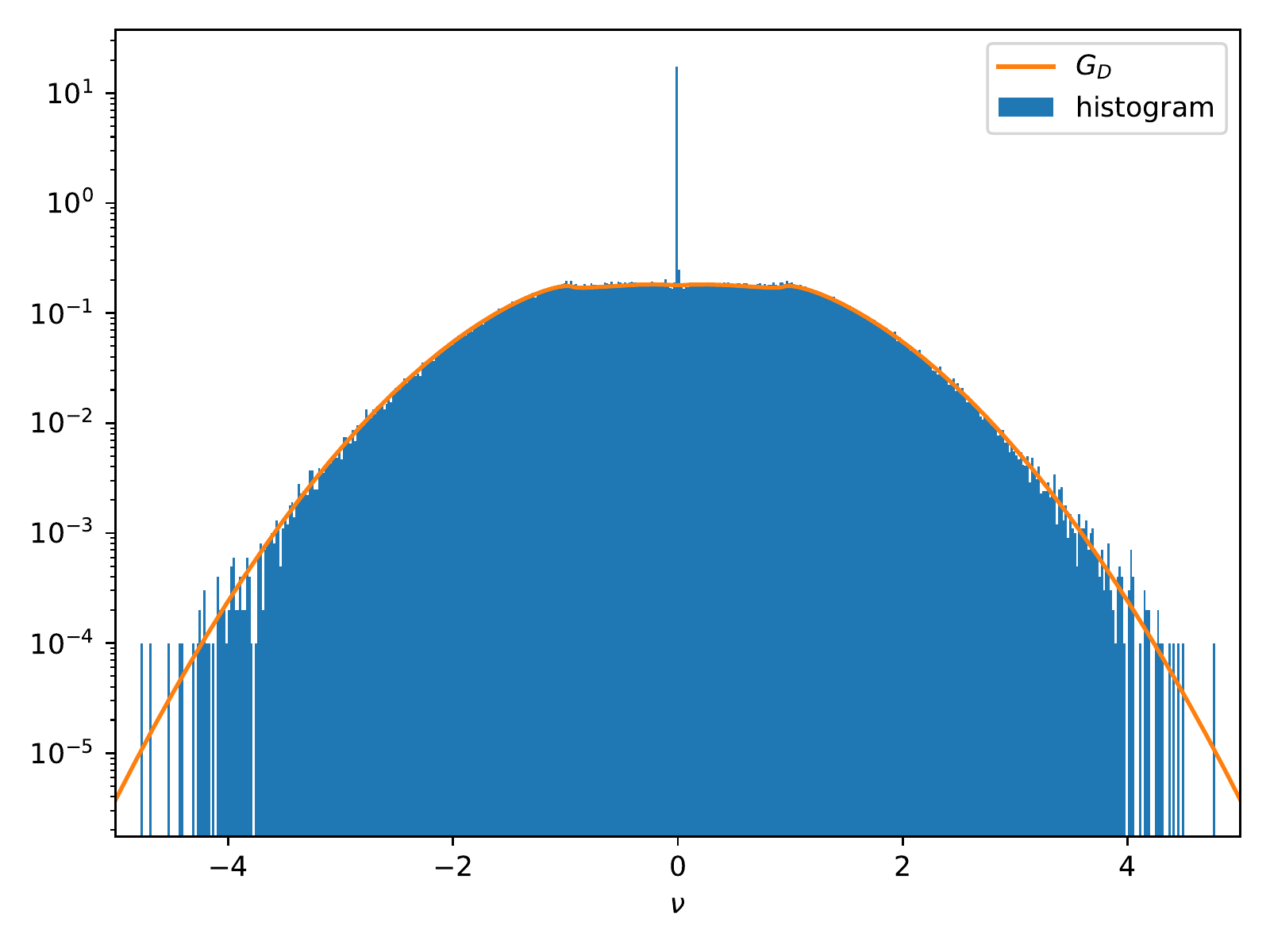}
\par\end{centering}
}
\par\end{centering}
\begin{centering}
\subfloat[$N=5\times10^{6};$ $K=1.64$]{\begin{centering}
\includegraphics[width=1\linewidth]{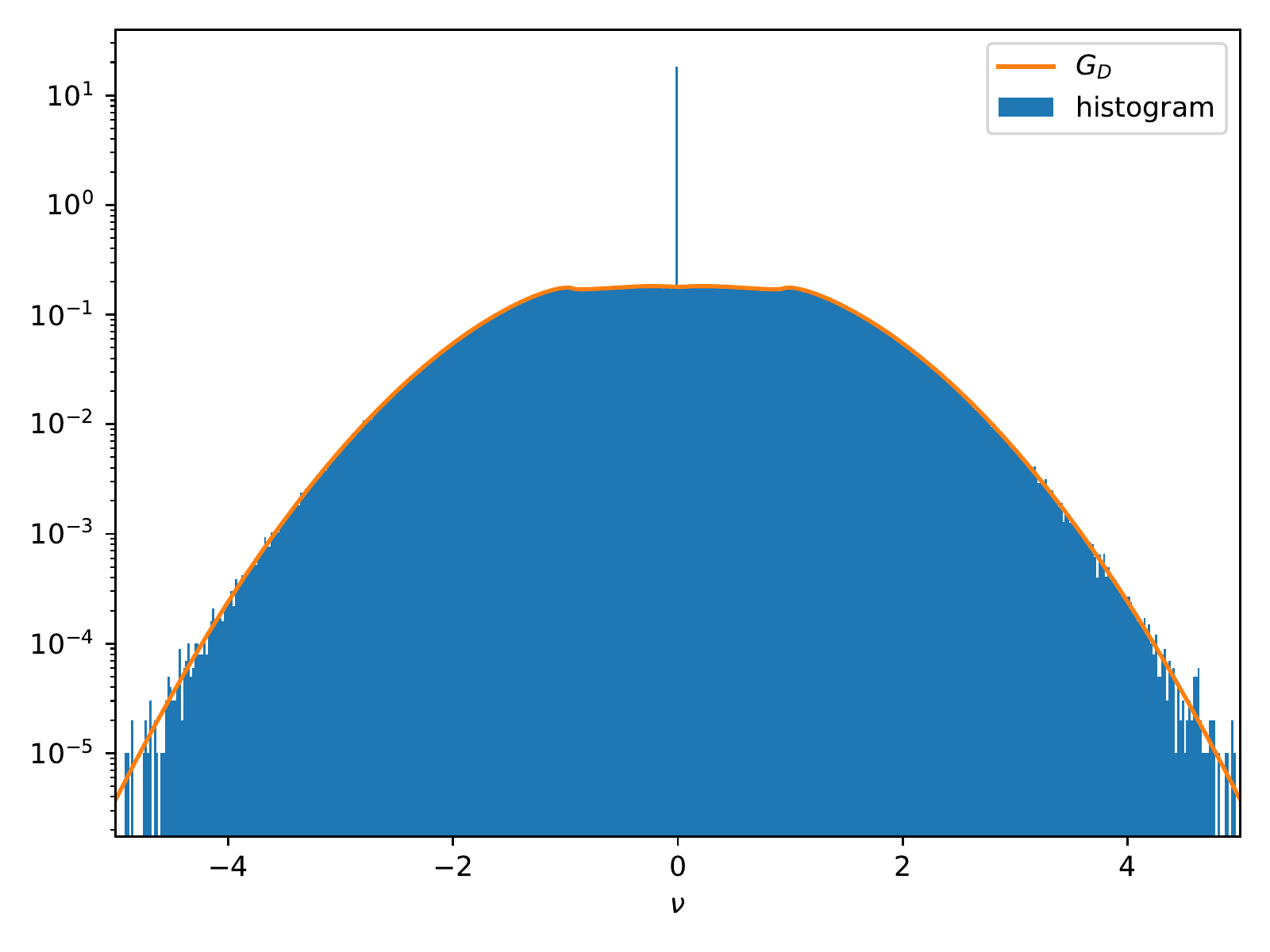}
\par\end{centering}
}
\par\end{centering}
\caption{Comparison (in logarithmic scale) between the graph of $G_{D}$ and
histograms of instantaneous frequencies. The tails of $G_{D}$ fit
better the histograms when network size increases from $N=5\times10^{5}$
to $N=5\times10^{6}$.}
\label{fig:ThSim-Gauss-Tails}
\end{figure}

Figures \ref{fig:ThSim-Gauss-Tails}(a) and \ref{fig:ThSim-Gauss-Tails}(b)
show, in logarithmic scale, the graph of $G_{D}$ and histograms of
instantaneous frequencies obtained from simulation data. 
The coupling strength has the same value used in Fig.~\ref{fig:ThSim-Gauss}(d),
$K=1.64$. The histograms are created with $N=5\times10^{5}$ (Fig.
\ref{fig:ThSim-Gauss-Tails}(a)) and $N=5\times10^{6}$ (Fig.~\ref{fig:ThSim-Gauss-Tails}(b)).
In both of them, the synchronization peak is clearly visible. Large
instantaneous frequency occurrences (rare events) are more difficult
to observe. However, by increasing the number of oscillators,
rare events are more common, and the tails of $G_{D}$ fit better
the histograms.

\subsection{Beta case}

\begin{figure*}

\begin{centering}
\subfloat[$K=0.2$]{\begin{centering}
\includegraphics[width=0.33333\linewidth]{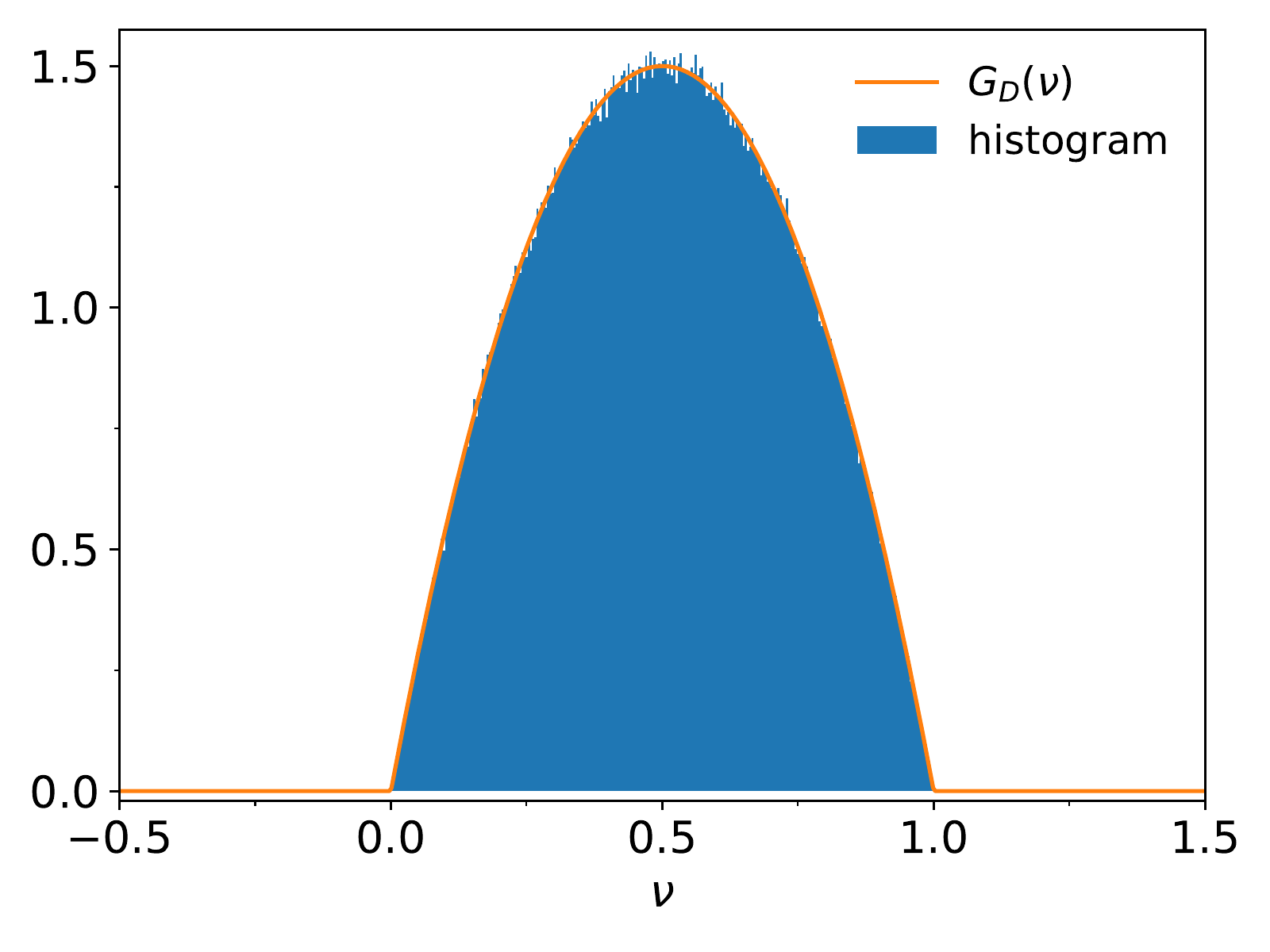}
\par\end{centering}
}\subfloat[$K=0.42442$]{\centering{}\includegraphics[width=0.33333\linewidth]{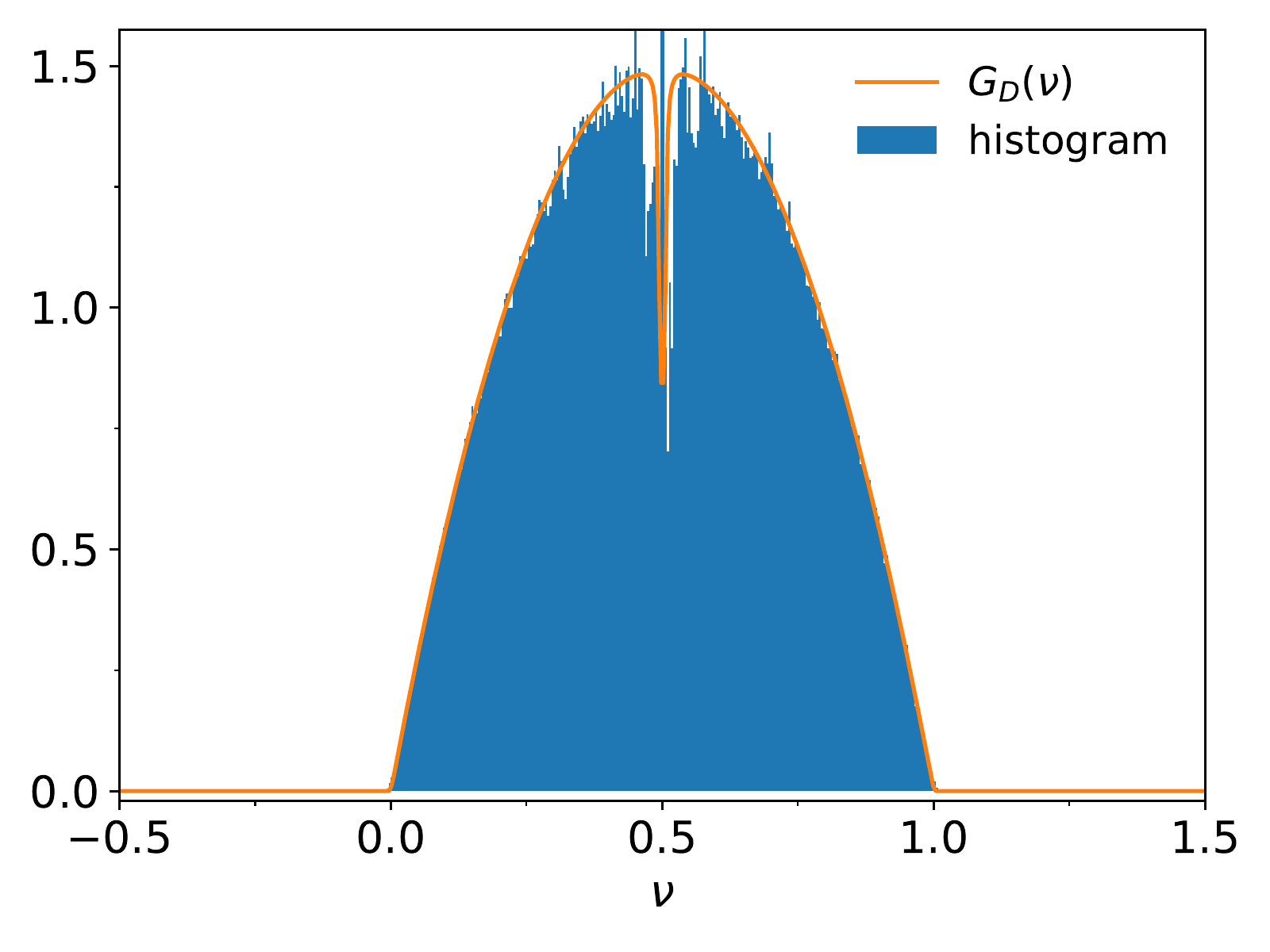}

}\subfloat[$K=0.43$]{\begin{centering}
\includegraphics[width=0.33333\linewidth]{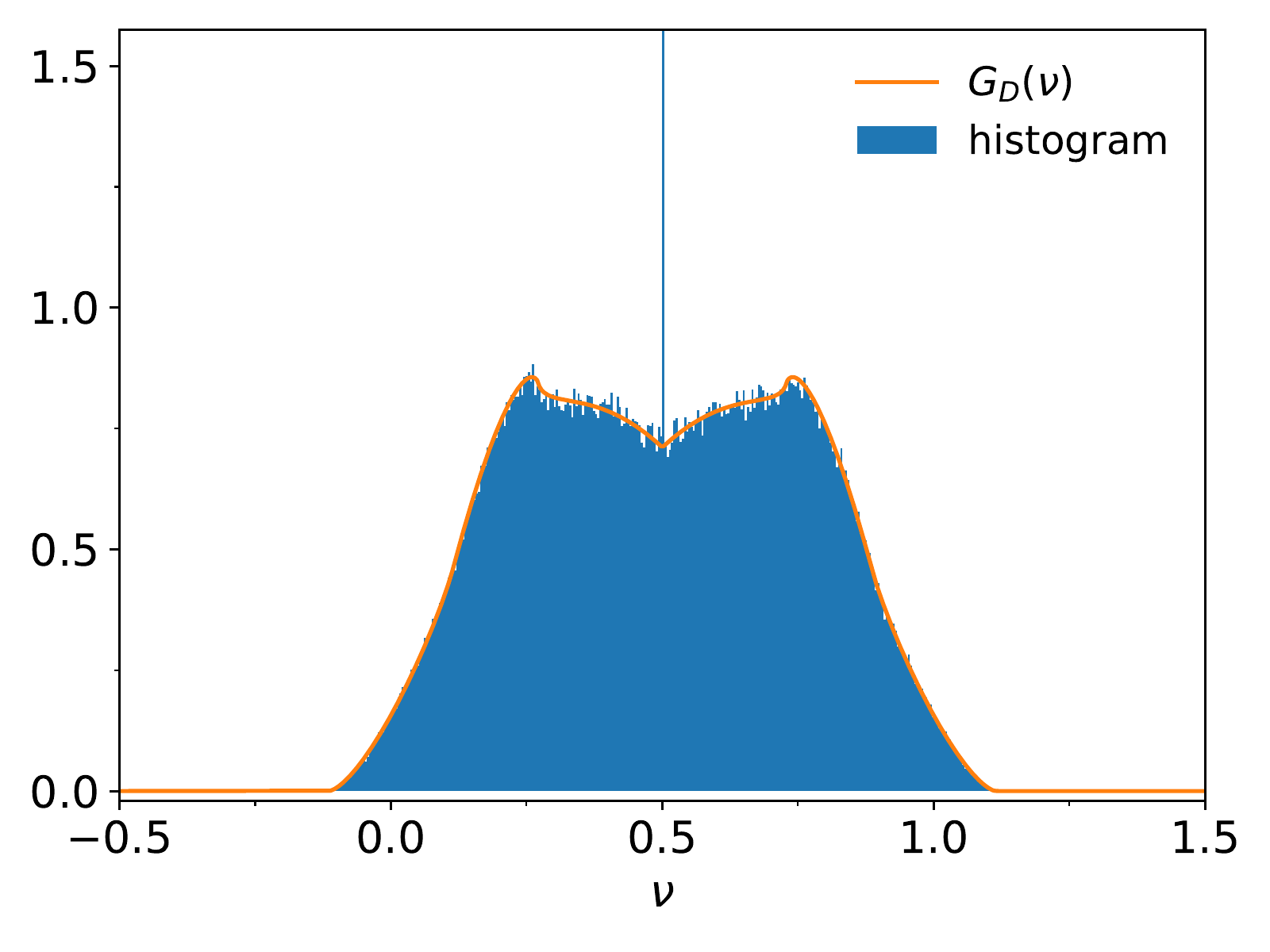}
\par\end{centering}
}
\par\end{centering}
\begin{centering}
\subfloat[$K=0.435$]{\centering{}\includegraphics[width=0.33333\linewidth]{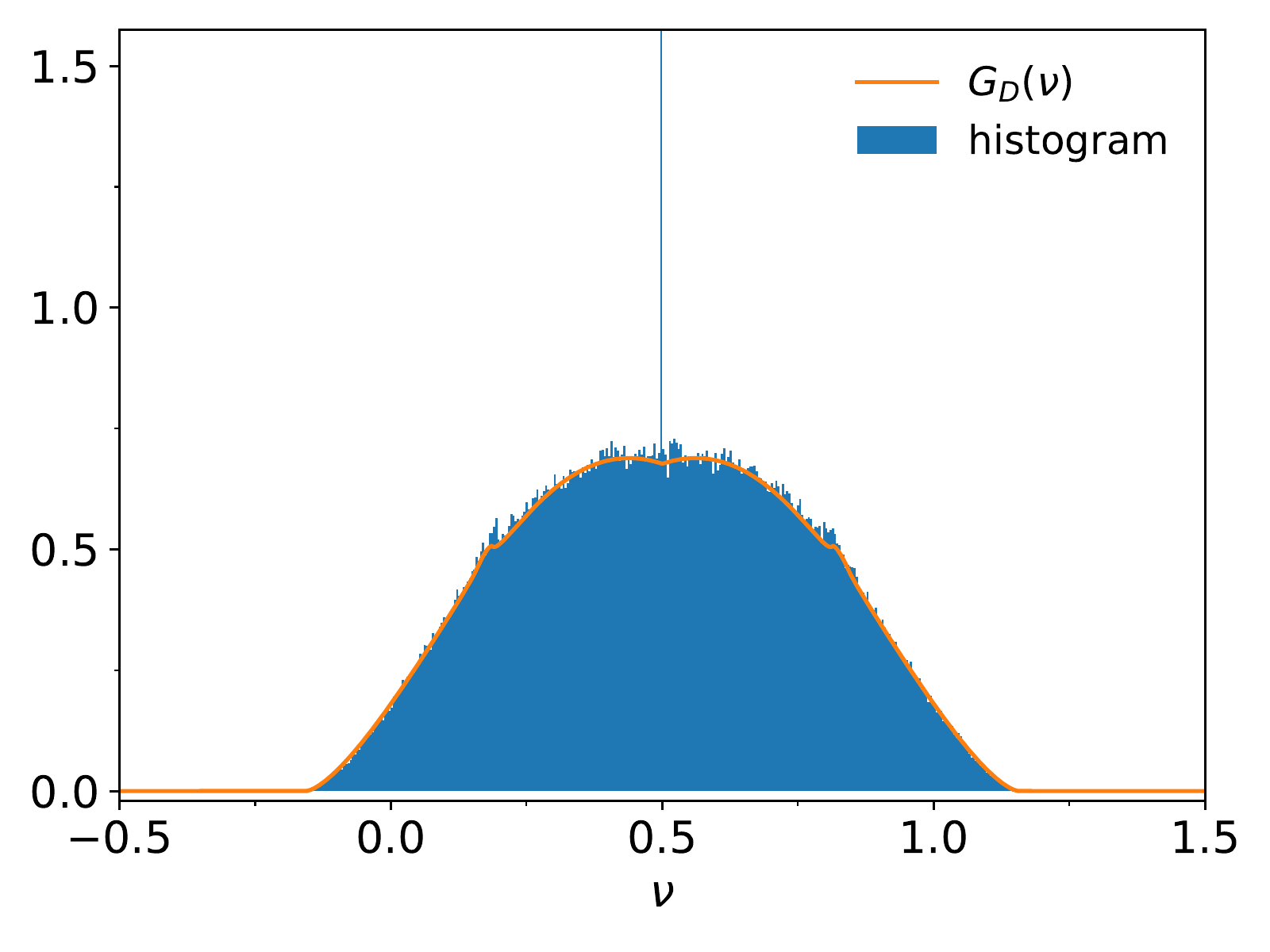}

}\subfloat[$K=0.45$]{\begin{centering}
\includegraphics[width=0.33333\linewidth]{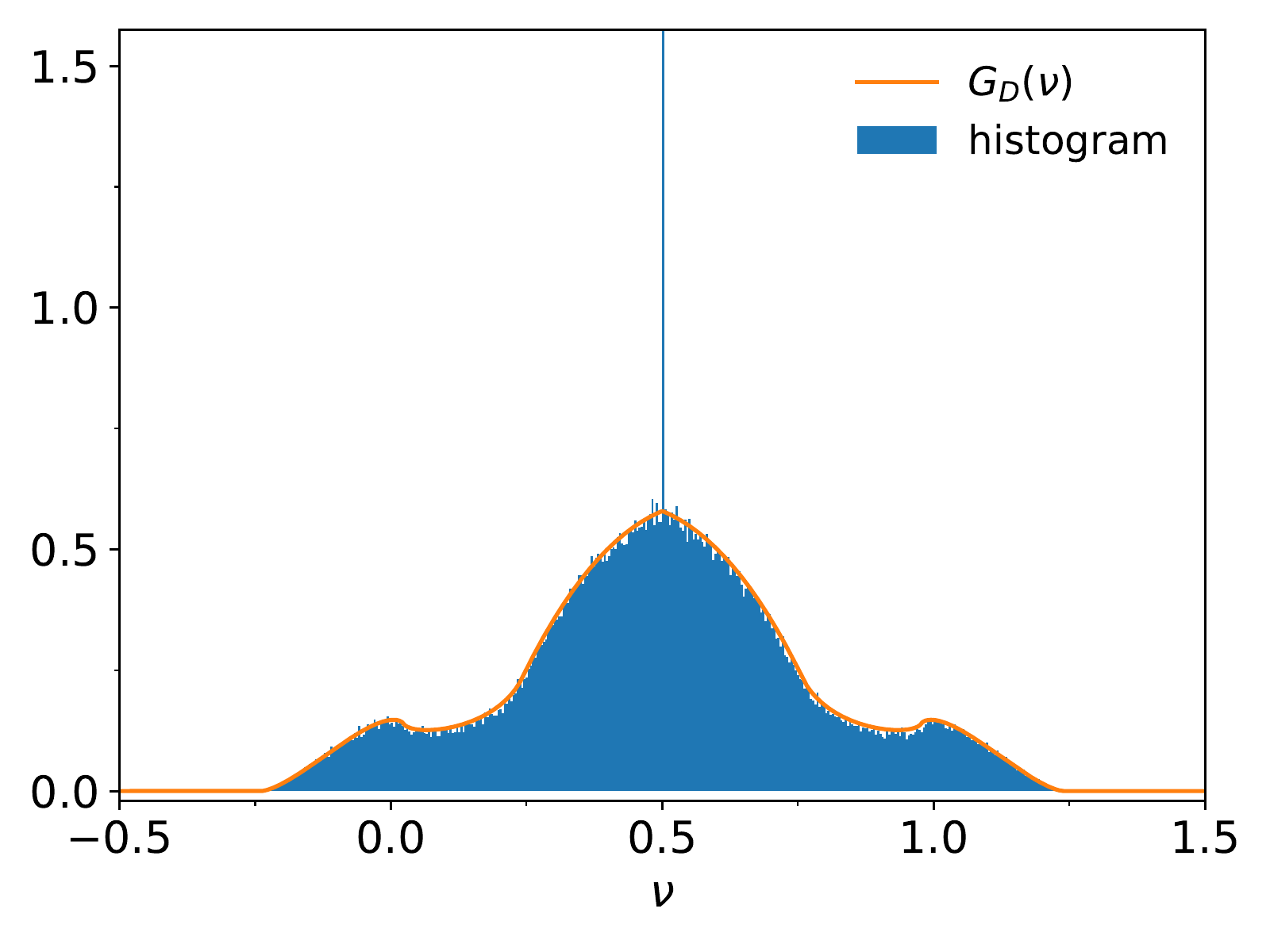}
\par\end{centering}
}\subfloat[$K=0.5$]{\centering{}\includegraphics[width=0.33333\linewidth]{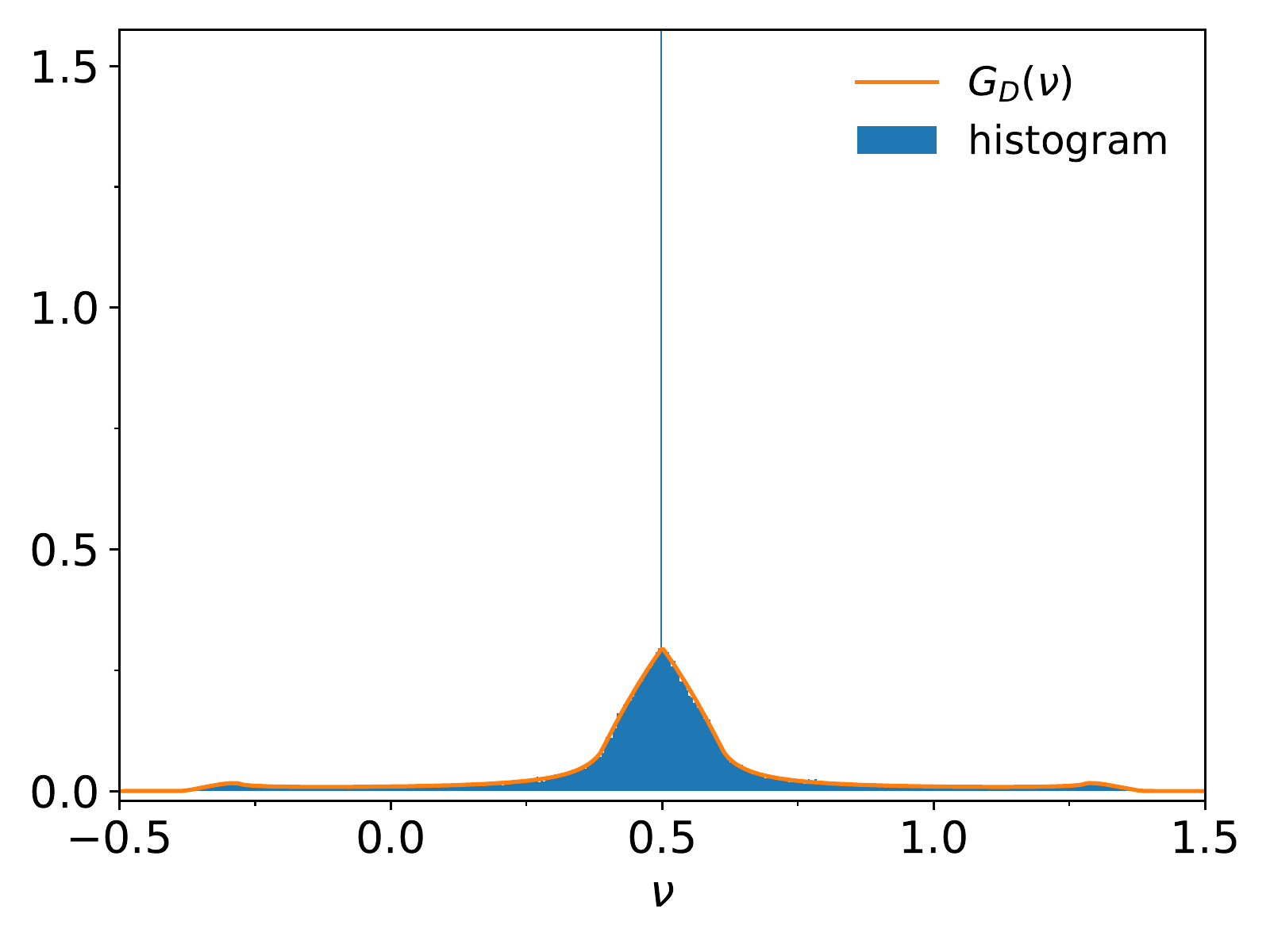}

}
\par\end{centering}

\caption{Comparison of $G_{D}$ graphs to instantaneous frequency histograms.
$g$ is a Beta$(2,2)$ distribution, and the histograms were obtained
from simulations of the Kuramoto model with $N=1\times10^{6}$. The
graphs provide better fits than near the transition ($K=0.42442$).}
\label{fig:ThSim-Beta}
\end{figure*}

\begin{figure}
\begin{centering}
\includegraphics[width=\columnwidth]{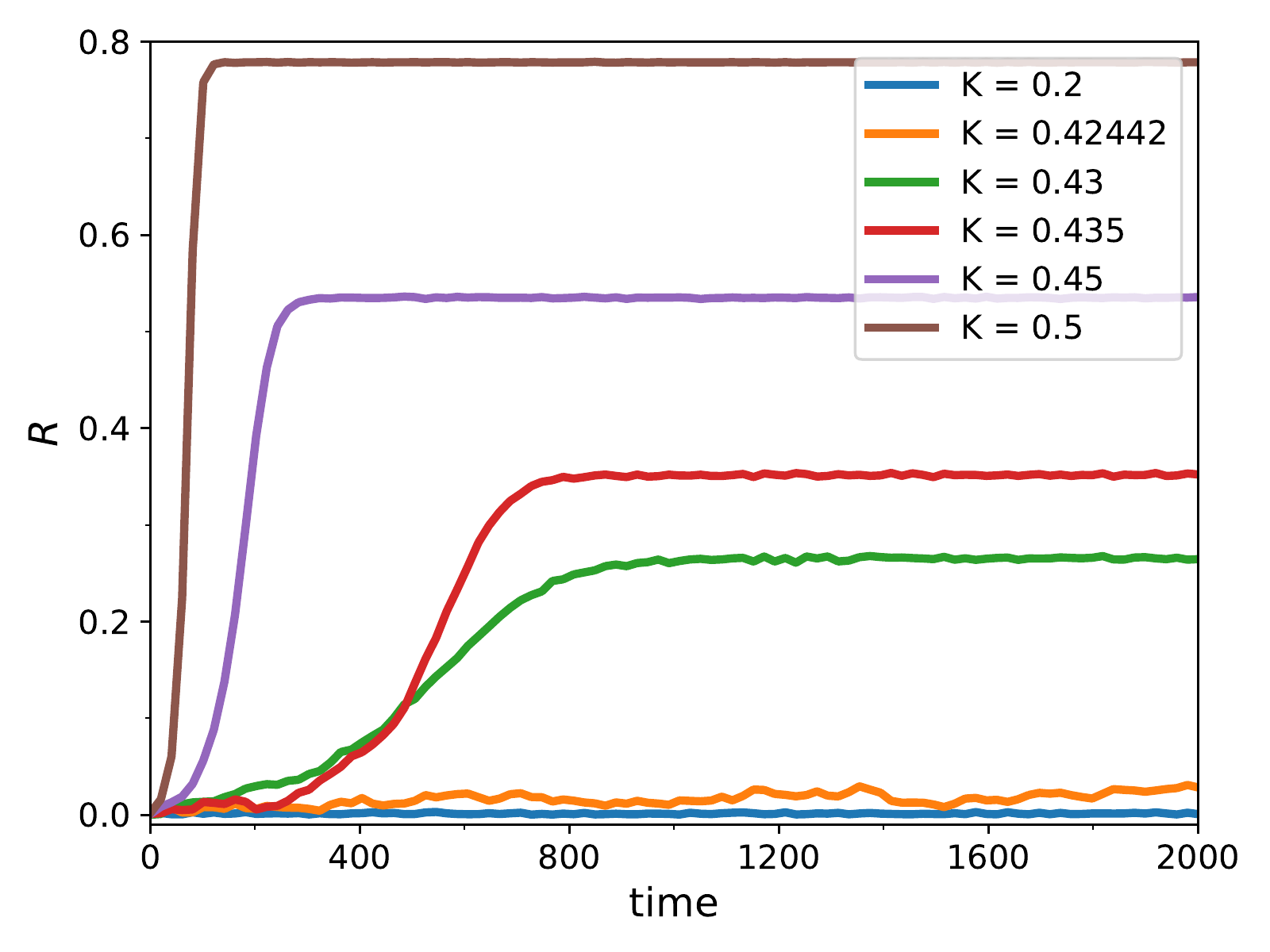}
\par\end{centering}
\caption{Time-dependence of the numerical order parameter: low fluctuations
after long time, except near the synchronization transition.}
\label{fig:sigma-t-beta}
\end{figure}

In Figs. \ref{fig:ThSim-Beta}(a)-(f), we compare instantaneous frequency
histograms to graphs of $G_{D}$ with $g$ defined as the Beta$(2,2)$ distribution,
given by Eq.~(\ref{eq:Beta(2,2)-dist}).
Simulations were performed with $N=1\times10^{6}$ oscillators, and integration
time was $2\times10^{3}$. Histograms
were created by using the instantaneous frequency set $\{\dot{\theta_{i}}(t_{f})\}_{i=1}^{N}$.
Corresponding time series of the numerical order parameter are shown in Fig.~\ref{fig:sigma-t-beta}.
Similarly to the case of normally-distributed natural frequencies, stronger
order-parameter fluctuations are observed near the transition ($K=0.42442$), away from
which our analytical result describes better the histograms.

\begin{figure}
\begin{centering}
\subfloat[$N=10^{6}$; $K=0.435$]{\begin{centering}
\includegraphics[width=1.0\linewidth]{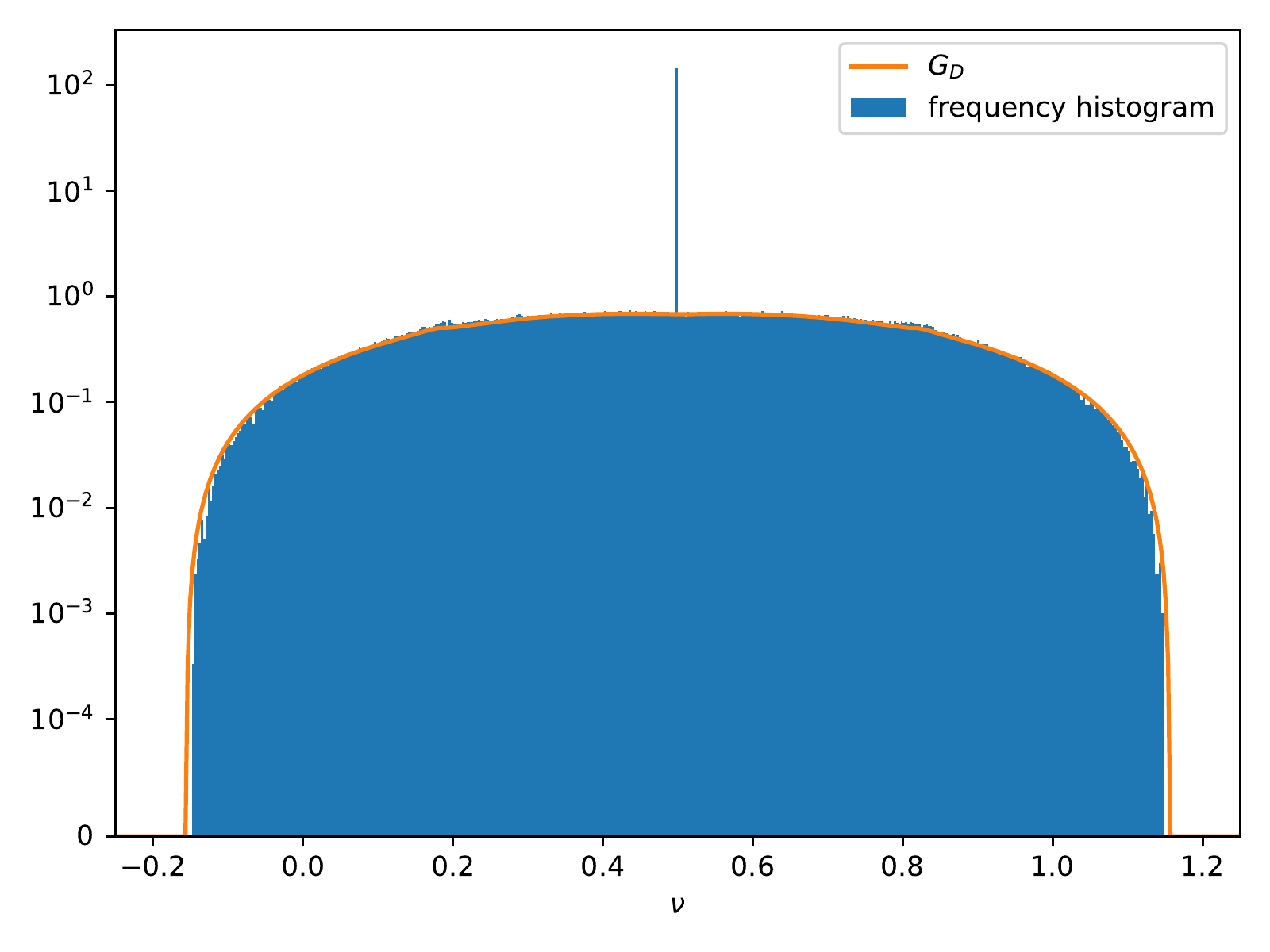}
\par\end{centering}
}
\par\end{centering}
\begin{centering}
\subfloat[$N=2\times10^{6}$; $K=0.435$]{\begin{centering}
\includegraphics[width=1.0\linewidth]{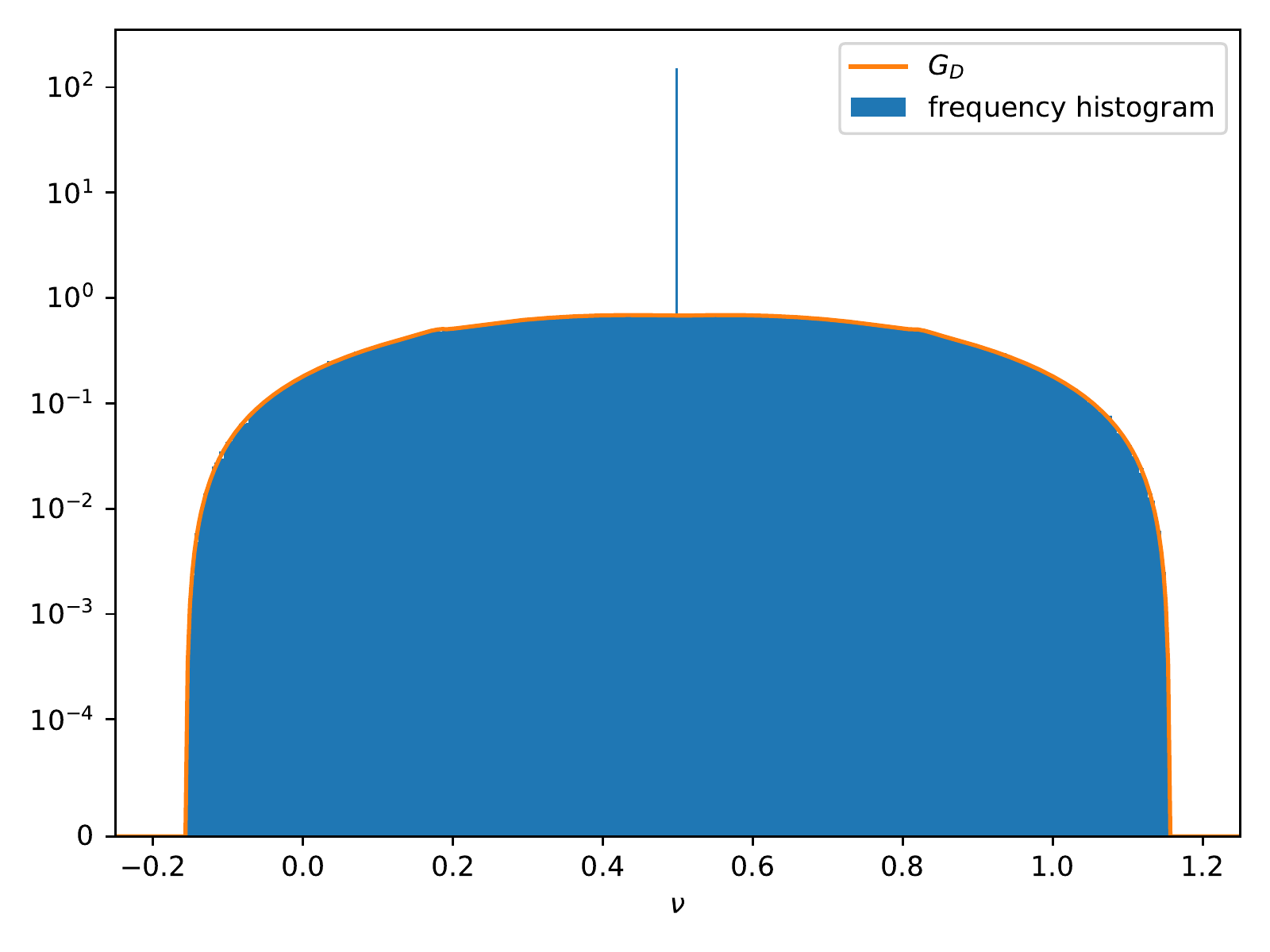}
\par\end{centering}
}
\par\end{centering}
\caption{Graph of $G_{D}$ and instantaneous frequency histograms. For the
vertical axis, we use logarithmic scale for values greater than $10^{-4}$
and linear scale those between $0$ and $10^{-4}$. By increasing
the number of oscillators, a better fit is attainable in the tails.}
\label{fig:ThSim-Beta-Tails}
\end{figure}

Behavior near the tails is shown in Fig.~\ref{fig:ThSim-Beta-Tails}, where
we compare again $G_{D}$ to instantaneous frequency histograms.
 The vertical axis has logarithmic scale for values greater than $10^{-4}$
and linear scale between $0$ and $10^{-4}$. We use 
the same coupling strength as in Fig.~\ref{fig:ThSim-Beta}(d). As expected:
i) rare events cannot be easily observed in the histograms; ii) by
increasing the number of oscillators from $N=1\times10^{6}$ to $N=2\times10^{6}$,
these events are more common, and a better fit is attainable in the
tails.

\end{document}